\documentclass [prb,preprint,superscriptaddress,floatfix,amsmath,amssymb] {revtex4-1}
\usepackage{amsmath}
\usepackage{graphicx}
\usepackage{hyperref}
\usepackage{color}
\usepackage{amssymb}
\usepackage{bm}
\newcommand{\beq}{\begin{equation}}
\newcommand{\eeq}{\end{equation}}
\newcommand{\bea}{\begin{eqnarray}}
\newcommand{\eea}{\end{eqnarray}}
\newcommand{\be}{\begin{equation}}
\newcommand{\ee}{\end{equation}}

\renewcommand{\[}{\left[}
\renewcommand{\]}{\right]}

\definecolor{darkgreen}{RGB}{0,170,0}

\begin{document}
\title {Interplay between superconductivity and non-Fermi liquid  at a quantum-critical point in a metal.\\
IV: The $\gamma$ model and its phase diagram at  $1<\gamma <2$.}
\author{Yi-Ming Wu}
\affiliation{School of Physics and Astronomy and William I. Fine Theoretical Physics Institute,
University of Minnesota, Minneapolis, MN 55455, USA}
\author{Shang-Shun Zhang}
\affiliation{School of Physics and Astronomy and William I. Fine Theoretical Physics Institute,
University of Minnesota, Minneapolis, MN 55455, USA}
\author{Artem Abanov}
\affiliation{Department of Physics, Texas A\&M University, College Station,  USA}
\author{Andrey V. Chubukov}
\affiliation{School of Physics and Astronomy and William I. Fine Theoretical Physics Institute,
University of Minnesota, Minneapolis, MN 55455, USA}
\date{\today}
\begin{abstract}
In this paper we continue with our analysis of the interplay between the pairing and the non-Fermi liquid behavior
  in a metal for a set of quantum-critical (QC) systems with an effective dynamical electron-electron interaction
 $V(\Omega_m) \propto 1/|\Omega_m|^\gamma$, mediated by a critical massless boson (the $\gamma$-model).
 In previous papers we considered the cases $0<\gamma <1$ and $\gamma \approx 1$. We  argued that
   the pairing by a gapless boson is fundamentally different from BCS/Eliashberg pairing by a massive boson
     as for the former there exists not one but an infinite discrete set of topologically distinct solutions for the gap function $\Delta_n (\omega_m)$ at $T=0$ ($n=0,1,2...$), each with its own condensation energy $E_{c,n}$.
   Here we extend the analysis to larger $1< \gamma <2$.  We argue that the discrete set of solutions survives, and  the spectrum of $E_{c,n}$ get progressively denser as $\gamma$ increases towards $2$ and eventually becomes continuous at $\gamma \to 2$.  This increases the strength  of
    "longitudinal" gap fluctuations, which tend to reduce the actual superconducting $T_c$ compared to the onset temperature for the pairing and give rise to a pseudogap region of preformed pairs. We also  detect two features on the real axis, which develop
     at $\gamma >1$  and also become critical at $\gamma \to 2$.  First, the density of states evolves towards a set of discrete $\delta-$functions. Second, an  array of dynamical vortices emerges in the upper frequency half-plane, near the real axis.   We argue that these two features come about because on a real axis, the real part of the dynamical electron-electron interaction, $V' (\Omega) \propto \cos(\pi \gamma/2)/|\Omega|^\gamma$,  becomes repulsive for $\gamma >1$, and  the imaginary $V^{''} (\Omega) \propto \sin(\pi \gamma/2)/|\Omega|^\gamma$, gets progressively smaller at $\gamma \to 2$.
      We speculate that the features on the real axis
     are consistent with the development of a continuum spectrum of the condensation energy, for which we used $\Delta_n (\omega_m)$ on the Matsubara axis.  We consider the case $\gamma =2$ separately in the next paper.
   \end{abstract}
\maketitle

\section{ Introduction.}

In this paper, we continue our analysis of the competition between non-Fermi liquid (NFL) physics and superconductivity (SC) near a quantum-critical point (QCP) of an itinerant correlated electron system,
whose low-energy properties are described by an effective, momentum-integrated
  dynamical four-fermion interaction $V(\Omega_m)$.   We consider a set of quantum-critical systems, for which the effective interaction has the form
 $V (\Omega_m) = {\bar g}^\gamma/|\Omega_m|^\gamma$ (the $\gamma$-model).
This interaction gives rise to NFL behavior in the normal state,
   with
$\Sigma (\omega_m) \propto \omega^{1-\gamma}_m$, and at the same time mediates an attraction in at least one pairing channel, specific to the underlying microscopic model.  The two tendencies compete with each other as
  a NFL self-energy reduces the pairing strength, while the feedback from the pairing reduces fermionic self-energy.
In
   Ref. \cite{paper_1} we listed quantum-critical systems, whose low-energy physics is described by the $\gamma$-model with different $\gamma$ and presented a long list of references to earlier publications on this
 subject. In that
and subsequent papers, Refs. \cite{paper_2,paper_3}, hereafter referred to as Papers I-III,
     respectively, our group  analyzed the behavior of the $\gamma$-model for $0 <\gamma <1$ (at
$T=0$ in Paper I and at a finite $T$ in Paper II), and at around $\gamma =1$ (in Paper III).
    In Paper I, we found that the
 system does become unstable towards pairing, i.e., in the ground state
 the  pairing gap $\Delta (\omega_m)$ is non-zero. However, in a qualitative distinction with BCS/Eliashberg theory of a pairing by a massive boson,  we found an infinite discrete set of solutions for the gap function, $\Delta_n (\omega_m)$, specified by an integer $n =0,1,2...$.  All solutions have the same spatial gap symmetry, but are topologically distinct in the sense that $\Delta_n (\omega_m)$ changes sign $n$ times as a function of Matsubara frequency (we discuss topological aspects in this paper). A gap function $\Delta_n (\omega_m)$ remains approximately $\Delta_n (0)$
       at $\omega_m \leq \Delta_n (0)$, oscillates $n$ times at $\Delta_n (0) < \omega_m < {\bar g}$, and at larger frequencies decays as $1/|\omega_m|^\gamma$.
  The magnitude of $\Delta_n (0)$ decreases with $n$ and at large  $n$ scales as
   $\Delta_n (0) \propto e^{-A n}$, where $A$ is a function of $\gamma$.
    In the limit $n \to \infty$, $\Delta_\infty$ is infinitesimally small and is the solution of the linearized gap equation.
      In Paper II, we found that $\Delta_n (\omega_m)$ emerge at different onset temperatures
       $T_{p,n}$.   The onset temperature and the condensation energy at $T=0$, $E_{c,n}$, is the largest for $n=0$,  yet the  existence of an infinite set of minima of the free energy  is a
        qualitatively new feature of the pairing at a QCP.
         In Paper III, we argued that the system behavior is continuous through $\gamma =1$, despite that the fermionic self-energy and the pairing vertex diverge at $T=0$ for $\gamma \geq 1$. The divergencies cancel out in the gap equation in a way similar to cancellation of contributions from non-magnetic impurities in a BCS superconductor.  As the consequence, the
           gap functions $\Delta_n (\omega_m)$ and  the
          condensation energies $E_{c,n}$   still form a discrete set for $\gamma \geq 1$, in which $E_{c,0}$ is the largest.
        \footnote{This does not hold for an extended $\gamma-$model with different interaction strength  in the particle-hole and the particle-particle channels.  In the latter case $E_{c,n}$  form a discrete set for $\gamma <1$, but the set becomes dense at $\gamma \to 1$ and transforms into a  continuous one
           for $\gamma \geq 1$.  This gives rise to a gapless branch of longitudinal gap fluctuations.}

     In this paper  we study the $\gamma$ model for $\gamma$ in the range $1<\gamma <2$.  Our goal is to understand whether the presence of an infinite number of gap functions has physical consequences.
      The most natural conjecture is that the existence of the set gives rise to a new branch of "longitudinal" gap  fluctuations, what increases the strength of phase fluctuations and may create a
       pseudogap, preformed pair  region between the onset temperature for the pairing and the actual  superconducting $T_c$.  However, this is by no means guaranteed first of all because each $E_{c,n}$ is proportional to the total number of particles in the system, and potential barriers between minima with different $n$ are infinitely high.   Strong fluctuating effects are only possible if the spectrum of  condensation energies  gets  dense and eventually becomes a continuous one.

       We argue that this happens when $\gamma$ is larger than one. We show that the spectrum of $E_{c,n}$
        gets denser  when $\gamma$ increases and for $\gamma \leq 2$
       splits into two sub-sets, one for $n  < n_{cr}$ and another for $n > n_{cr}$, where $n_{cr} \sim 1/(2-\gamma)$.  Condensation energies for $n < n_{cr}$ get progressively closer   to $E_{c,0}$ as $\gamma$ approaches $2$, while for $n > n_{cr}$, $E_{c,n}$  depends on the ratio of two large numbers $n$ and $n_{cr}$, which is a near-continuous quantity.  In the limit $\gamma \to 2$, $n_{cr}$ tends to infinity, and $E_{c,n}$ for all finite $n$ collapse into a single $E_{c,0}$, while  for  $n \to \infty$, the condensation energy
         becomes a continuous function of $n/n_{cr}$, whose value is determined by how the double limit $n \to \infty$ and $n_{cr} \to \infty$  (i.e. $\gamma \to 2$) is taken.
         This creates a continuous gapless spectrum of longitudinal fluctuations.

 We present  the numerical evidence for the collapse of the condensation energy obtained by analysing how  the functional form of the
      eigenfunctions $\Delta_n (\omega_m)$ with different finite $n$  evolve with $\gamma$   We argue that
   the frequency range, where $\Delta_n (\omega_m)$ changes sign $n$ times, shrinks, and moves to progressively smaller frequencies at $\gamma \to 2$, consistent our assertion that in this limit $E_{c,n}$ with finite $n$ tend to $E_{c,0}$.

 The transformation of the spectrum of $\Delta_n (\omega_m)$ and $E_{c,n}$ from a discrete to a continuous one is a rather non-trivial phenomenon, and we study it from different corners.  We find corroborative evidence by analyzing the gap function on the Matsubara axis.
We then analyze how the gap function evolves on the real axis, and in the upper half-plane of
  frequency. For a generic complex $z = \omega'+i \omega^{''}$ ($\omega^{''} >0$),   $\Delta_n (z)$ is a complex function:  $\Delta_n (z) = \Delta'_n (z) + i\Delta^{''}_n (z)$, and  can define its phase as $\eta_n (z) ={\text{Im}} \log{\Delta_n (z)}$.  Our reasoning to focus on the real axis for $\gamma$ between one and two
     stems from the fact that the interaction there, $V(\Omega) \propto  e^{i \pi \gamma/2}/|\omega|^\gamma$, is complex, and its real part,
       $V^{'}(\Omega) = \cos (\pi \gamma/2)/|\omega|^\gamma$, becomes repulsive for $\gamma >1$, while
              the imaginary part   $V^{''} (\Omega) = \sin(\pi \gamma/2)/|\omega|^\gamma$ vanishes at $\gamma \to 2$. Meanwhile, on the Matsubara axis, $V(\Omega_m) \propto 1/|\Omega_m|^\gamma$ remains attractive.

We show that this gives rise to  a dichotomy  between the behavior of the gap function  on the Matsubara and the real axis at frequencies larger than ${\bar g}$. Namely, on the Matsubara axis, all $\Delta_n (\omega_m)$ decrease as $1/|\omega_m|^\gamma$ ($n$ sign changes of $\Delta_n (\omega_m)$ occur at $\omega_m < {\bar g}$).  On the real axis,  $\Delta_n (\omega)$ display this power-law behavior only at large enough $\omega> \omega_{cr}$, where $\omega_{cr} \sim {\bar g}/(2-\gamma)^{1/2}$, while
   at  ${\bar g} < \omega < \omega_{cr}$, $\Delta_n (\omega)$ for  all $n$
    oscillate, and the phase of the gap  winds up by
         $2\pi m$,  where  $m$ is an integer. The value of $m$ depends on $\gamma$, but not on
     $n$, and approaches infinity
          when $\gamma \to 2$. We trace these oscillations to the  existence of an exponentially  small and seemingly irrelevant  oscillating term  in $\Delta_n (\omega_m)$  on the Matsubara axis. We argue that this term becomes the dominant one in the gap function upon analytical continuation to the real axis.   We compute the density of state (DoS)  and show that for $\gamma >1$ it develops a set of maxima and minima, and  for $\gamma \to 2$ it evolves towards a set of discrete $\delta-$functions, i.e., for $\gamma \to 2$, the energy levels available to paired quasiparticles effectively get quantized.

            We next consider the gap function in the upper frequency half-plane.
                We show that the phase winding by $2\pi m$  along the real axis
                 between ${\bar g}$ and $\omega_{cr}$
                 implies the presence of $m$ dynamical vortices in the same range of $|z|$ in the upper frequency half-plane.  We identify the vortices as the points where $\Delta (z) =0$  and show that each  vortex
          crosses into the upper frequency half-plane from the lower one
          at some $\gamma_i >1$.   The number of
  vortices is determined by $\gamma$ and is the same for all $n$. At $\gamma \to 2$,  the number of vortices increases, the location of vortex points $z_i$ becomes independent on $n$, and the array of vortices (points where $\Delta_n (z_i) =0$)
     extends to $|z| \to \infty$. Simultaneously,  oscillations of  $\Delta_n (\omega)$ on the real axis also extend to an infinite frequency. This implies that at $\gamma \to 2$, the value of $\Delta_n (|z| \to \infty)$ depends on the path, i.e., there is an  essential singularity at
      $|z| \to \infty$.

  The presence of an essential singularity is crucial by the following reason. Once the set of vortices at $z_i$ becomes infinite and the points $z_i$ accumulate at $|z| \to \infty$, one can use it as an input and obtain
   $\Delta_n (\omega_m)$ by analytically continuing from the set onto the Matsubara axis. If an essential singularity was not there, the analytic continuation would be unique and one would obtain  $\Delta_n (z) =0$ everywhere.  A finite $\Delta_n (z)$  emerges only because
        of essential singularity, whose presence also makes the analytic continuation non-unique. The outcome here
         is that the gap functions $\Delta_n$, which  form a continuum spectrum on the Matsubara axis and give rise to a continuum, massless spectrum of the condensation energies, originate from an essential singularity at $|z| = \infty$ on the complex frequency half-plane. At $\gamma <2$, the  number of vortices remains finite, and the spectrum of $\Delta_n (\omega_m)$ remains discrete.

   The structure of the paper is the following.  In Sec. \ref{sec:model} we briefly review the model and the gap equation. In Sec. we clarify our assertion that the gap functions $\Delta_n (\omega_m)$ with different $n$ are topologically distinct.  We  show that each nodal point of $\Delta_n (\omega_m)$ is
 a dynamical vortex with circulation $2\pi$, such that $\Delta_n (\omega_m)$
 is  a gap function with $n$ dynamical vortices and is a dynamical analog of a nodal topological superconductor.  This holds for any $\gamma >0$.
In Sec. \ref{sec:matsubara}  we analyze the  structure of  $\Delta_n (\omega_m)$ along the
 Matsubara axis for $1< \gamma <2$.
  In Sec. \ref{infnMatsubara} we present  the exact solution of the linearized gap equation, $\Delta_\infty (\omega_m)$ for $\gamma >1$ and show that it has a small oscillating component, not present for smaller $\gamma$.
   In Sec. \ref{sec:n=0}
  we analyze the sign-preserving solution $\Delta_{n=0}(\omega_m)$.
 We show that  at vanishing $\omega_m$, $\Delta_0 (\omega_m)$  tends to a finite $\Delta_0 (0) \sim {\bar g}$, and at $\omega_m > {\bar g}$, it decreases as $1/|\omega_m|^\gamma$.
  In Sec.  \ref{sec:finite_n}
    we  analyze $\Delta_n (\omega_m)$  with a  finite $n$ and the corresponding
    condensation energy $E_{c,n}$.
      We show that at $\omega_m < {\bar g}$,
         $\Delta_{n} (\omega_m)$ oscillates $n$ times as function of $\log{|\omega_m|}$ and then saturates at $\Delta_n (0)$, which decrease exponentially with $n$, while at  $\omega_m > {\bar g}$, it decreases as $1/|\omega_m|^\gamma$ for all $n$.
         We show that as $\gamma$ increases towards $2$, the spectrum of $E_{c,n}$ progressively splits into two subsets, one for $n < n_{cr} \sim 1/(2-\gamma)$ and  the other for $n > n_{cr}$.
In Sec.  \ref{sec:near_2} we discuss the evolution of $\Delta_n (\omega_m)$ and the corresponding onset temperatures $T_{p,n}$ at $\gamma \to 2$.

     In Sec.\ref{sec:real_axis} we analyze the gap equation along real frequency axis.
     We present the gap equation in Sec. \ref{sec:real_axis_Eliashberg} and solve it for
     $n = \infty$ in Sec. \ref{sec:real_n_infinite}, for $n=0$ in Sec. \ref{sec:real_n_0}, and for a finite $n$ in Sec. \ref{sec:real_finite_n}.  We show
         a complex $\Delta_n (\omega)$ oscillates $n$ times at $\omega < {\bar g}$. This holds for all $\gamma$.  Beyond that, for $\gamma >1$,  $\Delta_n (\omega)$ with any $n$ oscillates at ${\bar g} < \omega < \omega_{cr}$, and the phase of the gap function winds up by an integer number of $2\pi$ in this region.

      In Sec. \ref{sec:vortices} we analyze the gap function $\Delta (z)$ in the upper frequency half-plane, at $z = \omega' + i \omega^{''}$. We
 again
consider separately the cases $n = \infty$ (Sec. \ref{sec:vortex_n_infinite}), $n=0$ (Sec. \ref{sec:vortex_n_0}) and a finite $n$ (Sec. \ref{sec:vortex_finite_n}). We show that besides $n$ vortices on the Matsubara axis, which are present for all $\gamma$,  another set of vortices appears  for $\gamma >1$. These new dynamical  vortices are located near the real axis, at  $ {\bar g} < |z|< \omega_{cr}$, and  cause oscillations on the real axis in the same range of $\omega$.
       Each vortex from this new set moves into the upper  half-plane of frequency from the lower one
        at a particular discrete $\gamma_i$.
             The number of these vortices increases as $\gamma \to 2$ and they line up along a particular
 path in the upper frequency half-plane.

We  present our conclusions  in Sec. \ref{sec:conclusions}. Some technical details of calculations are presented in the Appendix.  The case $\gamma =2$ requires special consideration and we will analyze it in the next paper, where we also discuss in detail the behavior of the superfluid  stiffness and the interplay between the pair formation and the true superconductivity.

\section{$\gamma$-model}
\label{sec:model}

We consider  itinerant fermions at a QCP, interacting by exchanging  fluctuations of a
 critical order parameter.  At a QCP, the propagator of a soft boson becomes massless and mediates singular interaction between fermions. We assume that this interaction is attractive in at least one pairing channel and that a pairing boson can be treated as slow mode compared to a fermion, i.e., at a given momentum $q$, typical fermionic frequency is much larger than typical bosonic frequency. In this situation, the ratio of typical bosonic and fermionic frequencies is a small parameter, analogous to the Migdal parameter in the electron-phonon case. This small parameter allows one to neglect vertex corrections and explicitly integrate feromionic and bosonic propagators over the momentum components. As a result, the problem reduces  to a  set of coupled integral equations for frequency dependent fermionic self-energy and the pairing vertex, in which an input is the effective, singular, frequency dependent  interaction
  $V(\Omega_m) = {\bar g}^\gamma/|\Omega_m|^\gamma$. The exponent $\gamma$ is different for different microscopic systems.
This model, nicknamed the $\gamma-$model,  has been  introduced in Paper I and we refer a reader to the list of references in that paper to earlier works on the justification of the model and its relation to various microscopic quantum-critical systems with momentum and frequency-dependent interaction.

   The local interaction $V(\Omega_m)$ contributes to the fermionic self-energy  $\Sigma (k_F, \omega_m) = \Sigma (\omega_m)$ and the pairing vertex $\Phi (\omega_m)$.
  The coupled equations for  $\Sigma (\omega_m)$ and $\Phi (\omega_m)$  have the same structure as  Eliashberg equations for a dispersion-less phonon, albeit with the exponent $\gamma$,
 instead of  $2$,    and for shortness we will be calling them ``Eliashberg equations''.

The  superconducting gap function $\Delta (\omega_m)$ is defined as
\beq
 \Delta (\omega_m) = \omega_m  \frac{\Phi (\omega_m)}{{\tilde \Sigma} (\omega_m)} = \frac{\Phi (\omega_m)}{1 + \Sigma (\omega_m)/\omega_m}
  \label{ss_1}
  \eeq
   At a finite $T$ the Eliashberg gap equation is
   \beq
   \Delta (\omega_m) = {\bar g}^\gamma \pi T \sum_{m' \neq m} \frac{\Delta (\omega'_{m}) - \Delta (\omega_m) \frac{\omega'_{m}}{\omega_m}}{\sqrt{(\omega'_{m})^2 +\Delta^2 (\omega'_{m})}}
    ~\frac{1}{|\omega_m - \omega'_{m}|^\gamma}.
     \label{ss_11}
  \eeq
    The gap function is defined up to an overall $U(1)$ phase factor, which we set to zero.
The gap equation for infinitesimally small $\Delta (\omega_m)$
  \beq
   \Delta (\omega_m) = {\bar g}^\gamma \pi T \sum_{m'} \frac{\Delta (\omega'_{m}) - \Delta (\omega_m) \frac{\omega'_{m}}{\omega_m}}{|\omega'_{m}| |\omega_m - \omega'_{m}|^\gamma}.
     \label{ss_11_l}
  \eeq
  determines the onset temperature for the pairing.
    As in previous papers, we label this temperature $T_p$. Within Eliashberg theory, it is the same as superconducting $T_c$, but in the presence of fluctuations $T_p$  is generally larger than  $T_c$.

 \section{Gap function on  the Matsubara axis, $T=0$}
 \label{sec:matsubara}

In Paper I we showed that at $T=0$ and $\gamma <1$,  Eq. (\ref{ss_11}) has an infinite number of topologically distant solutions, $\Delta_n (\omega_m)$, all with the same spatial gap symmetry.
 A gap function with a given $n$  tends to a finite value  at $\omega_m =0$ and decays as $1/|\omega_m|^\gamma$ at $\omega_m \geq {\bar g}$, but in between
 it changes sign $n$ times.
  In Paper III we showed that
   each of the two terms in the r.h.s. of  (\ref{ss_11})  (separate contributions  from the pairing vertex and the fermionic self-energy) diverges for $\gamma \geq 1$ as $\int d\Omega_m/|\Omega_m|^\gamma$.
   However, the two divergencies cancel out, and
   $\Delta_n (\omega_m)$ at $T=0$ evolve smoothly
  through $\gamma =1$.  We now extend the analysis to $\gamma$ between $1$ and $2$.  We show that
    the forms of $\Delta_n (\omega_m)$ remain qualitatively the same as for smaller $\gamma$, yet  the spectrum of  condensation energies become progressively more dense.  We also argue that when $\gamma$ increases  towards $2$, the spectrum of condensation energy $E_{n,c}$ becomes more and more dense, and in the limit $\gamma\to 2$  all $E_{n,c}$ with finite $n<n_{cr}$ become almost equal to $E_{0,c}$,  while $E_{n>n_{cr},c}$ form a continuous spectrum.

 \subsection{Nodal points on the Matsubara axis as dynamical vortices}
\label{sec:matsubara_move}
 Before we discuss $\Delta_n (\omega_m)$, we pause for a moment and elaborate on the notion that $\Delta_n (\omega_m)$ are topologically different.  We argue that each nodal point is a dynamical vortex, i.e., $\Delta_n (\omega_m)$ is an $n-$vortex state -- a dynamical analog of a nodal topological superconductor.

\begin{figure}
	\includegraphics[width=12cm]{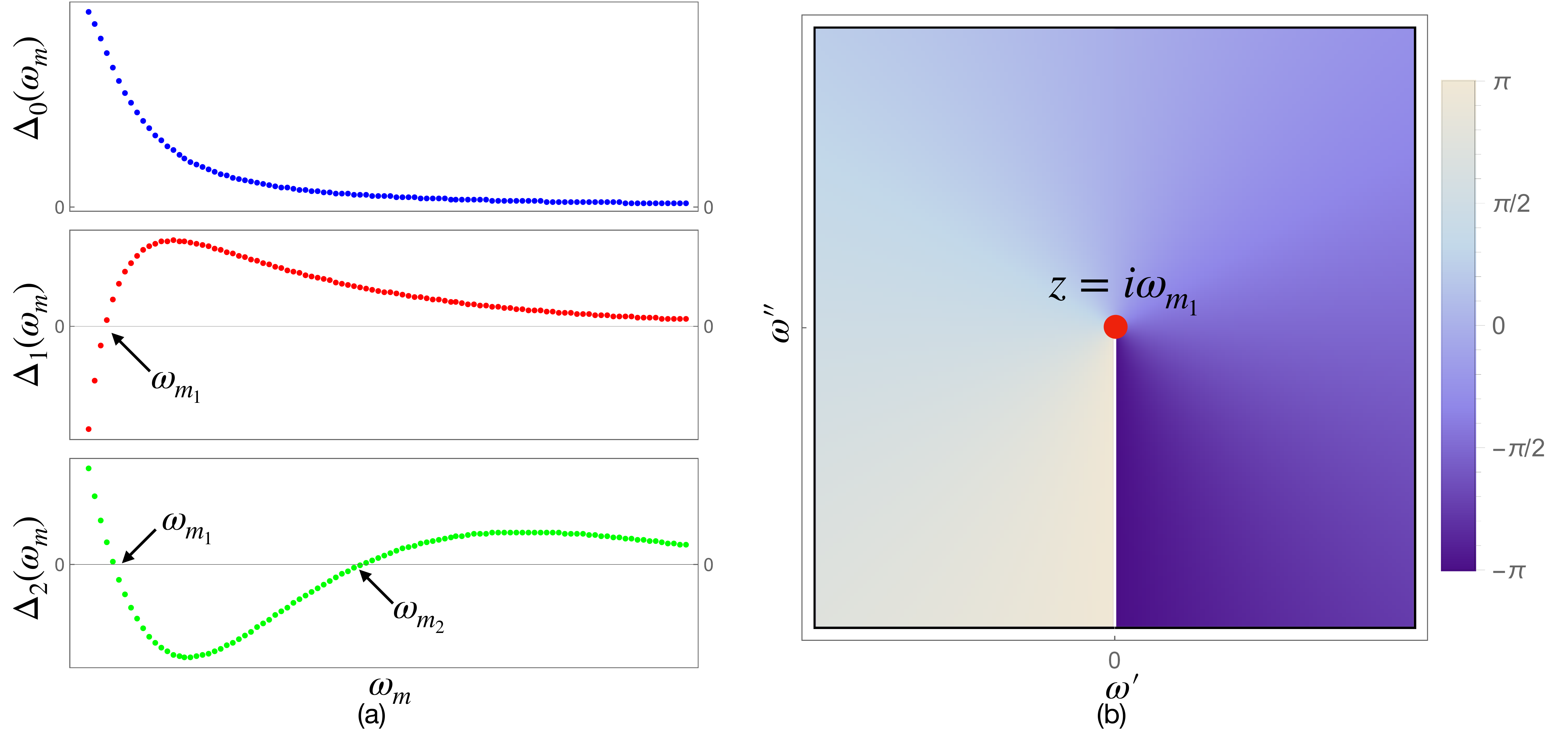}
	\caption{Left panel: The schematic forms of the first 3 solutions of the gap equation $\Delta_0 (\omega_m)$, $\Delta_1 (\omega_m)$ and $\Delta_2 (\omega_m)$. A gap function $\Delta_n (\omega_m)$ changes sign $n$ times at $\omega_m >0$.
    Right panel: Variation of the phase of $\Delta_1 (z)$ ($z = \omega' + i \omega^{''}$) around the nodal point at
     $\omega_m = \omega_{m1}$ .  Anticlockwise circulation of the phase around $i \omega_{m1}$ is $2\pi$. The same holds for all other nodal points.  }\label{fig:demo}
\end{figure}

For definiteness, let's compare the behavior of sign-preserving $\Delta_0 (\omega_{m})$ and of
$\Delta_1 (\omega_{m})$, and $\Delta_2 (\omega_{m})$, which change sign once and twice between $\omega_m =0$ and $\omega_m = \infty$, respectively.
 We  show these functions  in the left panel of Fig.\ref{fig:demo}.
Suppose that $\Delta_1 (\omega_m)$ changes sign at $\omega_m = \omega_1$.  Near this frequency,
$\Delta_1 (\omega_m) = - c (\omega_m - \omega_1)$,
 where $c >0$ for consistency with the figure.
  Let us analytically continue
 $\Delta_1 (\omega_m)$ to a near vicinity of the Matsubara axis, to $z = \omega' + i \omega^{''}$ (on the Matsubara axis, $z = i \omega_m$).  Because
 $\Delta_1 (\omega_m)$ is non-singular, $\Delta_1 (z) = \Delta_1 (i\omega_m \to \omega' + i \omega^{''})$.   For any non-zero $\omega'$,  $\Delta_1 (z)$ is a complex function: $\Delta_1 (z) = \Delta'_1 (z) + i \Delta^{''}_1 (z)$, and we can introduce the phase of $\Delta_1 (z)$ as
  \beq
     \eta_1 (z) ={\text{Im}} [\log{\Delta_1 (z)}].
     \label{4_1}
     \eeq
 We plot the variation of $\eta_1 (z)$ around $z = i\omega_1$ in the right panel of Fig.\ref{fig:demo}.
  We see that the phase varies by $2\pi$ upon anticlockwise circulation around $\omega_1$. This  implies that the nodal point at $\omega_1$ is in fact the center of a dynamical vortex.  One can  verify that
    $\Delta_n (\omega_m)$ with a generic $n$  contains $n$ vortices, each with anticlockwise circulation $2\pi$.

 \begin{figure}
 	\includegraphics[width=16cm]{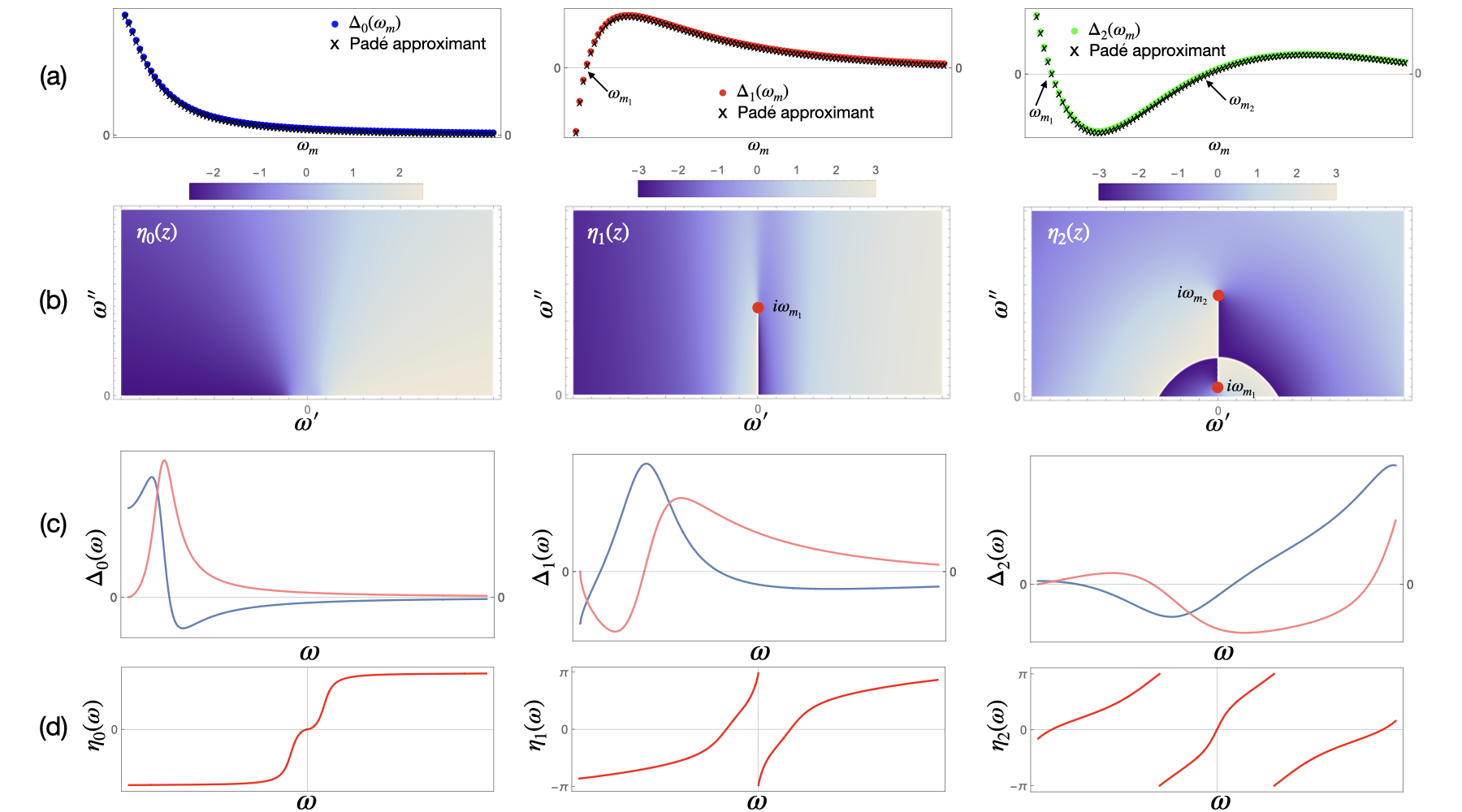}
 	\caption{Analytic continuation of $\Delta_n(\omega_m)$ in Fig.\ref{fig:demo} to the upper half plane
 ($i \omega_m \to z = \omega' + i \omega^{''}$) using Pad\'e approximant. Panel (a):
   the accuracy of fitting $\Delta_n(\omega_m)$ by Pad\'e approximant.  Panel (b):
  the phase of the gap function in the upper frequency half-plane,  $\eta_n(z)$ = Im$[\log{\Delta (z)}]$.
   The locations of the  vortices are marked by red dots. Panel (c): the gap function $\Delta_n(\omega)$ along the real axis. Panel (d): variation of $\eta_n (\omega)$ along the real axis.  Each vortex gives rise to $2\pi$ phase variation.
    For convenience of presentation, we confined $\eta_n(\omega)$ to $(-\pi,\pi)$.
     In this convention, $\eta_n (\omega)$
winds by $2\pi n$
due to vortices.}
 \label{fig:pade}
 \end{figure}

One can straightforwardly verify that $n$ vortices  on the Matsubara axis give rise to $2\pi n$ phase variation on the real axis, between $-\infty$ and $\infty$. To see this, one should  compute
 $\int_{-\infty}^\infty d \omega  \partial \eta_1 (\omega)/\partial \omega$ by closing the integration contour over the upper half-plane. The function
 $\partial \eta_1 (\omega)/\partial \omega$
is analytic in the upper half-plane except for the nodal points
 where it has simple poles.
Modifying the contour to circle out each nodal point, one obtains $2\pi n$ phase variation from the vortices. In addition, there is
 also the
$\pi \gamma$   from the integral over the upper half-circle, due to the fact that at the largest frequencies, $\Delta_n (z) \propto 1/(-z^2)^{\gamma/2}$. This form is consistent with $\Delta_n (\omega_m) \propto 1/|\omega_m|^{\gamma}$, as one can verify by Cauchy relation.

In Fig. \ref{fig:pade} we show $\Delta_n (z)$, where $z = \omega' + i \omega^{''}$, for $n=0,1$, and $2$,
 obtained by analytic continuation  from $\Delta_n (\omega_m)$ in Fig.\ref{fig:demo} using Pad\'e approximants method\cite{Beach2000}.    We see that $\eta_n (\omega+i0^+)$ indeed
winds by $2\pi n$.

We now analyze the full frequency dependence of $\Delta_n (\omega_m)$  for $1<\gamma <2$. We first consider separately the opposite limits $n = \infty$ and $n = 0$ and then discuss a finite $n$.

\subsection{Frequency dependence of $\Delta_n (\omega_m)$. The case $n = \infty$}
\label{infnMatsubara}

        The gap function $\Delta_{\infty} (\omega_m)$ is a potential solution of the linearized gap equation
   \beq
   \Delta_\infty (\omega_m) = \frac{{\bar g}^\gamma}{2} \int d \omega'_m \frac{\Delta_\infty (\omega'_{m}) - \Delta_\infty (\omega_m) \frac{\omega'_{m}}{\omega_m}}{|\omega'_{m}|}
    ~\frac{1}{|\omega_m - \omega'_{m}|^\gamma}.
     \label{ss_11_l0}
  \eeq
We found that the solution exists and obtained the exact form of $\Delta_\infty (\omega_m)$.
  We emphasize that the solution exists despite that  $T=0$ is not a critical point for superconductivity. We will see below that non-linear gap equation at $T=0$ has solution(s)  with a finite gap magnitude.
    The solution exists only at a QCP, when the  pairing  is mediated by a gapless boson.

\begin{figure}
	\includegraphics[width=12cm]{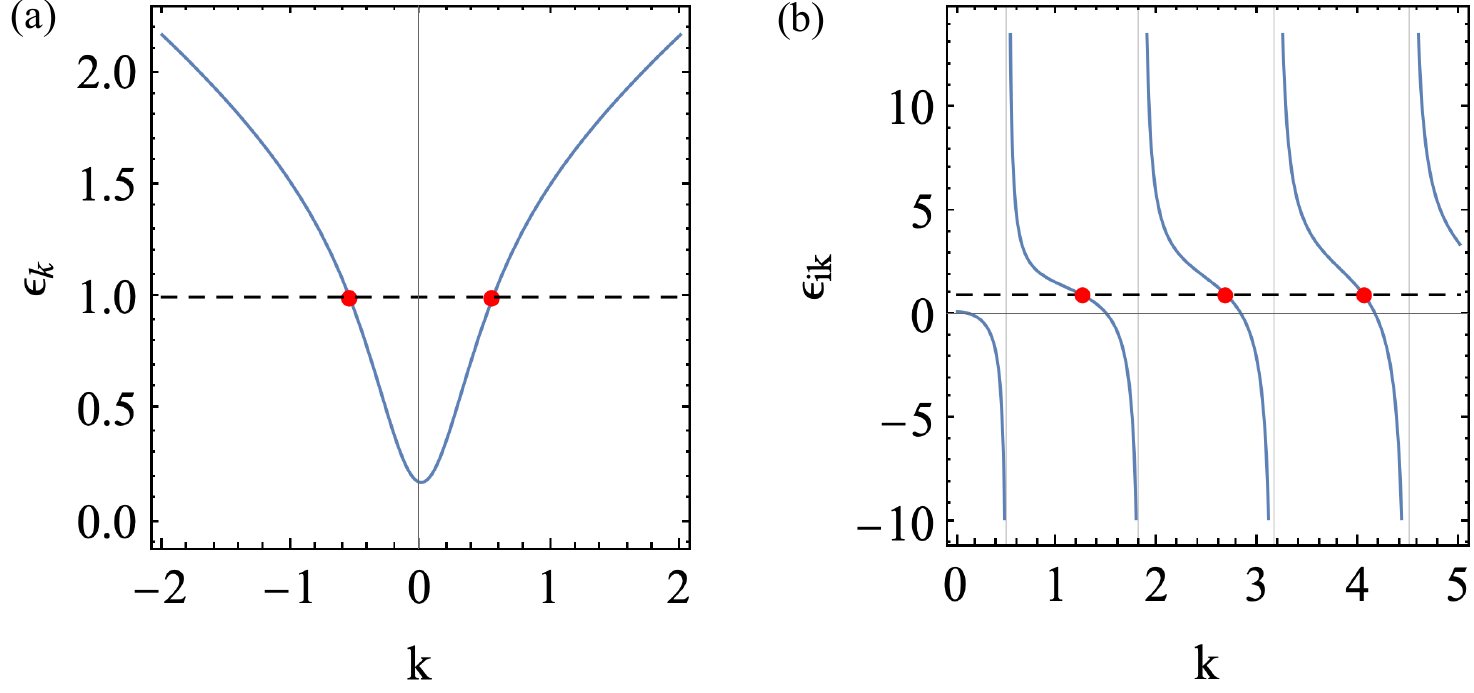}
	\caption{ Functions $\epsilon_{k}$ (a) and $\epsilon_{ik}$ (b) for $\gamma=1.5$. The red dots show the solutions $\epsilon_{\beta}=1$ and $\epsilon_{i \beta_m}=1$ ($m=0,1,2,...$).}
\label{fig:epsilon}
\end{figure}

  We present technical details  of the analysis of Eq. (\ref{ss_11_l0}) in Appendix \ref{app:exact} and here
quote the result.
 It is convenient to introduce
   \beq
  y = \left(\frac{|\omega_m|}{\bar g}\right)^\gamma,
  \label{nn_6_1}
  \eeq
  The gap function $\Delta (\omega_m) = \Delta_\infty (y)$
 \beq
 \Delta_\infty (y) = \int_{-\infty}^\infty dk b_k e^{-ik \log{y}}.
 \label{nn_2}
 \eeq
is the Fourier transform of $b_k$ given by
 \beq
  b_k = \frac{e^{-i (I_k + k\log{(\gamma-1)})}}{\left[\cosh(\pi (k-\beta))\cosh(\pi (k+\beta))\right]^{1/2}}.
  \label{nn_2_1}
  \eeq
 Here
\beq
  I_k = \frac{1}{2} \int_{-\infty}^\infty dk' \log{|\epsilon_{k'} -1|} \tanh{\pi (k'-k)},
  \label{nn_2_2}
  \eeq
\beq
\epsilon_{k'} = \frac{1-\gamma}{2} \frac{\Gamma\left(\frac{\gamma}{2}\left(1 + 2i k'\right)\right)\Gamma\left(\frac{\gamma}{2}\left(1 - 2i k'\right)\right)}{\Gamma(\gamma)} \left(1+ \frac{\cosh{\pi  \gamma k'}}{\cos{\pi \gamma/2}}\right),
\label{nn_3}
\eeq
and $\beta>0$ is the solution of $\epsilon_{\beta} =1$.
We plot $\epsilon_{k'}$  in Fig.~\ref{fig:epsilon} (a). The value of $\beta$ depends on  $\gamma$ and evolves between $\beta \approx 0.79$ for $\gamma =1$ and $\beta \approx 0.39$ for $\gamma = 2$.

\begin{figure}
	\includegraphics[width=12cm]{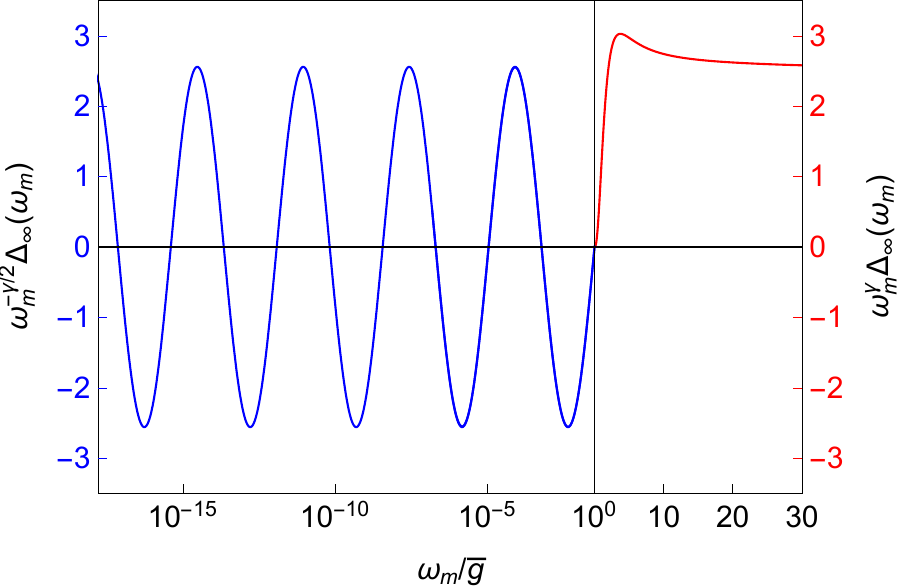}
	\caption{The exact solution $\Delta_{\infty}(\omega_m)$ for $\gamma = 1.91$.
	To better display the distinctive power-law behavior, we use different scales for $\omega > {\bar g}$ and $\omega < {\bar g}$.
}
\label{fig:Delta_infty_Mats}
\end{figure}

The gap function $\Delta_\infty (y)$ can be straightforwardly obtained numerically. We show the results in Fig.~\ref{fig:Delta_infty_Mats}
for $\gamma=1.91$.
 At small $y$, i.e., at  $\omega_m < {\bar g}$, the gap function oscillates down to the lowest frequencies with the period set by $\log{y}$.  At $y >1$, i.e., at
 $\omega > {\bar g}$, oscillations end, and $\Delta_\infty (y)$ gradually decreases with increasing $y$.
By practical reasons,  we set the lowest frequency in Fig.~\ref{fig:Delta_infty_Mats}
at $10^{-18} {\bar g}$. There are 10 zeros of $\Delta_\infty (\omega)$ (10 vortex points)  above this frequency.

 The limiting forms of $\Delta_\infty (y)$ at small and large $y$
  can be obtained analytically.  At $y \ll 1 $ we find
 \beq
  \Delta_\infty (y) \sim y^{1/2} \cos{\left(\beta\log{y} + \phi\right)},
  \label{nn_3_1}
  \eeq
  where $\phi$ is a $\gamma-$dependent number. At  $y \gg 1$
 \beq
  \Delta_\infty (y) \sim \frac{1}{y}
  \label{nn_3_2}
  \eeq

 The corrections to (\ref{nn_3_1}) hold in powers of $y$ and form two series,  local and non-local, depending on whether they come from fermions with running frequencies of order $\omega_m$ or much larger  frequencies, of order $ {\bar g}$.
 The full $\Delta_\infty (y)$ for $y <1$ is
  \beq
   \Delta_\infty (y) =  {\text Re}~ \sum_{n=0}^{\infty } e^{(i \beta \log{y} + \phi)} C^{<}_n~ y^{n+1/2}
+  \sum_{n,m=0}^{\infty }D^{<}_{n,m} y^{(n + \beta_m +1/2)}
\label{nn_6}
\eeq
where the first  term describes the sum of the leading term and the local corrections, and the second term describes the  non-local corrections. In the latter,  $\beta_m >0$ are the solutions of $\epsilon_{i\beta_m} =1$ for imaginary argument(same as in Fig.\ref{fig:epsilon}).  There is  an infinite set of such $\beta_m$, specified by integer $m =0,1,2..$ and located at  $1/2 + 2m/\gamma <\beta_m < 1/2 + 2(m+1)/\gamma$, see Fig.~\ref{fig:epsilon} (b).

For $y >1$, the corrections to (\ref{nn_3_2}) hold in powers of $1/y$ and also form two series:
 \begin{equation}
 \Delta_\infty (y)= \sum_{n=0}^\infty C^{>}_n~ \left(\frac{1}{y}\right)^{(n+1)} +
 \sum_{n,m=0}^{\infty }D^{>}_{n,m} \left(\frac{1}{y}\right)^{(n+1 + 2(m+1)/\gamma)}
\label{nn_7}
\end{equation}
Eq. (\ref{nn_3_2}) is the $C^>_0$ term in this series.

\subsubsection{Distinction between $\Delta_\infty (y)$ at $\gamma <1$ and $\gamma >1$.}

The limiting forms of $\Delta_\infty (y)$ at small and large $y$ and the general structure of $\Delta_\infty (\omega_m)$ for $\gamma >1$ are the same as we obtained earlier for $\gamma <1$.
However, on a more careful look,  we found that there is  a qualitative
distinction between the two cases for $y >1$. Namely, for $\gamma <1$,
  the series in (\ref{nn_7}) are convergent, while
  for $\gamma >1$,  series expansion in (\ref{nn_7}) holds only up to a certain $n_{max} \sim y^{1/(\gamma-1)}$, while instead of power series with larger $n$, there appears an oscillating term (see below).  The value $n_{max}$ is large for $y \gg 1$, such that this term has an exponentially small prefactor  $e^{-y^{1/(\gamma-1)}}$.
  Yet, we will see below that once $\Delta_\infty$  is analytically continued from Matsubara to  real axis, the exponential smallness is reduced, particularly near $\gamma =2$, and the oscillating term becomes the dominant contribution to $\Delta_\infty (\omega)$ over a wide frequency range and gives rise to the emergence of a new vortex structure in the upper half-plane of frequency.

 The difference between  $\Delta_\infty (y \gg 1)$ for $\gamma <1$ and $\gamma >1$
 originates from change of the  behavior of $\epsilon_{k'}$ in Eq. (\ref{nn_2_2}).
 At $\gamma <1$,
  $\epsilon_{k'}$ vanishes at $k' \to \infty$, while at $\gamma >1$, $\epsilon_{k'}$ diverges at large $k'$ as
 \beq
 \epsilon_{k'} = (A_\gamma|k'|)^{\gamma-1} (\gamma-1),  ~~A_\gamma = \gamma \left(\frac{\pi}{2\Gamma(\gamma) \cos \[\pi (2-\gamma)/2 \] }\right)^{\frac{1}{\gamma-1}}
  \label{nn_3_3}
 \eeq
 Substituting this form into (\ref{nn_2_2}) and integrating over $k'$, we find that at large $k$,
 \beq
 I_k + k\log{(\gamma-1) = - (\gamma-1) k \log{\frac{A_\gamma k}{e}}}
  \label{nn_3_3_1}
 \eeq
  The integral comes from $k' \sim k$, what justifies using the form of $\epsilon_{k'}$ at large $k'$.
 Substituting (\ref{nn_3_3_1}) into (\ref{nn_2_1}) and approximating  the denominator in (\ref{nn_2_1}) by its value at large $k$, we obtain
 \beq
 b_k \approx  e^{-\pi |k|} e^{i(\gamma-1) k \log{\frac{A_\gamma k}{e}}}
 \label{nn_3_4}
 \eeq
 Substituting  this $b_k$  into (\ref{nn_2}), we find
  the  contribution to $\Delta_\infty (y)$  in the form
 \beq
 \Delta_\infty (y) \sim \int_{k_{min}}^\infty e^{-\pi k} \cos{\left((\gamma -1) k \log{\frac{y^{1/(\gamma-1)} e}{A_\gamma k}}\right)}
  \label{nn_3_5}
 \eeq
 where $k_{min} = O(1)$.
 The argument of the $\cos$ has a maximum at $k = k^* = y^{1/(\gamma-1)}/A_\gamma$. For large $y$, this
 $k^*$ is well inside the range of the integration.  Expanding to quadratic order near the maximum and substituting into
 (\ref{nn_3_5}), we find after some algebra (see Appendix \ref{app:exact}) that the integral in (\ref{nn_3_5}) contains the
 universal  contribution, which we label as $ \Delta^{u}_{\infty}$:
  \beq
  \Delta^{u}_{\infty} (y) \sim y^{\frac{1}{2(\gamma-1)}} e^{-\frac{\pi }{A_\gamma}y^{\frac{1}{(\gamma-1)}}} \cos{\left[ y^{\frac{1}{(\gamma-1)}}~ \frac{(\gamma-1)}{A_\gamma} \left(1- \frac{\pi^2}{2(\gamma-1)^2}\right) - \frac{\pi}{4}\right]}
  \label{nn_3_6}
 \eeq
 In the actual Matsubara frequencies this formula reads
      \beq
  \Delta^{u}_{\infty} (\omega_m) \propto |\omega_m|^{\frac{\gamma}{2(\gamma-1)}} e^{-\frac{\pi (|\omega_m|/{\bar g})^{\gamma/(\gamma-1)}}{A_\gamma}}\cos{\left[  \left(\frac{|\omega_m|}{{\bar g}}\right)^{\gamma/(\gamma-1)}  \frac{(\gamma -1)}{A_\gamma} \left(1- \frac{\pi^2}{2(\gamma-1)^2}\right) -\frac{\pi}{4}\right]}
  \label{nn_3_6_1}
   \eeq
The oscillating contribution is exponentially small at large $y$. It vanishes at $\gamma =1$ and does not exist  at smaller $\gamma$.  We will return to this contribution in {Sec.~\ref{sec:real_n_infinite}},
where we discuss $\Delta^{u}_\infty (\omega)$ on the real axis.

\subsection{Frequency dependence of $\Delta_n (\omega_m)$. The case $n=0$}
\label{sec:n=0}
We now consider the solutions of the non-linear gap equation  (\ref{ss_11}) at $T=0$.
 We begin with  sign-preserving solution $\Delta_0 (\omega)$.
    To set the stage,  we first quickly review the results for $\gamma <1$.  For such $\gamma$,  the term with $m'=m$ in the r.h.s. of (\ref{ss_11}) is non-divergent and
      can be neglected as it  renormalizes the prefactor for $\Delta_0 (\omega_m)$ in the r.h.s. by at most $O(1)$. Without this term, we have
   \beq
   \Delta_0 (\omega_m) = \frac{{\bar g}^\gamma}{2} \int_{-\infty}^\infty \frac{d \omega'_m  \Delta_0(\omega'_m)}
   {|\omega_m - \omega'_m|^\gamma \sqrt{\Delta^2_0 (\omega'_m)+ (\omega'_m)^2}},
   \label{4_5}
   \eeq
    At  small $\omega_m$, $\Delta_0 (\omega_m) \approx \Delta_0 (0)$, and we have
   \beq
    \Delta_0 (0) = {\bar g}^\gamma \int_0^\infty d \omega'_m  \frac{\Delta_0(\omega'_m)}{
   |\omega'_m|^\gamma \sqrt{\Delta^2_0 (\omega'_m)+ (\omega'_m)^2}}
  \label{4_5_1}
   \eeq
    The integral is determined by $\omega'_m \sim \Delta_0 (\omega'_m)$. For such $\omega'_m$,  we can approximate $\Delta_0(\omega'_m)$ by $\Delta_0 (0)$ up to $O(1)$ corrections. Evaluating the integral in (\ref{4_5_1}), we then obtain $\Delta_0 (0) \sim {\bar g}$. In the opposite limit $\omega_m \gg {\bar g}$, we can pull out $1/|\omega|^\gamma$ from the r.h.s. of (\ref{4_5}) and obtain
    \beq
    \Delta_0 (\omega_m) = \left(\frac{{\bar g}}{|\omega_m|}\right)^\gamma \int_0^\infty d \omega'_m  \frac{\Delta_0(\omega'_m)}{
   \sqrt{\Delta^2_0 (\omega'_m)+ (\omega'_m)^2}}
   \eeq
   The integral is determined by $\omega'_m = O({\bar g})$ and is $O(\bar g)$, such that at high frequencies,
  $\Delta_0 (\omega_m) \sim {\bar g} ({\bar g}/|\omega_m|)^\gamma$.    The low-frequency and high-frequency forms of $\Delta_0 (\omega_m)$ then match at $\omega_m  \sim {\bar g}$.

 Now we do the same analysis for $\gamma >1$.
  For $\gamma >1$, there is an identity
\beq
 \int_{-\infty}^\infty d\omega'_m \frac{1 - \frac{\omega'_m}{\omega_m}}{|\omega_m-\omega'_m|^\gamma} =0.
 \label{new_5}
 \eeq
 We use  it to re-express Eq. (\ref{ss_11}) as
 \bea
 &&\Delta_0(\omega_m) \left[1 -
  \frac{{\bar g}^\gamma}{2} \int_{-\infty}^\infty d \omega'_m \frac{1 - \frac{\omega'_m}{\omega_m}}{|\omega_m - \omega'_m|^\gamma} \left(\frac{1}{\sqrt{\Delta^2_0 (\omega'_m)+ (\omega'_m)^2}}- \frac{1}{\Delta_0 (\omega_m)} \right)\right] \nonumber \\
&& =  \frac{{\bar g}^\gamma}{2} \int_{-\infty}^\infty d \omega'_m  \frac{\Delta_0(\omega'_m)-\Delta_0 (\omega_m)}{|\omega_m - \omega'_m|^\gamma \sqrt{\Delta^2_0 (\omega'_m)+ (\omega'_m)^2}}
\label{new_6}
\eea
Each integral in (\ref{new_6})  is infra-red convergent and determined by $\omega'_m \sim \Delta_0 (\omega'_m)$.

 In the limit $\omega_m \to 0$, Eq. (\ref{new_6}) reduces to
  \bea
&&\Delta_0 (0) \left[1 - {\bar g}^\gamma (\gamma -1) \int_{0}^\infty d \omega'_m  \frac{\sqrt{\Delta^2_0 (\omega'_m)+ (\omega'_m)^2} - \Delta_0 (0)}{\Delta_0 (0) \sqrt{\Delta^2_0 (\omega'_m)+ (\omega'_m)^2}|\omega'_m|^\gamma}\right] \nonumber \\
&& = {\bar g}^\gamma \int_{0}^\infty \frac{d \omega'_m ~ \left(\Delta_0(\omega'_m)-\Delta_0 (0)\right)}{\sqrt{\Delta^2_0
 (\omega'_m)+ (\omega'_m)^2}|\omega'_m|^\gamma}
\label{new_7}
\eea
To estimate the magnitude of $\Delta_0 (0)$,  assume that  $\Delta_0 (\omega'_m)$ remains  approximately equal to $\Delta_0 (0)$ for relevant
 $\omega'_m \leq \Delta _0 (\omega'_m)$.

\begin{figure}
	\includegraphics[width=15cm]{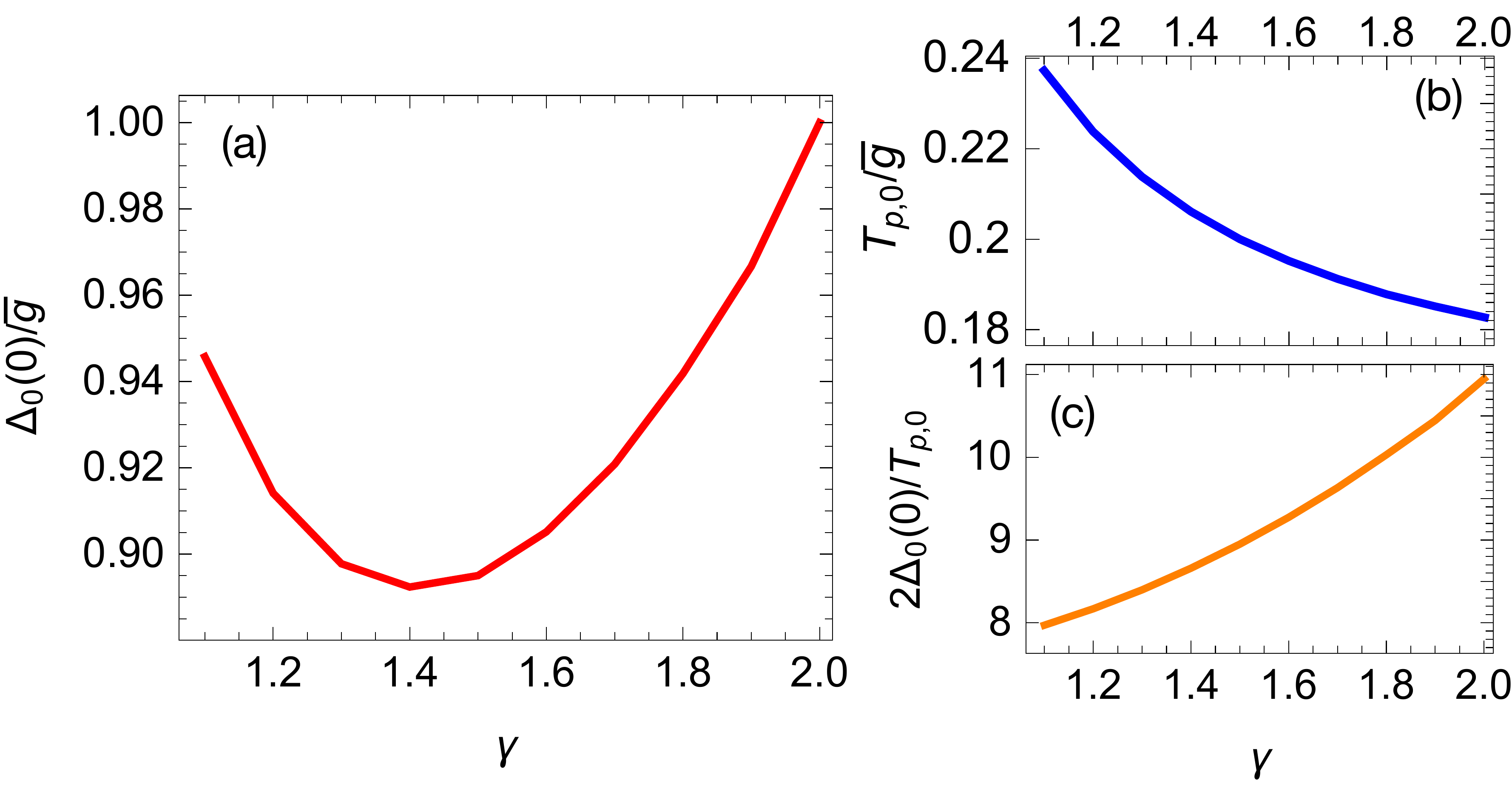}
	\caption{Panel (a):
the solution of Eq.\eqref{new_8} for $\Delta_0(\omega_m =0)$;
  Panel (b): the onset temperature of the pairing $T_{p,0}$ vs $\gamma$. Panel (c): the ratio $2\Delta_0(0)/T_{p,0}$ vs $\gamma$.  The ratio is slightly different from the one obtained numerically in Ref. \cite{Wu_19} as here we use Eq.\eqref{new_8} for $\Delta_0(\omega_m =0)$. }\label{fig:Delta0}
\end{figure}

 The r.h.s. of (\ref{new_7}) then vanishes, and we obtain
  \beq
  1= {\bar g}^\gamma (\gamma -1) \int_{0}^\infty d \omega'_m  \frac{\sqrt{\Delta^2_0 (0) + (\omega'_m)^2} - \Delta_0(0)}{\Delta_0 (0) \sqrt{\Delta^2_0 (0) + (\omega'_m)^2}|\omega'_m|^\gamma}
=\frac{1-\gamma }{2\sqrt{\pi }}\left(\frac{\bar{g}}{\Delta_0 (0)} \right)^{\gamma }\Gamma \left(\frac{1-\gamma}{2}\right) \Gamma \left(\frac{\gamma }{2}\right)
\label{new_8}
\eeq
We plot $\Delta_{0}(0)/\bar{g}$ as a function of $\gamma $ in Fig.\ref{fig:Delta0} (a).
We see that $\Delta_0 (0)$ is of order ${\bar g}$ and weakly depends on $\gamma$.
  One can further expand $\Delta_0 (\omega_m)$ in frequency and find that the expansion is analytic and holds in powers of $\omega^2_m$.   For completeness, we also
    show the onset temperature $T_{p,0}$ and the ratio $2\Delta_0 (0)/T_{p,0}$, both as functions of $\gamma$.
   The results are consistent with earlier calculations~\cite{Marsiglio_91,Wang2016,Kotliar2018,Wu_19}.

\begin{figure}
   \includegraphics[width=8cm]{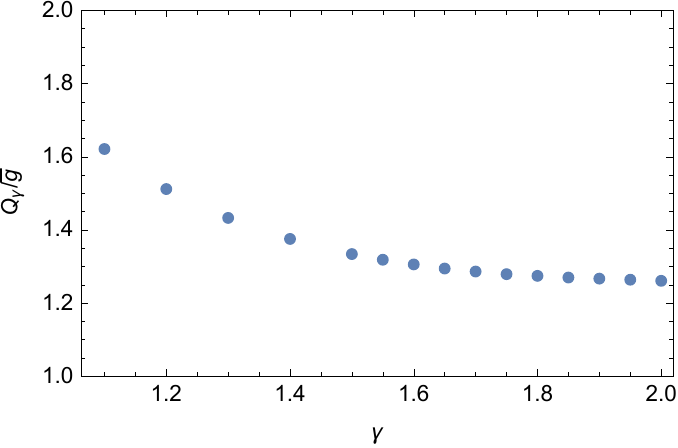}
   \caption{$Q_\gamma$, defined in Eq. (\ref{eq:Qgamma}), as a function of $\gamma$.  To obtain this function, we solved numerically the non-linear gap equation for sign-preserving
 $\Delta_0 (\omega_m)$ at a small temperature.}
 \label{fig:Qgamma}
\end{figure}

 In the opposite limit of large $\omega_m$, the prefactor for $\Delta_0 (\omega)$ in the l.h.s. of (\ref{new_6}) is approximately $1$, and  $1/|\omega_m|^\gamma$ can be pulled out from the integral in the r.h.s.  This yields
  \beq
 \Delta_0 (\omega_m) \approx Q_\gamma \left(\frac{{\bar g}}{|\omega_m|}\right)^\gamma
 \label{4_4}
 \eeq
 where
\begin{equation}
    Q_\gamma=\int_{0}^\infty \frac{d \omega'_m \Delta_0 (\omega'_m)} {\sqrt{\Delta^2_0 (\omega'_m)+ (\omega'_m)^2}}\label{eq:Qgamma}
\end{equation}
The integral is again determined by frequencies at which $\Delta_0 (\omega'_m) \sim \omega'_m$ and  is of order
 $\Delta_0 (0)$. Then $\Delta_0 (\omega_m) \sim \Delta_0 (0) ({\bar g}/|\omega_m|)^\gamma$.
 In Fig.\ref{fig:Qgamma} we show  $Q_\gamma$, obtained by solving numerically the nonlinear gap equation for various $\gamma$. We see that $Q_\gamma$ is indeed  of order
 $Q_{\gamma }\sim \Delta_0 (0) \sim \bar g$.

The high-frequency behavior of $\Delta_0 (\omega_m)$ in (\ref{4_4}) is the same as of $\Delta_\infty (\omega_m)$. This is expected because at such frequencies $\Delta_0 (\omega_m) \ll \omega_m$.  This implies that $\Delta_0 (\omega_m)$ has the same exponentially small oscillating term as in Eq. (\ref{nn_3_6}).  In Sec.\ref{sec:real_n_infinite}
 we will see how this term evolves upon analytical continuation to the real axis.

 \subsection{Frequency dependence of $\Delta_n (\omega_m)$.  Finite $n$.}
\label{sec:finite_n}

\subsubsection{A generic $\gamma$ from $1<\gamma <2$.}

To analyze the behavior of $\Delta_n (\omega_m)$ at a finite $n$, one has to solve the non-linear gap equation at $T=0$ for a class of functions which change sign $n$ times.
  At $\omega_m > {\bar g}$,
$\Delta_n (\omega_m)$ decreases as $1/|\omega_m|^\gamma$   and has an exponentially small oscillating piece,  the same as in (\ref{nn_3_6}). At $\omega_m <{\bar g}$,  $\Delta_n (\omega_m)$
 oscillates $n$ times and saturates at the smallest $\omega_m$ at $\Delta_n (0)$, which decreases with increasing $n$.
 To estimate the magnitude of $\Delta_n (0)$  and, more generally, understand why there is a discrete set of solutions, we use the solution of the linearized gap equation,  Eq. (\ref{nn_3_1}), as an input,  and treat the non-linear terms by expanding  in powers
 of  $\Delta_n (\omega_m)/\omega_m$.  The gap function, obtained this way, is represented by the series
 \beq
 \Delta_n (y) =  {\bar g} y^{1/2} \sum_{m=0}^\infty C_n^{2m+1} f_{m} (y), ~~ y = \left(\frac{|\omega_m|}{\bar g}\right)^\gamma
 \label{nn_5_1}
 \eeq
  where $f_0 (y) = \cos{(\beta \log{y} + \phi)}$,
  $f_{m \geq 1} (y)$ are obtained in the order-by-order expansion in $\Delta_n (\omega_m)/\omega_m$, and  $C_n$  is yet unknown factor.
    Evaluating the few first $f_m (y)$, we obtain that they are also oscillating functions of
    $\log{y}$ with $y-$dependent prefactors of order
      $(1/y^{2m})^{1/\gamma - 1/2}$. The perturbative expansion holds as long as $C_n < y^{(1/\gamma-1/2)}$.  In this range,  one can approximate  the r.h.s. of (\ref{nn_5_1}) by the $m=0$ term for order-by-magnitude estimates.
      The function $f_0 (y)$ changes sign $n$ times between $y = O(1)$ and $y_{min} \sim e^{-n\pi/\beta}$.  It is natural to associate $y_{min}$ with the lower boundary of the perturbative expansion. Doing this,
        we obtain a discrete set of $C_n$:
      \beq
      C_n \sim e^{\frac{-n\pi}{\beta} \left(1/\gamma -1/2\right)}
      \label{nn_5_2}
 \eeq
  For large $n$,  $\Delta_n (y) \approx  {\bar g}  y^{1/2}  C_n
 \cos{(\beta \log{y} + \phi)}$. For $n = O(1)$, $C_n = O(1)$, and one needs to keep the full series
 in (\ref{nn_5_1}). This reasoning also yields $\Delta_n (0) \sim  y^{1/2}_{min} C_n \sim (y_{min})^{1/\gamma} \sim {\bar g}  e^{-n\pi/(\beta \gamma)}$.

 \subsubsection{$\Delta_n (\omega_m)$ for $\gamma$ close to $2$.}
\label{sec:near_2}

\begin{figure}
  \includegraphics[width=8cm]{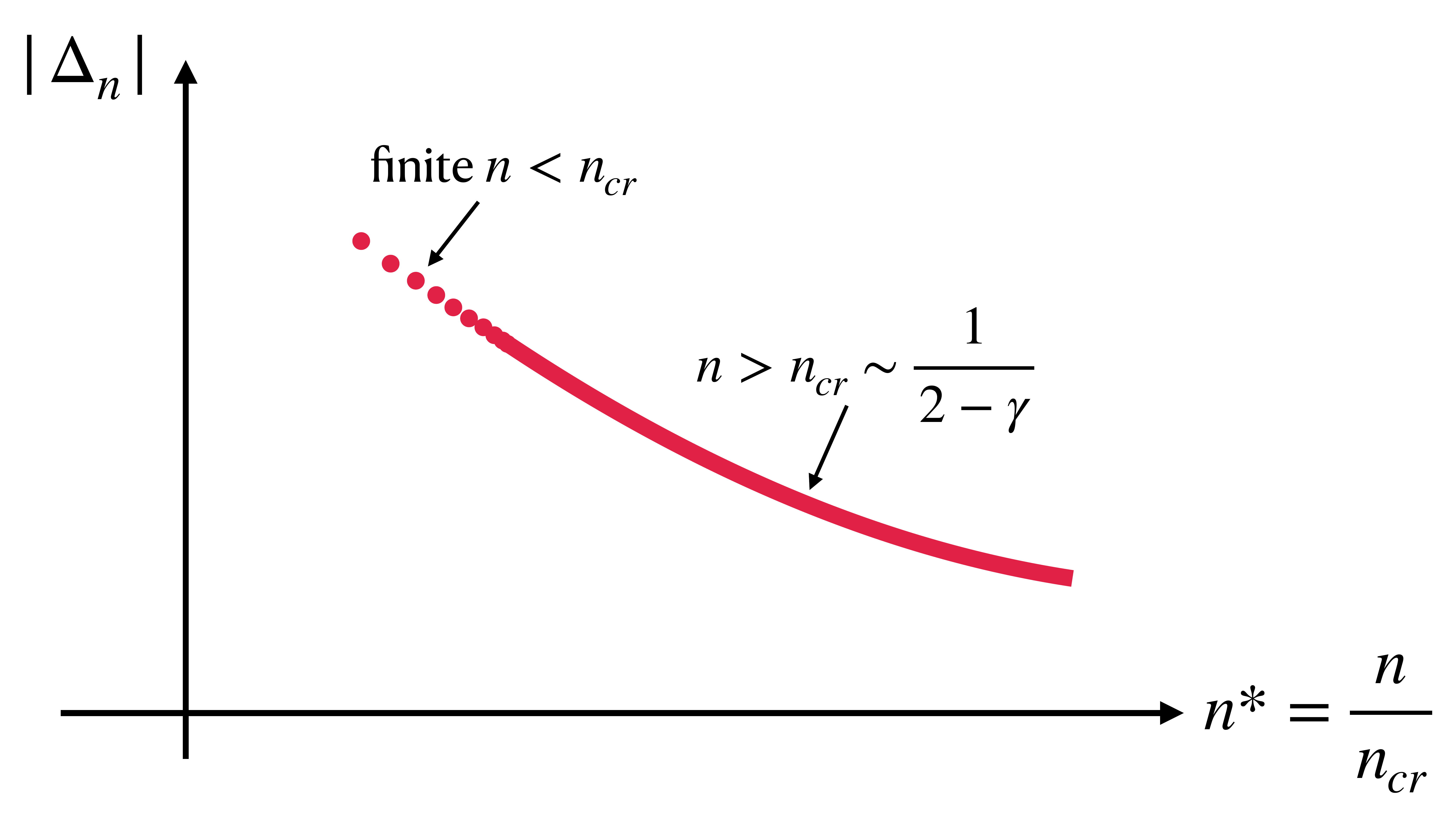}
  \caption{A schematic dependence of $\Delta_n$ on $n$ at $\gamma\to2$,  at some fixed $\omega_m$. All $\Delta_n$ with finite $n$ tend to the same value, while the magnitude of $\Delta_{n\to\infty}$ depends on the ratio $n^*=n/n_{cr}$, where $n_{cr} \sim 1/(2-\gamma)$.}\label{fig:sketch_Dn}
\end{figure}

\begin{figure}
   \includegraphics[width=10cm]{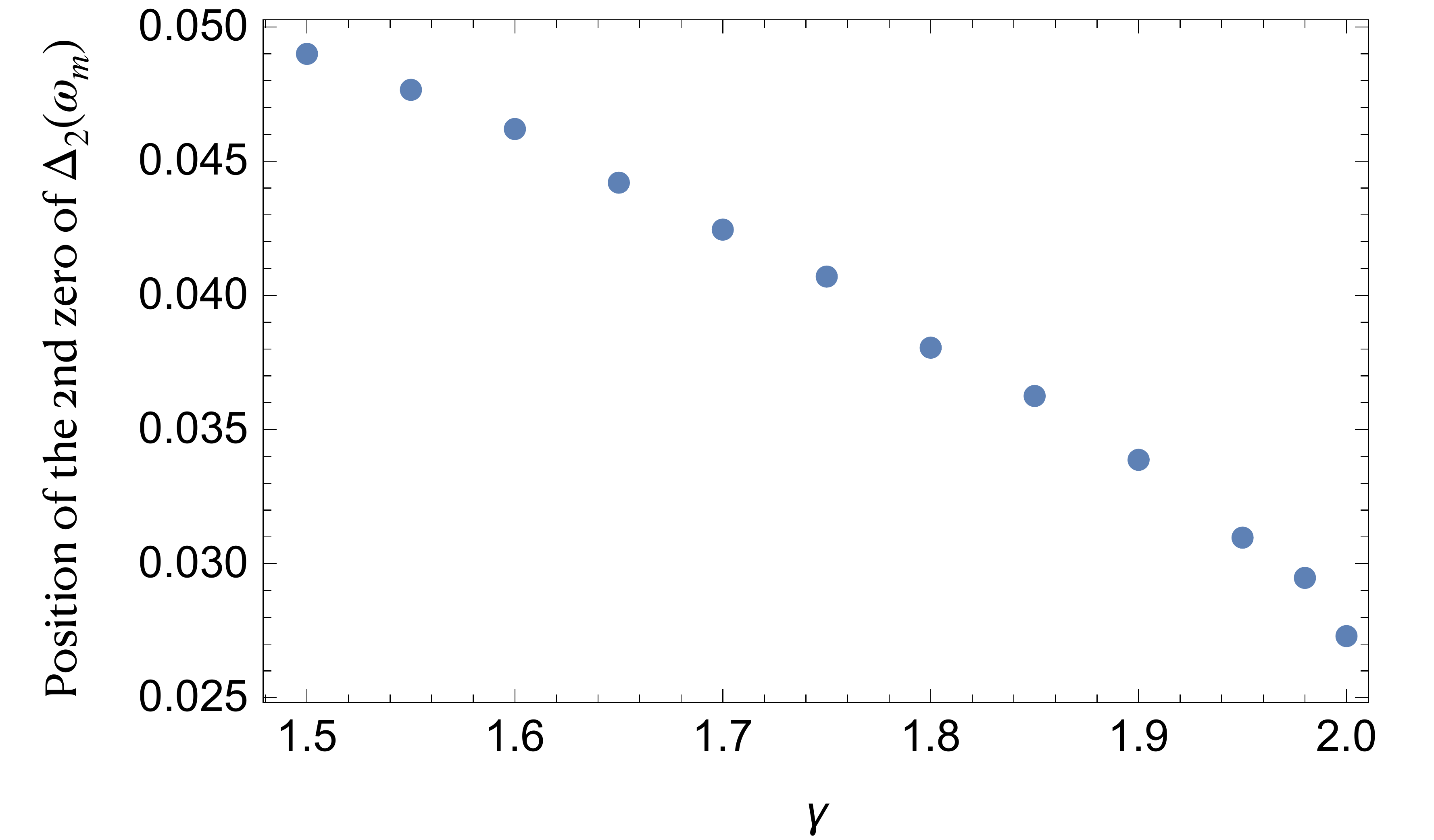}
   \caption{The positions of the second zero of $\Delta_2 (\omega_m)$,  as a functions of $\gamma$.  This  frequency decreases as $\gamma \to 2$, implying that sign change at the highest $\omega_m$ occurs at progressively smaller frequency. The gap function $\Delta_{2} (\omega_m)$ has been obtained by solving the linearized gap equation at the corresponding $T_{p,2}$. We conjecture that the functional form of
   $\Delta_{2} (\omega_m)$ does not change substantially between $T_{p,2}$ and $T=0$.}\label{fig:Tp1_1}
 \end{figure}

\begin{figure}
   \includegraphics[width=15cm]{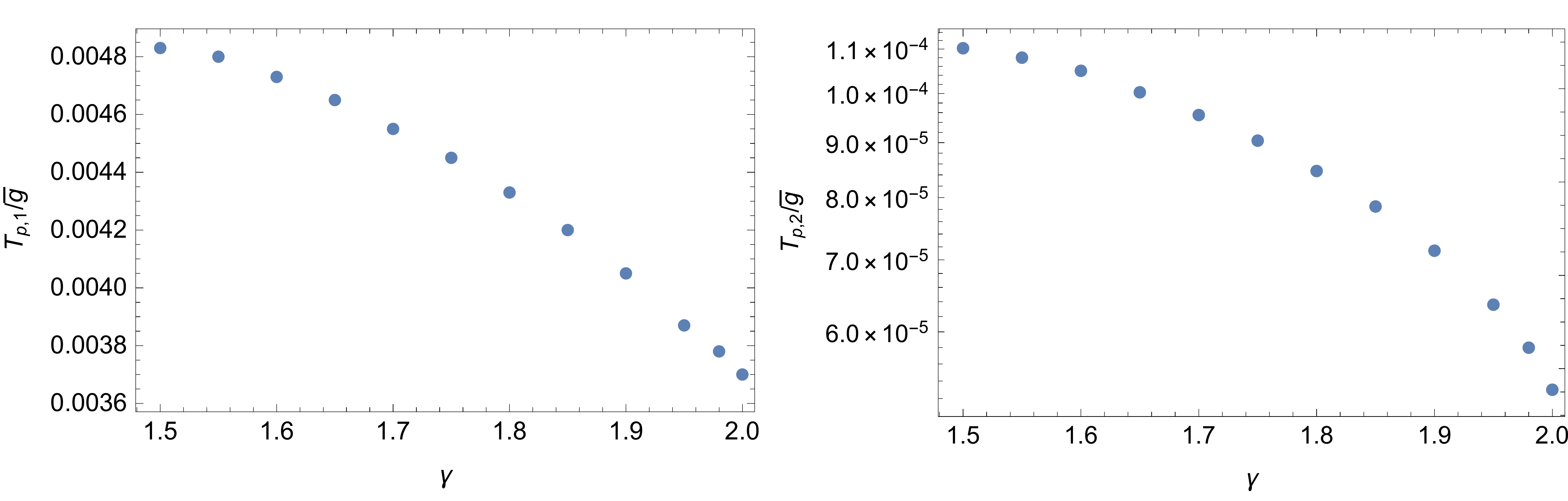}
   \caption{The onset temperature $T_{p,1}$(left) and $T_{p,2}$(right) vs $\gamma$. Within our accuracy, $T_{p,1}$ and $T_{p,2}<T_{p,1}$ vanish at $\gamma =2$.
     Other $T_{p,n}$ with $n >2$ are smaller than $T_{p,2}$ and also vanish at $\gamma \to 2$.}\label{fig:Tp1}
 \end{figure}

 Eq. (\ref{nn_5_2}) shows that for a generic $\gamma$, $C_n = O(1)$ for $n = O(1)$, and decays exponentially at larger $n$.
 However, for $\gamma$ close to $2$, the dependence on $n$  in (\ref{nn_5_2})  is actually via the product
 $n^*=n (2-\gamma) \sim n/n_{cr}$.
  In the limit $\gamma \to 2$, $n^* \to 0$ for any finite $n$. The corresponding  $C_n$
 then become independent on $n$ and coincide with $C_0$ up to
  corrections, which become relevant only at the smallest $y$.  At the same time,  at $n \to \infty$, $n^*$
     becomes a continuous variable, whose value depends on how the double limit $n \to \infty$ and $\gamma \to 2$ is taken.  Then $C_{n\to \infty}$ and $\Delta_{n \to \infty}$
      become continuous functions of $n^*$.
   \footnote{This is similar to the behavior of $\Delta_n$  for $\gamma \approx  1$  in the extended
   $\gamma$ model, when the extension parameter $N \neq 1$.}  We illustrate this in schematically in Fig.\ref{fig:sketch_Dn}.
     We show the numerical evidence for this behavior in Figs.\ref{fig:Tp1_1} and \ref{fig:Tp1}. We use the fact that each $\Delta_n (\omega_m)$ terminates at its own $T_{p,n}$ and analyze the functional forms of the eigenvalues $\Delta_n (\omega_m)$ at $T_{p,n}$,   assuming that these forms do not change significantly between $T_{p,n}$ and $T=0$.
 In Fig.\ref{fig:Tp1_1} we plot the largest frequency at which $\Delta_{1,2} (\omega_m)$ change sign   We see that these frequencies get progressively smaller as $\gamma$ approaches $2$, while at larger $\omega_m$ the functional forms of $\Delta_{1} (\omega_m)$ and $\Delta_{2} (\omega_m)$ are the same as of $\Delta_0 (\omega_m)$.
   We expect the same to hold for other finite $n$.
 On a more careful look, we find that the  shift of oscillations to smaller frequencies is related to the reduction of the corresponding $T_{p,n}$. Namely, we verified that the largest frequency, at which $\Delta_n (\omega_m)$ changes sign,   scales with the corresponding $T_{p,n}$.   In Fig. \ref{fig:Tp1} we
 show $T_{p,n}$ for $n=1,2$ as a function of $\gamma$.
We see that $T_{p,n}$ decreases  as  $\gamma$ increases toward $2$ and, within our accuracy,
 vanishes at $\gamma =2$.

Now, each gap function $\Delta_n (\omega_m)$ generates a certain condensation energy $E_{c,n}$.
For a generic $\gamma <2$, the spectrum of $E_{c,n}$ is discrete, and $E_{c,0}$  is the largest by magnitude. Then, at  low $T \ll T_{p,0}$, only the $n=0$ solution matters, while the existence of other $E_{c,n}$ affects the system behavior  only at $T \leq T_{p,0}$.
As $\gamma$ increases towards two,
 the spectrum of $E_{c,n}$ becomes denser, and
 $(E_{c,0}- E_{c,n})/E_{c,0}$ progressively gets smaller for any finite $n$.   At $\gamma \to 2$,  the spectrum of $E_{c,n}$ can be viewed  as almost continuous spectrum with a small gap, i.e., there emerges a branch of low-energy "longitudinal" gap fluctuations.  These fluctuations affect the system behavior beginning at a  progressively smaller $T$, as $\gamma$ approaches two.  At $\gamma = 2-0$,
   the spectrum of condensation energies becomes a gapless continuous function $E_{c} (n^*)$ with $E_c (0) = E_{c,0}$ and
    $E_c (\infty) = E_{c, \infty}$.
      In the next Paper V, where we specifically analyze the case $\gamma =2$, we  show there that gapless longitudinal fluctuations give rise to
     singular downward renormalization of the stiffness.

\section{Gap function on the real axis}
\label{sec:real_axis}

The analysis on the Matsubara axis is sufficient for the computation of the condensation energy at $T=0$ and  thermodynamic properties at $T>0$.   For the analysis of transport and spectroscopic properties, one needs to know the gap function on the real frequency axis.  In this section we obtain $\Delta_n (\omega)$ for real $\omega$.
 The gap function on the real axis is complex: $\Delta_n (\omega) = \Delta'_n (\omega) + i \Delta^{''} (\omega)$. It is convenient to introduce its $\omega-$dependent phase as $\Delta_n (\omega) = |\Delta_n (\omega)|e^{i\eta (\omega)}$. As usual, $\eta (\omega)$ is defined up to a constant.

 \subsection{Eliashberg gap equation on the real axis}
\label{sec:real_axis_Eliashberg}

Eliashberg gap equation in real frequencies cannot be obtained by
simply replacing $\omega_m$ by $-i\omega$ in (\ref{ss_11}) as  $\Delta_n (\omega_m \to -i\omega)$  would have branch cuts
 in the upper frequency half-plane, while the actual $\Delta (z)$ must be an analytic function.

 Below we follow the approach, suggested in Refs~\cite{Karakozov_91,Marsiglio_91,combescot,Wu_19_1}
  for electron-phonon problem, i.e., convert to real axis using the spectral representation $f(i\omega_m) = (1/\pi) \int dx f^{''} (x)/(x -i\omega_m)$ and, where possible, use $\Delta (\omega_m)$, which we treat as the known function. For our $\gamma$ model this procedure yields
\begin{equation}
\Delta(\omega) B(\omega) = A(\omega) + C(\omega)
\label{4_7}
\end{equation}
where, as before, $D(\omega )=\Delta (\omega )/\omega $, and
	\begin{equation}
		\begin{aligned}
			A(\omega)&=\pi T\sum_{\omega_m>0}\frac{D(\omega_m)}{\sqrt{1+D^2(\omega_m)}}\left(V(\omega_m+i\omega)+V(\omega_m-i\omega)\right)\\
			B(\omega)&=1+\frac{i\pi}{\omega} T\sum_{\omega_m>0}\frac{1}{\sqrt{1+D^2(\omega_m)}}\left(V(\omega_m+i\omega)-V(\omega_m-i\omega)\right)\\
			C(\omega)&=\frac{i}{2}\int_{-\infty}^\infty d\Omega V^{''}(\Omega)\frac{D(\omega-\Omega)-D(\omega)}{\sqrt{1-D^2(\omega-\Omega)}}\left(\tanh\frac{\omega-\Omega}{2T}+\coth\frac{\Omega}{2T}\right)
		\end{aligned}
\label{last_1}
	\end{equation}
The functions $A(\omega)$ and $B(\omega)$ are obtained by just rotating the gap equation from $\omega_m$ to $-i\omega$, and the function $C(\omega)$ contains  the extra term that cancels out a parasitic contribution from the branch cuts.

The interaction $V(\Omega)$ on the real axis is $V(\omega) = V' (\Omega) + i V^{''} (\Omega)$, where
\beq
V' (\Omega) = \left(\frac{{\bar g}}{|\Omega|}\right)^\gamma \cos{\frac{\pi \gamma}{2}},~~
V^{''} (\Omega) = \left(\frac{{\bar g}}{|\Omega|}\right)^\gamma \sin {\frac{\pi \gamma}{2}}
 {\text{sign}} \omega
\eeq
Observe that on the Matsubara axis, $V(\Omega_m) = ({\bar g}/|\Omega_m|)^\gamma$ positive (attractive in our notations), on a real axis $V^{'} (\omega)$ is negative (repulsive) for $\gamma >1$.

Below we focus on the case of zero temperature.
At $T=0$, the expressions for $A, B$ and $C$  reduce to
\begin{equation}
		A(\omega)=\bar{g}^\gamma\frac{2}{\pi}\sin\frac{\pi\gamma}{2}\int_0^\infty d\omega_m\frac{D(\omega_m)}{\sqrt{1+D^2(\omega_m)}}\int_0^\infty\frac{d\Omega}{\Omega^{\gamma-1}}\frac{\Omega^2-
\omega^2+\omega^2_m}{[(\Omega-\omega)^2+\omega^2_m][(\Omega+\omega)^2+\omega^2_m]}
\label{4_12_1}
	\end{equation}
	\begin{equation} B(\omega)= \left(1 + \bar{g}^\gamma\frac{4}{\pi}\sin\frac{\pi\gamma}{2}\int_0^\infty\frac{d\omega_m}{\sqrt{1+D^2(\omega_m)}}
\int_0^\infty\frac{d\Omega}{\Omega^{\gamma-1}}\frac{\omega_m}{[(\Omega-\omega)^2+\omega^2_m]
[(\Omega+\omega)^2+\omega^2_m]}\right)
\label{4_12_2}	
\end{equation}
\begin{equation}
	C(\omega) = i {\bar g}^\gamma~\sin{\frac{\pi \gamma}{2}}~ {\text{sign}} \omega \int_{0}^{|\omega|}  \frac{d\Omega}{\Omega^\gamma}~
~\frac{D(|\omega| - \Omega) -D(|\omega|)}{\sqrt{1 - D^2(|\omega| - \Omega)}}
\label{4_12}
\end{equation}
At large $\omega$,
 \beq
B(\omega) \approx 1 + \left(\frac{\bar g}{|\omega|}\right)^\gamma I_\gamma ,
\label{4_14_2}
 \eeq
  where
  \beq
  I_\gamma = \frac{i}{2} \int_0^\infty
\frac{(x-i)^\gamma -(x+i)^\gamma}{(x^2+1)^\gamma}
  \eeq
  In the two limits,  $I_1 = \pi/2, I_2 = 1$.

\subsection{Frequency dependence of $\Delta_n (\omega)$.  The case $n=\infty$.}
\label{sec:real_n_infinite}

\begin{figure}
	\includegraphics[width=12cm]{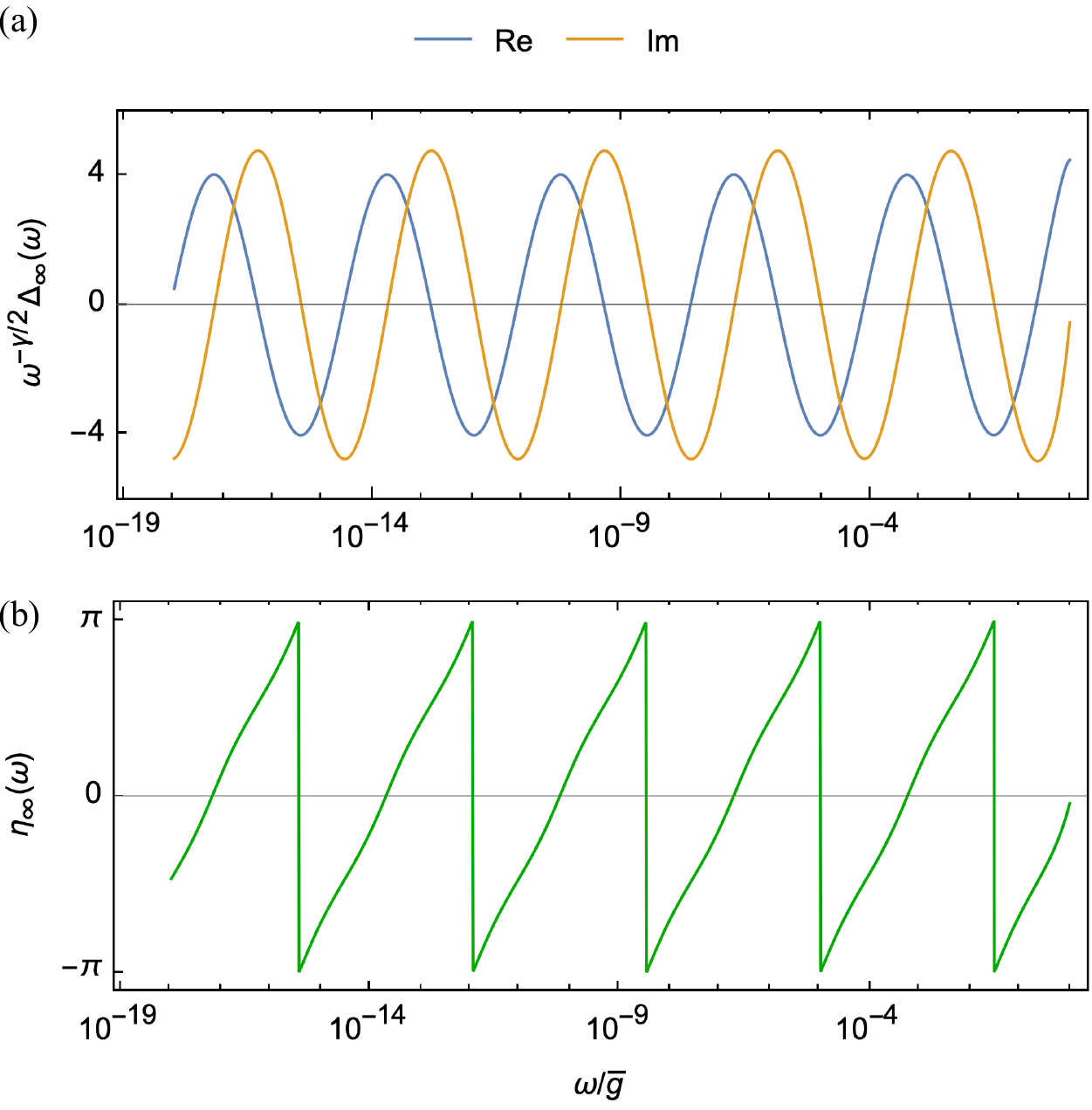}
	\caption{The complex gap function $\Delta_{\infty} (\omega)$ along the real frequency axis and its  phase $\eta_{\infty} (\omega)$, constrained to $(-\pi,\pi)$, for $\gamma=1.91$.
  $\Delta_{\infty} (\omega)$ oscillates an infinite number of times between $\omega=0$ and $\omega/{\bar g}=O(1)$. For better understanding of the total phase variation $\delta \eta$ , we count oscillations starting from a small but finite $\omega_{min} \sim10^{-17}\bar g$. For this choice there are 10 oscillations at  $\omega < {\bar g}$ (panel a).
Accordingly,
$\eta_\infty(\omega)$ changes by  $2\pi$ five times
(panel b) such that the total phase variation
is $10 \pi$. }
\label{fig:Delta_eta_n16}
\end{figure}

We verified that the exact $\Delta_\infty (\omega)$  can be obtained by converting Eq. (\ref{nn_2}) from Matsubara to real axis by a rotation, i.e., by replacing $\omega_m$ by $-i \omega+ \delta$. Under such rotation,  $e^{-ik \log{y}} = e^{-ik \gamma \log{|\omega_m|/{\bar g}}}$  transforms into $e^{-ik \gamma \log{|\omega|/{\bar g}}} e^{-(k \pi \gamma/2) \text{ sgn} \omega }$.
At small $\omega < {\bar g}$, $\Delta'_\infty (\omega)$ and $\Delta^{''}_\infty (\omega)$ oscillate as functions of $\log|\omega|$ down to the lowest frequencies.
  In explicit form, we have in this regime
 \beq
 \Delta_\infty (\omega) = C_\infty \left(\frac{|\omega|}{\bar g}\right)^{\gamma/2} e^{-\frac{i\pi \gamma}{4} \text{sgn} \omega} \cos{\left(\beta \gamma \left(\log{\frac{|\omega|}{\bar g}} - i\frac{\pi}{2} \text{sgn} \omega\right) + \phi\right)}
\label{nn_6_1_a}
\eeq
  where $C_\infty$ is infinitesimally small.  Separating real and imaginary parts, we obtain
 \bea
&& \Delta'_\infty (\omega) = C_\infty \left(\cos {\frac{\pi \gamma}{4}} \cos{D_1} \cosh{D_2} + \sin {\frac{\pi \gamma}{4}} \sin{D_1} \sinh{D_2} {\text{sign}} \omega\right) \nonumber \\
&& \Delta^{''}_\infty (\omega) = C_\infty \left( \cos {\frac{\pi \gamma}{4}} \sin{D_1} \sinh{D_2}  -
\sin {\frac{\pi \gamma}{4}}  \cos{D_1} \cosh{D_2} {\text{sign}} \omega \right)
\label{e_4}
\eea
 where
 \bea
&&D_1 = \beta \log{\left(\frac{|\omega|}{\bar g}\right)^\gamma}  + \phi, \nonumber \\
&&D_2 = \frac{\pi \beta \gamma}{2} {\text{sign}} \omega
\label{e_4_1}
 \eea
We show $\Delta'_\infty (\omega)$, $\Delta^{''}_\infty (\omega)$, and the phase $\eta_\infty (\omega)$ for $\omega < {\bar g}$
 in Fig. \ref{fig:Delta_eta_n16} for $\gamma = 1.91$.
 We set the lowest frequency in  at $10^{-17} {\bar g}$. In this case the total phase variation up to
  $O({\bar g})$ is $10 \pi$. This is consistent with 10 vortices on the Matsubara axis
  (Fig.~\ref{fig:semicircle}) as each vortex gives rise to a phase
change
of $\pi$ for positive $\omega$ on the real axis and another phase
change
of $\pi$ for negative $\omega$.

 We now move to $\omega > {\bar g}$ and use Eq. (\ref{4_7}).
 At the largest $\omega$,  we have
 $B(\omega) \approx 1$ and
 \bea
 && A(\omega) =  \left(\frac{\bar g}{|\omega|}\right)^\gamma Q_{\gamma,\infty} \cos{\frac{\pi \gamma}{2}} \nonumber \\
 &&  C(\omega) \approx i  \left(\frac{\bar g}{|\omega|}\right)^\gamma \sin{\frac{\pi \gamma}{2}} {\text{sign}} \omega {\bar Q}_{\gamma,\infty}
 \label{nn_6_2}
 \eea
 where
 \bea
&&Q_{\gamma,\infty} = \int_0^\infty d\omega_m \frac{\Delta_{\infty} (\omega_m)}{\omega_m}, \nonumber \\
&&{\bar Q}_{\gamma,\infty} = \int_0^\infty  d \omega \frac{\Delta_\infty(\omega)}{\omega}.
 \label{nn_6_3_1}
 \eea
The two integrals in (\ref{nn_6_3_1}) are actually identical.  To prove this, we recall that
$D_\infty (\omega)=
\Delta_\infty (\omega)/\omega$ satisfies KK relation
 $D^{''} (\omega) = -(2/\pi) P\int_0^\infty dx D^{'} (x)x/(x^2-\omega^2)$, where $P$ stands for principle part,   and that
  $D_\infty(\omega_m) = (2/\pi) \int_0^\infty dx D^{'} (x)x/(x^2 + \omega^2_m)$.
   Using these relations, we obtain
    $\int_0^\infty  d \omega_m  D_\infty (\omega_m) =  \int_0^\infty  d \omega D^{'} (\omega)$. Using further that $\int_0^\infty  d \omega D^{''} (\omega) = (2/\pi) \int_0^\infty dx x D^{'} (x) P \int_0^\infty d \omega/(x^2-\omega^2) =0$  we find
 $ {\bar Q}_{\gamma,\infty} = \int_0^\infty d \omega D_\infty(\omega) =
    \int_0^\infty d \omega  D^{'}(\omega) = \int_0^\infty  d \omega_m  D_\infty (\omega_m)
     = Q_\gamma$.  Hence, at the largest $\omega$,
     \beq
     A(\omega) + C(\omega) = e^{i(\pi \gamma/2) {\text sgn} \omega} \left(\frac{\bar g}{|\omega|}\right)^\gamma
     \eeq
     Substituting into (\ref{4_7}) and using \eqref{4_14_2}, we obtain
  \beq
 \Delta_\infty (\omega) \propto \left(\frac{\bar g}{|\omega|}\right)^\gamma   e^{(i\pi \gamma/2) {\text{sign}} \omega} Q_{\infty \gamma}
 \label{4_14_3}
 \eeq
 This result could also be obtained by a direct rotation of $\Delta_\infty (\omega_m) \propto 1/|\omega_m|^\gamma$ from Matsubara to real axis.

 \begin{figure}
	\includegraphics[width=12cm]{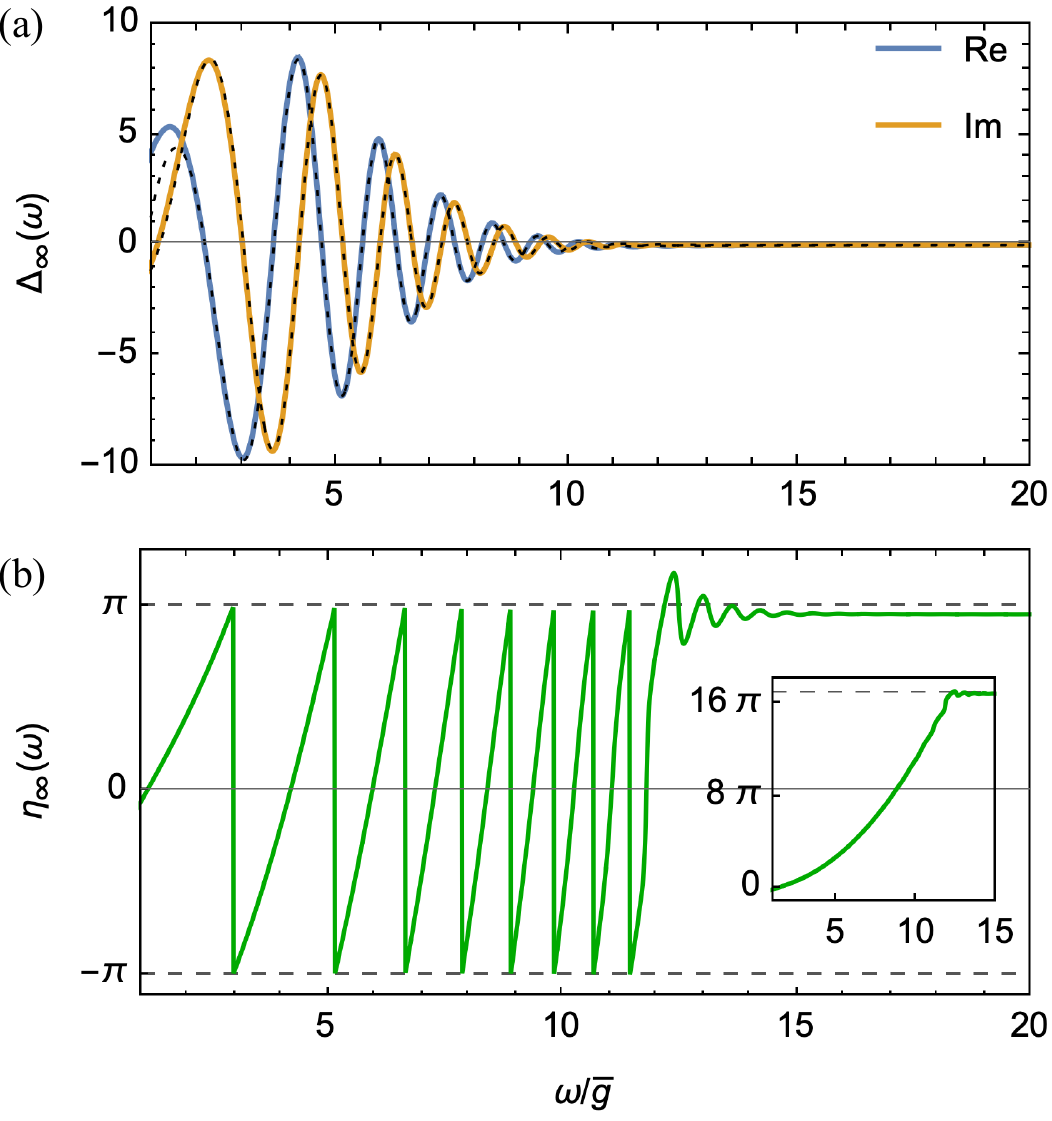}
	\caption{ (a) Comparison between the exact $\Delta_{\infty} (\omega)$ along the real frequency axis (blue and orange thick lines) and the universal contribution to $\Delta_\infty(\omega)$ from Eq.~(\ref{nn_6_6_1}) (black dotted lines) for $\gamma=1.91$. The agreement is nearly perfect. (b) Variation of the phase of the gap function, $\eta_\infty (\omega)$, between $\omega \sim {\bar g}$ and $\omega = \infty$. For convenience of presentation we confined $\eta_\infty(\omega)$ to $(-\pi,\pi)$, up to small variations.   In the inset we plot the continuous $\eta_\infty(\omega)$.   We see that the total phase variation between $\omega\sim\bar{g}$ and $\omega=\infty$ is $16\pi+\pi\gamma/2$. }
\label{fig:Delta_real_compare}
\end{figure}

 For $\gamma \geq 1$ there is a single crossover between the two limiting forms of
 $\Delta_\infty (\omega)$, Eqs. (\ref{nn_6_1}) and (\ref{4_14_3}).
  However, for $\gamma \leq 2$, the new intermediate regime emerges at $\omega > {\bar g}$, as we now demonstrate. In this regime,  $\Delta'_\infty (\omega)$ and $\Delta^{''}_\infty (\omega)$ again oscillate, but with the period set by a power of $\omega$ rather than by $\log|\omega|$.

To show this, we recall that in Sec. \ref{infnMatsubara} we found an exponentially small  oscillating component
   of $\Delta_\infty (\omega_m)$ on the Matsubara axis at $\omega_m > {\bar g}$.  We now convert this component
    onto the real axis.  Replacing $\omega_m$ by $-i\omega + \delta$,  we obtain  for $\omega >0$:
     \beq
     \Delta_\infty (\omega) = \int_{k_{min}}^\infty dk \left(e^{-\pi k (1 + \gamma/2)} e^{-ik S_k(\omega)} +
      e^{-\pi k (1 - \gamma/2)} e^{ik S_k(\omega)}\right)
      \label{nn_6_3}
      \eeq
      where  $k_{min} = O(1)$ and  $S_k (\omega) = \log{\left[(|\omega|/{\bar g})^\gamma (e/A_\gamma k)^{\gamma-1}\right]}$
    Comparing this result with Eq. (\ref{nn_3_4}) we see that the exponentially small factor $e^{- \pi |k|}$
      splits into $e^{- \pi k (1+\gamma/2)}$ and $e^{-\pi k(1-\gamma/2)}$.  The first term is smaller and the second one is larger than the original term. Keeping only the larger term and evaluating the integral over $k$ in the same way as in Sec. \ref{infnMatsubara}, we obtain the oscillating term on the real axis:
\begin{equation}
\Delta_\infty (\omega) \propto
|\omega|^{\frac{1}{2}\frac{\gamma}{\gamma-1}}
\exp \left(- \frac{2-\gamma}{A_\gamma}  \left(\frac{|\omega|}{\bar{ g}} \right)^{\frac{\gamma }{\gamma -1}}\left[
\frac{\pi }{2}
-i \frac{\gamma-1}{2-\gamma } \left(1- \frac{\pi^2}{2(\gamma-1)^2}\right)
 \right]
\right)
  \label{nn_6_4}
\end{equation}
  For $\gamma \leq 2$, the exponential part of the prefactor  is small in $2-\gamma$, and the power-law part
     increases with $\omega$.  As the result, this oscillating contribution  exceeds the one from Eq. (\ref{4_14_3}) for frequencies between $O({\bar g})$ and  $\omega_{cr}$, where
      \beq
 \omega_{cr} \sim {\bar g} \left(\frac{|\log{(2-\gamma)}|}{2-\gamma}\right)^{1/2} \gg {\bar g}.
 \label{nn_6_1_1}
\eeq
As a result,  at ${\bar g} < \omega < \omega_{cr}$,  $\Delta^{'} (\omega)$ and
  $\Delta^{''} (\omega)$ oscillate as functions of $(|\omega|/{\bar g})^{\gamma/(\gamma-1)}$ with weakly, yet exponentially decaying prefactor. To the leading order in $2-\gamma$, Eq. (\ref{nn_6_4}) gives
    \beq
  \Delta_\infty (\omega) \propto |\omega| e^{-(1-\gamma/2) (|\omega|/{\bar g})^2} e^{i (|\omega|/{\bar g})^2/\pi}
  \label{nn_6_5}
 \eeq
We used $A_2 = \pi$.
For completeness, we computed the subleading term under $e^{i ...}$.
It changes
 Eq. (\ref{nn_6_5}) into
    \beq
  \Delta_\infty (\omega) \propto |\omega| e^{-(1-\gamma/2) (|\omega|/{\bar g})^2} e^{i \left((|\omega|/{\bar g})^2 + \log{((|\omega|/{\bar g})^2)}\right)/\pi}
  \label{nn_6_6_1}
 \eeq
In Fig.~\ref{fig:Delta_real_compare} (a), we compare the exact $\Delta_\infty (\omega)$ with Eq. (\ref{nn_6_6_1})  for $\gamma = 1.91$.
 We see that the agreement is quite good, and the range of $\omega^2$ oscillations is quite wide for this $\gamma$.

In Fig.~\ref{fig:Delta_real_compare} (b), we show the variation of the phase $\eta (\omega)$ between $\omega \sim {\bar g}$ and $\omega = \infty$. We see that
the phase changes by $2\pi m$,
where $m$ is an integer, and the total phase variation is $2\pi m + \pi \gamma/2$.  The last piece just follows from (\ref{4_14_3}), and the first one is due to oscillations given by (\ref{nn_6_5}).  The integer $m$ increases one-by-one
 as $\gamma$ increases towards $2$. In Sec. \ref{sec:vortices}  below we associate $2\pi m$ phase variation with the emergence of $m$ dynamical vortices in the upper
 half-plane of frequency.

  \begin{figure}
	\includegraphics[width=12cm]{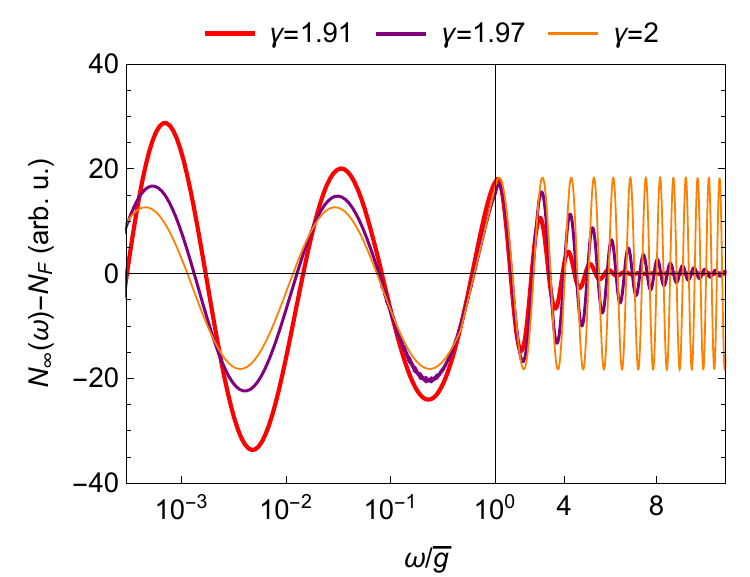}
	\caption{The density of states $N_\infty(\omega)$, counted from the normal state value $N_F$. To better display oscillations we used different scales at $\omega>\bar{g}$ and $\omega<\bar{g}$.}
\label{fig:dos_infty}
\end{figure}

In Fig.~\ref{fig:dos_infty}, we plot the DoS $N_\infty(\omega)$ for several $\gamma$.
 The DoS is defined as  $N(\omega) = (-1/\pi) {\text{Im}} G_l (\omega)$, where
 \beq
 G_l (\omega) = -i \pi \sqrt{\frac{\omega^2}{\omega^2 - \Delta^2 (\omega)}}
\label{nn_6_6_2}
  \eeq
 is the retarded
 Green's function, integrated over the dispersion.
 \footnote{ One can verify that the extension of (\ref{nn_6_6_2}) to the upper frequency half-plane yields
  the correct $G(\omega_m) = - i \pi {\text sgn} \omega_m  \sqrt{\omega^2_m/(\omega^2_m + \Delta^2 (\omega))}$.  We also  caution that $\int d \omega N(\omega)$ is
  not equal to the number of particles. The latter is expressed via a local time-ordered Green's function.  }
 Because $\Delta_\infty (\omega)$ is vanishingly small,
   \beq
   N_\infty (\omega) = N_F \left (1 - {\text Re} \frac{\Delta^2_\infty (\omega)}{\omega^2} \right)
   \label{nn_6_6_3}
   \eeq
  where $N_F$ is the DoS in the normal state. We see that $N_\infty(\omega)$  oscillates around
 $N_F$ up to $\omega \sim \omega_{cr}$. At $\omega < {\bar g}$, the period of oscillations is set by $\log {\omega}$, and at ${\bar g} < \omega < \omega_{max}$, the period is set by $\omega^{\gamma/(\gamma-1)}$.

\subsection{Frequency dependence of $\Delta_n (\omega)$.  The case $n=0$.}
\label{sec:real_n_0}
We now consider the opposite limit $n=0$. We remind that on the Matsubara axis $\Delta_0 (\omega_m)$ is sign-preserving.
 We show that on the real axis both real and imaginary parts of $\Delta_{0}$ again oscillate
    in a finite frequency range at $\omega > {\bar g}$,
    and the phase of complex $\Delta_0 (\omega)$ is winding by $2\pi m$ in this range.

 At high frequencies  $\omega \gg {\bar g}$, the form of $\Delta_0 (\omega)$  can be readily obtained from (\ref{4_12}) in the same way as it was done in the previous section.  For the case $n=0$ we have
 \bea
&& A(\omega) =  \left(\frac{\bar g}{|\omega|}\right)^\gamma Q_\gamma \cos{\frac{\pi \gamma}{2}} \nonumber \\
&&  C(\omega) \approx i  \left(\frac{\bar g}{|\omega|}\right)^\gamma Q_\gamma\sin{\frac{\pi \gamma}{2}} {\text{sign}} \omega,
 \label{4_14_1}
\eea
 where $Q_\gamma$ is given by \eqref{eq:Qgamma}.
  Substituting
  these
   forms along with $B(\omega) \approx 1$  into (\ref{4_7}), we obtain at $\omega \gg {\bar g}$,
 \beq
 \Delta_0 (\omega) \approx \left(\frac{\bar g}{|\omega|}\right)^\gamma   e^{(i\pi \gamma/2) {\text{sign}} \omega} Q_\gamma
 \label{4_14_3_1}
 \eeq
 This is consistent with
$\Delta_0 (\omega_m) \propto ({\bar g}/|\omega_m|)^\gamma$ on the Matsubara axis.

 At small $\omega$, $\Delta_0 (\omega)  = (A(\omega) + C(\omega))/B(\omega)  \approx \Delta_0 (0)$ where $\Delta_0 (0)$ is the same as we found  in the $\omega_m \to 0$ limit on
 the  Matsubara axis.
   We note, however,  that  the relative strength of $A(\omega), B(\omega)$, and $C(\omega)$   changes between
   $\gamma <1$ and $\gamma >1$.  For $\gamma <1$, $A(\omega)$ and $ B(\omega)$ tend to finite values
    at $\omega \to 0$, while $C(\omega) \propto \omega^{1-\gamma}$ vanishes.
     Then $\Delta_0 (0) = A(0)/B(0)$.
    For $\gamma >1$, $B(\omega)$ and $C(\omega)$  scale as $\omega^{1-\gamma}\sin (\pi \gamma/2)/(2-\gamma)$ at $\omega \to 0$, while $A(\omega) \propto \omega^{1-\gamma}  \sin (\pi \gamma/2)/(1-\gamma)$ is smaller, at least for $\gamma \leq 2$,
 and, moreover,  its sign changes to  negative at $\gamma >1$. The
     gap still
      tends to a finite positive value at $\omega =0$ because once $A(\omega) \ll C(\omega)$,
      $\Delta_0 (0) = \lim_{\omega \to 0} C(\omega)/B(\omega) ={\text const}$.  However, we see  that
        $\Delta_0 (0)$ now scales with $C(\omega)$, which, we recall, is  present in
         the gap equation on the real axis because of the need to cancel out parasitic contributions from the branch cut.
           This hints that for $\gamma >1$ the behavior of $\Delta_0 (\omega)$ on the real axis may be quite different from that of $\Delta_0 (\omega_m)$.

      We now show that this is indeed the case.  We focus on frequencies
       $\omega > {\bar g}$,  where we detected the new behavior  at $n = \infty$.
   Eq. (\ref{4_14_1})  for $A(\omega)$ is valid for all $\omega > {\bar g}$ as typical $\omega_m$ in
    (\ref{4_12_1}) are of order ${\bar g}$.  Similarly, $B(\omega) \approx 1$ for  all $\omega > {\bar g}$.
For $C(\omega)$, we have to be more careful and include
     not only the contribution in (\ref{4_12}) from $\omega - \Omega = O({\bar g})$, which yields (\ref{4_14_2}), but
      also contributions from $\Omega$ immediately below $\omega$ (i.e., $\omega - \Omega  \ll \bar g$).  These  contributions are expressed in terms of derivatives of $D_0(\omega) = \Delta_0 (\omega)/\omega$.
The leading term here is the one with the first derivative: $ D_0(\omega-\Omega) - D_0(\omega) \approx - \Omega {\dot D}_0 (\omega)$. For $\gamma \leq 2$, the integral $\int_0^\omega d \Omega/\Omega^{\gamma-1} = \omega^{2-\gamma}/(2-\gamma)$ is determined by small $\Omega$, and
        the prefactor $1/(2-\gamma)$ compensates the smallness of the overall factor  $\sin (\pi \gamma/2) \approx (\pi/2) (2-\gamma)$ in (\ref{4_12}). This term then gives
       \beq
       -i \frac{\pi}{2}  \omega^{2-\gamma}  \frac{{\dot D}_0 (\omega)}{\sqrt{1-D^2_0 (\omega)}}
       \eeq
       We will also need terms of order $2-\gamma$. For this,  we keep the subleading terms
         with $({\dot D}_0)^2$ and ${\ddot D}_0$.
         Evaluating the prefactors, we obtain
         differential gap equation  in the form ($\omega >0$)
             \beq
 - i \frac{\pi}{2} {\bar g}^\gamma \frac{\omega^{2-\gamma}}{\sqrt{1 -D^2_0(\omega)}}
\left[{\dot D}_0 (\omega) - (2-\gamma) \omega\left(\frac{1}{2}{\ddot D}_0(\omega) +\frac{D_0(\omega) ({\dot D}_0 (\omega))^2}{1 -D^2_0(\omega)}\right) \right] = D_0(\omega)\omega  - Q_\gamma\left(\frac{\bar g}{\omega}\right)^\gamma e^{i\pi \gamma/2}
\label{4_16}
\eeq
  We follow Refs. \cite{Karakozov_91,combescot} and introduce
  $D_0(\omega) = 1/\sin(\phi (\omega))$. Both $D_0(\omega)$ and $\phi (\omega)$ are complex functions
   of $\omega$.  Substituting into (\ref{4_16}) we  obtain
\beq
 {\dot \phi} - \frac{(2-\gamma)}{2} \omega \left(\ddot \phi+ ({\dot \phi})^2 \tan{\phi}\right)  = \frac{2}{\pi{\bar g}^\gamma} \left(\omega^{\gamma-1} - Q_\gamma\frac{\bar g^\gamma}{\omega^2} e^{i\pi \gamma/2} \sin{\phi}\right)
 \label{4_17}
 \eeq
 This equation is similar to the one  for $\gamma =2$ and a finite $T$, analyzed by Combescot in Ref. \cite{combescot}.

 At the highest frequencies, the gap function must obey Eq. (\ref{4_14_3}). This gap function is reproduced if we choose
  \bea
 && \phi' (\omega) = 2\pi m + \frac{\pi}{2} (\gamma +1)  \label{4_17_1} \\
 &&\phi^{''} (\omega) = \log{\frac{2{\bar g}}{Q_\gamma}} + (\gamma+1) \log{\frac{\omega}{{\bar g}}} \label{4_17_2}
 \eea
  where $m$ is integer. We see that at large enough $\omega$, $\phi^{''} (\omega) \gg \phi^{'} (\omega)$.
 On the other hand, at $\omega \geq {\bar g}$,  $\phi' (\omega) \approx (2/\pi \gamma) (\omega/{\bar g})^{\gamma}$, and $\phi^{''}$ is small, of order $(2-\gamma)$.
   We use this to set the
  boundary condition at $\omega = {\bar g}$ as
 \beq
 \phi'  = \frac{2}{\pi \gamma},~~\phi^{''} = a (2-\gamma),
 \label{e_1}
 \eeq
 where $a = O(1)$. We will argue that the solution of (\ref{4_17}) is largely independent on $a$, as long as $a (2-\gamma) \ll 1$.
 To simplify the calculation, below we neglect $\ddot \phi$ term in Eq. \eqref{4_17} and use the boundary condition \eqref{e_1} as the initial condition for the  first order differential equation. We show that the solution of  (\ref{4_17}) without $\ddot \phi$  by itself satisfies the boundary condition (\ref{4_17_1},\ref{4_17_2}).
 We discuss the role of $\ddot \phi$ and the validity of dropping it at the end of this section.

 A simple analysis of the Eq. \eqref{4_17} without  $\ddot \phi$ term shows that $\phi^{''} (\omega)$  rapidly increases  shortly before $\phi' (\omega)$ reaches $\pi/2$. To see this,
   we neglect momentarily  the $Q_\gamma$ term in (\ref{4_17}), which gives rise to a small
    initial $\phi^{''}$ and solve the remaining equation as a quadratic equation on $\dot \phi$. We obtain
      \beq
   \dot \phi = \frac{1-\sqrt{1-\frac{4}{\pi }(2-\gamma )\left(\frac{\omega }{\bar{g}} \right)^{\gamma }\tan \phi (\omega )}}
{(2-\gamma) \omega \tan \phi (\omega)}.
  \label{e_1_1}
  \eeq
  (the sign is chosen to satisfy the initial condition).
  An elementary analysis shows  that $\phi^{''} (\omega)$ emerges at $\omega = \omega_{a}$, where
     when
     \beq
     \left(\frac{\omega_a}{\bar g}\right)^\gamma  \tan (\phi' (\omega_a)) = \frac{\pi}{4} \frac{1}{2-\gamma},
     \label{e_1_1_a}
     \eeq
     This $\omega_a$ is smaller than the one at which $\phi' (\omega)$ reaches  $\pi/2$.  This is essential as in the absence of $\phi^{''}$, the behavior near $\phi' (\omega) =\pi/2$ would be singular.
        Once $\phi^{''}$ is non-zero, the singularity is cut.   Keeping
         the $Q_\gamma$ term, we find that  $\phi^{''}$ is non-zero at all frequencies, where Eq.(\ref{4_17}) is valid, but still rapidly increases around $\omega_a$, specified by (\ref{e_1_1_a}).

 At larger $\omega$,  $\phi^{''} (\omega)$ increases, and eventually  $e^{\phi^{''}}$ becomes larger than one.  At such frequencies,
  $\tan {\phi} \approx i$, and (\ref{4_17}) simplifies to
  \bea
  && {\dot \phi}' (\omega) \approx \frac{2}{\pi} \frac{\omega^{\gamma -1}}{{\bar g}^\gamma}  \nonumber \\
  && {\dot \phi}^{''} (\omega) \approx \frac{2 (2-\gamma)}{\pi^2} \frac{\omega^{2\gamma-1}}{{\bar g}^{2\gamma}}
  \label{4_17_3}
  \eea
   Solving, we find $\phi' (\omega) \approx (2/\pi \gamma) (\omega/{\bar g})^\gamma$, $\phi^{''} (\omega) \approx (2-\gamma)/(\pi^2 \gamma) (\omega/{\bar g})^{2\gamma}$.  Note that these forms are universal and  do not depend on the boundary condition (i.e., on the factor $a$ in Eq.  (\ref{e_1})).  For $\gamma \approx 2$, $\phi' (\omega) \approx  (\omega/{\bar g})^2/\pi$.  To compare with $n=\infty$ case, we computed the subleading term. It comes from the second term  in $B(\omega)$ in (\ref{4_14_2}) and changes $\phi' (\omega)$ to
   \beq
   \phi' (\omega) = \frac{1}{\pi} \left(\left(\frac{\omega}{\bar g}\right)^2 + \log{\left(\frac{\omega}{\bar g}\right)^2}\right).
   \label{nn_6_6}
   \eeq
   Observe that the r.h.s. of (\ref{nn_6_6}) is the same function as we found for $n = \infty$, Eq. (\ref{nn_6_6_1}).
   \begin{figure}
  \includegraphics[width=14cm]{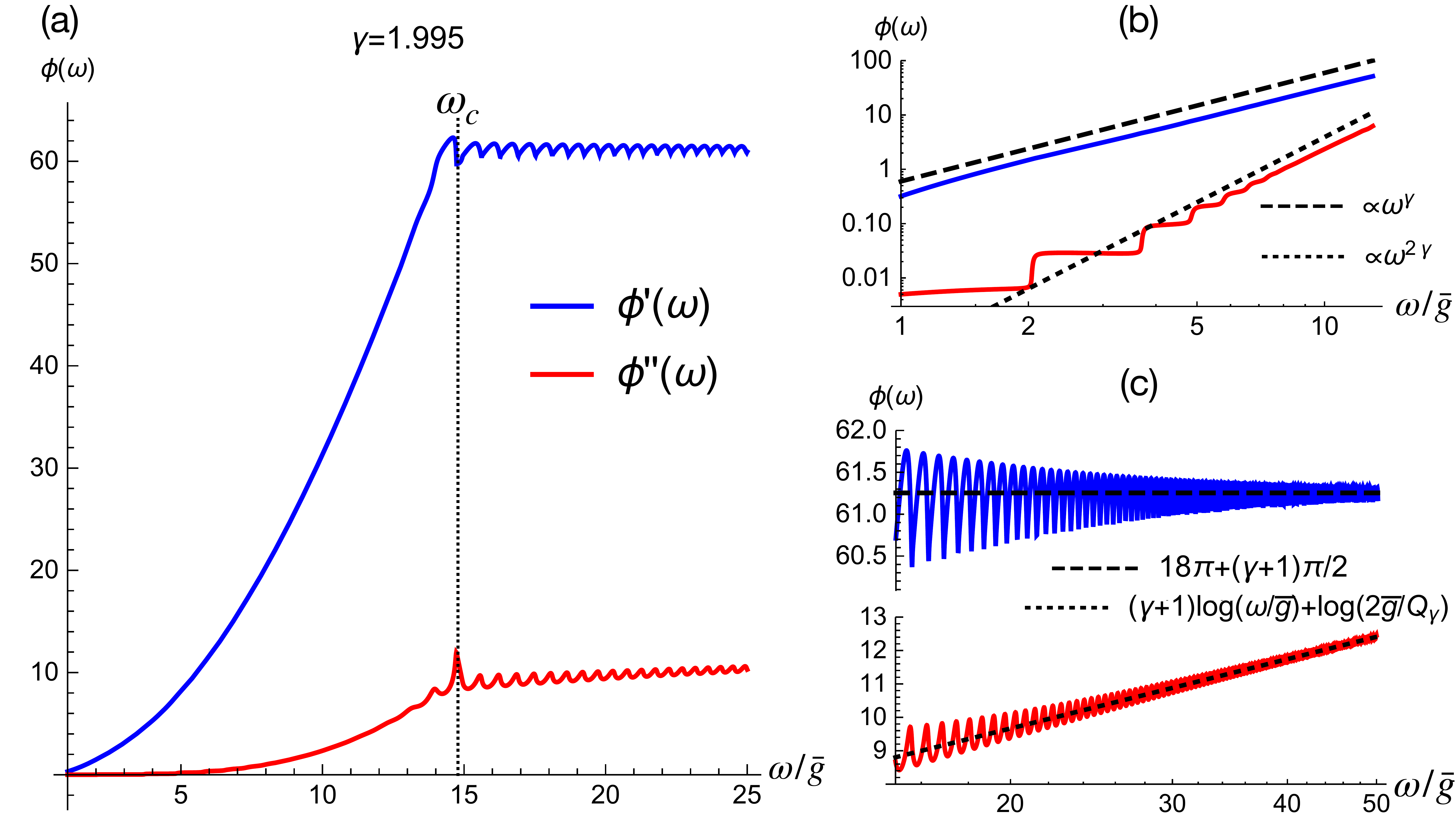}
  \caption{Panel (a). The numerical solution of Eq. \eqref{4_17} for $\gamma=1.995$. Frequency is in units of ${\bar g}$. The crossover at  $\omega_{cr} \approx 15 {\bar g}$ is clearly visible.
   Panels (b) and (c) -- zoom into regions $\omega < \omega_{cr}$ and $\omega > \omega_{cr}$.
   At $\omega<\omega_{cr}$, $\phi'$ grows as $\omega^\gamma$, while $\phi''$ first increases step-like, and then behaves as $\omega^{2\gamma}$. At $\omega>\omega_{cr}$, $\phi'$ saturates at $2\pi m+(\gamma+1)\pi/2$,  where $m=9$ for $\gamma=1.995$, while $\phi^{''}$ increases as $(\gamma +1) \log{\omega/\bar{g}}$. }
   \label{fig:phi}
\end{figure}
    Eq. (\ref{4_17_3}) holds up to $\omega$ at which $e^{\phi^{''}} \sim (\omega/{\bar g})^3$. At larger frequencies, the second term in the r.h.s. of (\ref{4_17}) cannot be neglected, and the  functional form of $\phi (\omega)$ changes.  In Fig.\ref{fig:phi} (a)
    we show the result of numerical solution of (\ref{4_17})
     We see that there is  a single crossover, at $\omega = \omega_{cr}  \sim {\bar g} (|\log{(2-\gamma)}|/(2-\gamma))^{1/(2\gamma)}$,  from $\phi (\omega)$ given by (\ref{4_17_3})  to $\phi (\omega)$ given by (\ref{4_17_1},\ref{4_17_2}).
      In Fig.\ref{fig:phi} (b) and (c)
       we show separately the behavior of $\phi' (\omega)$ and $\phi^{''} (\omega)$ at $\omega < \omega_{cr}$ and $\omega > \omega_{cr}$.  The fits to Eq. (\ref{4_17_3})  Eqs. (\ref{4_17_1},\ref{4_17_2}), respectively,
        are almost perfect.
         Deviations from the high-frequency forms decrease as $1/\omega^{\gamma}$ and oscillate
          as trigonometric functions of $(2/\pi \gamma) (\omega/{\bar g})^\gamma$.
              \begin{figure}
  \includegraphics[width=15cm]{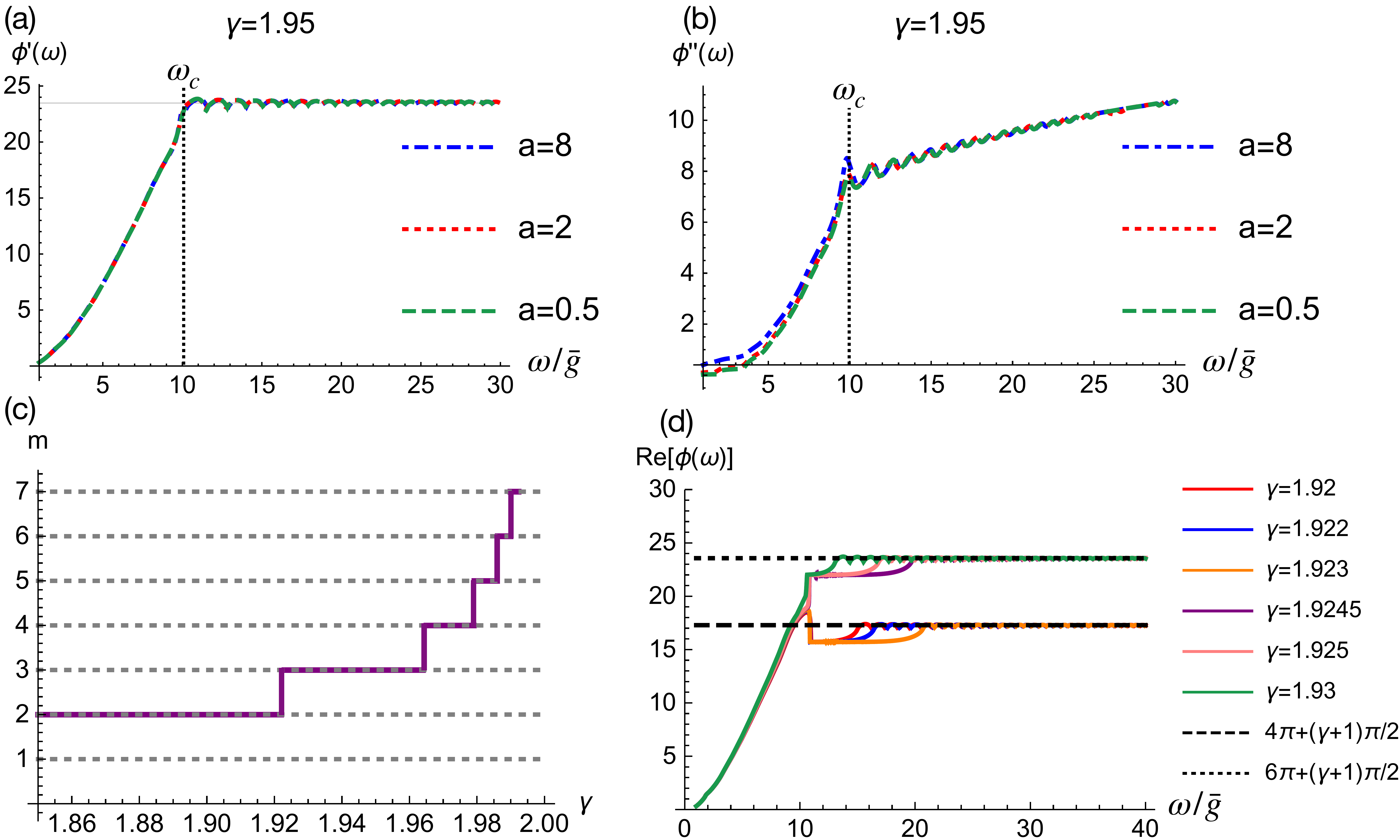}
  \caption{
(a)-(b) The functions  $\phi'(\omega)$ and $\phi''(\omega)$  for different boundary conditions, set by a parameter $a$ in eq.(\ref{e_1}).  The figure shows that the value $2\pi m+(\gamma+1)\pi/2$,  at which $\phi'(\omega)$ saturates is independent on $a$. The behavior of
   $\phi' (\omega)$ in the universal regime, where  both $\phi'$ and $\phi{''}$ are continuous functions of $\omega$, also does not depend on $a$. (c)  Variation of the integer $m$ with $\gamma$.  There is a discrete set of $\gamma_i$, at which $m$ changes by $1$. The set becomes progressively more dense at $\gamma \to 2$. (d) The behavior of $\phi'(\omega)$ near one of these $\gamma_i\approx 1.924$. The value of $\phi'$  at large $\omega$ jumps by $2\pi$ as $\gamma$ passes through $\gamma_i$ and a new vortex moves into the upper half-plane of frequency. }
   \label{fig:phi1_m}
\end{figure}
     In Fig.\ref{fig:phi1_m}((a) and (b))   we show that the value of $\phi'$  for $\omega > \omega_{cr}$ is independent on the parameter $a$ in the boundary
        condition (\ref{e_1}), as long as $a = O(1)$.  At the same time, the value of an integer $m$ in
        (\ref{4_17_1}) changes if we change $\gamma$, as is shown in Fig.\ref{fig:phi1_m}(c).
        Specifically, $m$ jumps to the nearest integer
         at a discrete set of $\gamma_i \leq 2$ (the smaller is $2-\gamma$, the larger is $m$). We demonstrate this in Fig.\ref{fig:phi1_m}(d).

\begin{figure}
  \includegraphics[width=12cm]{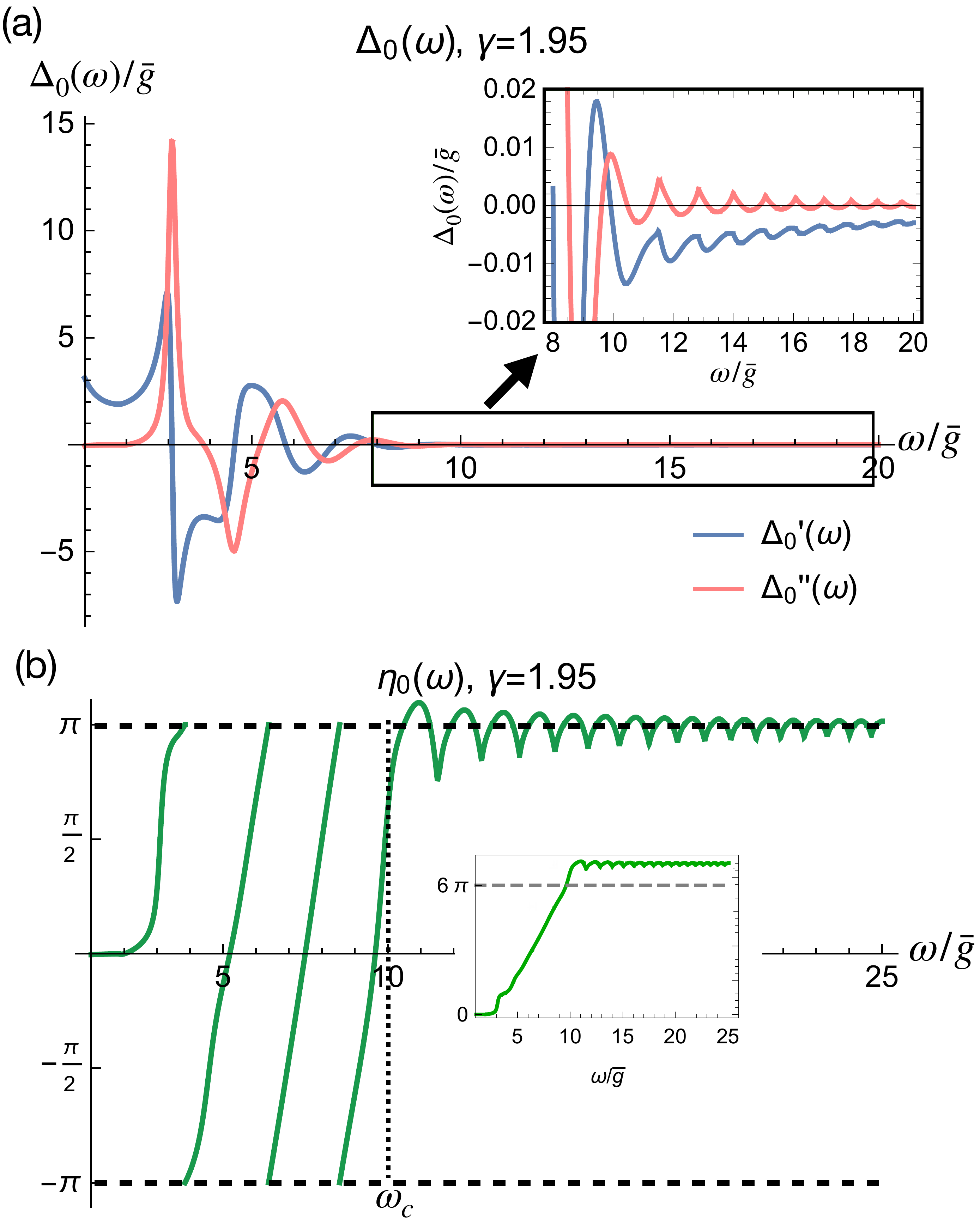}
  \caption{
Upper panel: frequency dependence  of the  gap function   $\Delta_0(\omega)=\Delta_0'(\omega)+i\Delta_0''(\omega)$, from
  Eq. \eqref{eq:Delta}. We set $\gamma=1.95$.
    For $\bar g<\omega<\omega_{cr}$, both $\Delta_0'(\omega)$ and $\Delta_0''(\omega)$ display oscillations with a decreasing magnitude.  For $\omega\geq\omega_{cr}$, $\Delta_0'(\omega)$ becomes negative and does not oscillate. Lower panel: Variation of the phase of the gap $\eta_0(\omega)$ as a function of $\omega$.  As before, we confined phase variation to $(-\pi,\pi)$, up to small variations.
The phase changes by $2\pi$ three times
between $\omega=O(\bar g)$ and $\omega_{cr}$. There are no more
$2\pi$ phase variations, despite that the phase shows wiggling at large frequencies. In the inset we plot the continuous $\eta_0 (\omega)$. We see that the total phase variation between $\omega_m=\bar{g}$ and $\omega_m=\infty$ is $6\pi+\pi\gamma/2$.}\label{fig:Delta_eta}
\end{figure}

      In Fig \ref{fig:Delta_eta}(a)
      we show real and imaginary parts of the gap function
      \beq
      \Delta_0 (\omega) = \frac{\omega}{\sin{\phi (\omega)}} =2\omega \frac{\sin{\phi'} \cosh{\phi^{''}} - i \cos{\phi'} \sinh{\phi^{''}}}{\cosh{2\phi^{''}} - \cos{2\phi'}}\label{eq:Delta}
      \eeq
       We see that  both $\Delta'_0 (\omega)$ and $\Delta^{''}_0 (\omega)$ oscillate
        between $\omega = O({\bar g})$ and $\omega_{cr}$ and display sign-preserving $1/\omega^\gamma$ behavior at $\omega > \omega_{cr}$.
          At $\omega \geq \bar g$, $\Delta' (\omega)$ varies roughly as
$\omega / \sin{((2/(\pi\gamma)) (\omega/{\bar g})^\gamma)}$ and $\Delta^{''} (\omega)$ has almost $\delta$-functional spikes in where $\sin (2/(\pi\gamma)) (\omega/{\bar g})^\gamma)$ is small. At larger $\omega \leq \omega_{cr}$, both $\Delta'$ and $\Delta^{''}$ oscillate with progressively decreasing magnitude.

    In Fig. \ref{fig:Delta_eta}(b)
     we show the variation of the phase of the gap function $\eta (\omega)$. The total variation of $\eta$ between $\omega =0$ and $\omega = \infty$ is $2\pi m + \pi \gamma/2$.
     We emphasize that this is the result for $\Delta_{0} (\omega)$, which is sign-preserving on the Matsubara axis.  We clearly see that there is a huge difference between the forms of the gap function on the real and the Matsubara axis in  between ${\bar g}$ and $\omega_{cr}$.

\begin{figure}
  \includegraphics[width=14cm]{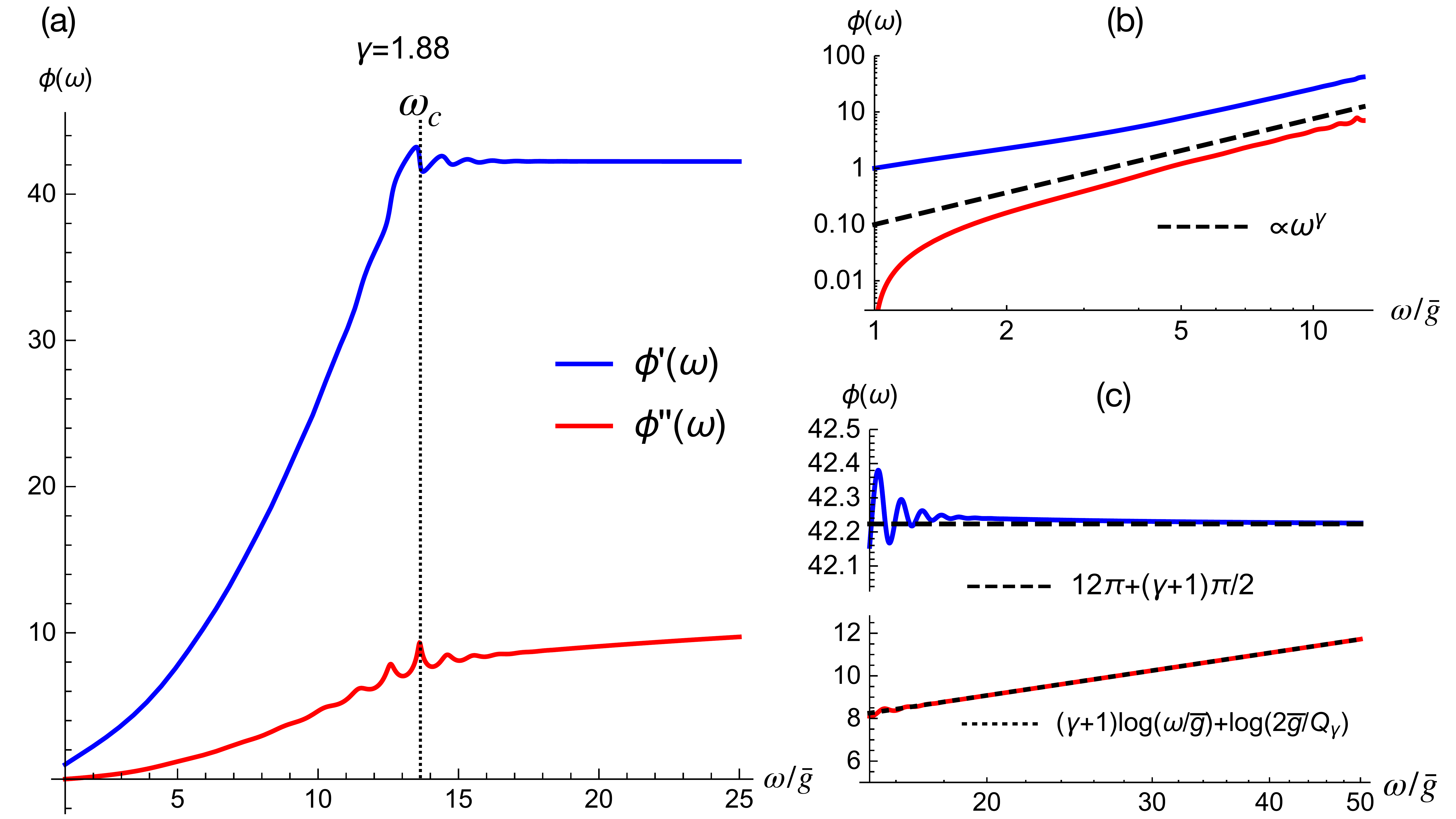}
  \caption{
Panel (a). The solution of Eq. \eqref{e_3_1} for $\gamma=1.88$.  Panels (b) and (c) - zooms into the regions below and above the crossover frequency. The behavior of  $\phi' (\omega)$ and $\phi^{''} (\omega)$ is  qualitatively similar to that in Fig. \ref{fig:phi1_m}, but differs in detail. In particular, both $\phi' (\omega)$ and $\phi^{''} (\omega)$ approach asymptotic forms
   $12\pi+(\gamma+1)\pi/2$ and $(\gamma +1) \log{\omega}$, respectively, without oscillations. }\label{fig:phi_2}
\end{figure}

\begin{figure}
  \includegraphics[width=17cm]{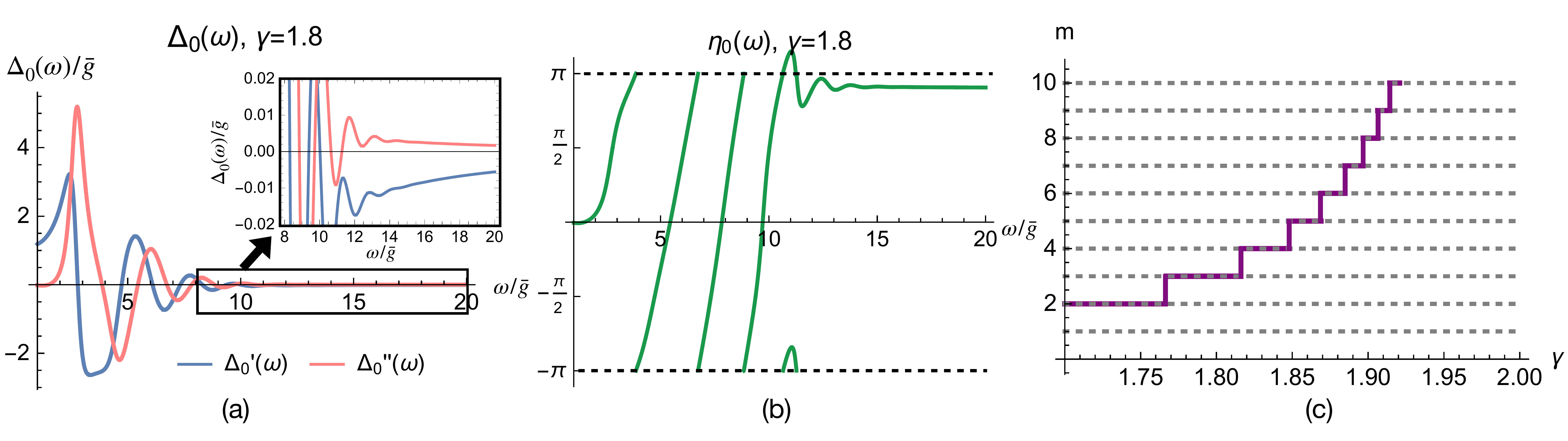}
  \caption{Panels (a) and (b) -- the same as in Fig. \ref{fig:Delta_eta}, but for $\phi'$ and $\phi^{''}$ from Eq.\eqref{e_3_1}.
  At large frequencies, $\Delta (\omega) \propto e^{i\pi\gamma/2}/\omega^\gamma$.  The total variation of the phase of $\Delta$ is $\delta \eta = 2 \pi m + \pi \gamma/2$.  Panel (c) -- variation of $m$ with $\gamma$. As before, there is a discrete set of $\gamma_i$, where $m$ jumps by 1.  The set becomes progressively more dense at $\gamma \to 2$.  }\label{fig:Delta_eta_m}
\end{figure}

   We now use more precise analysis to determine $\omega_{cr}$.  In Eq. (\ref{4_17}) we only included the first two terms in the expansion of $C(\omega)$ in powers of $\dot \phi$. Meanwhile, the
     expansion in derivatives holds in powers of ${\dot \phi} \omega/{\bar g}$ without additional $(2-\gamma)$ in the prefactor. This implies that higher-order terms are not negligible at $\omega > {\bar g}$.
    We now use the fact that before $\phi (\omega)$ crosses over to (\ref{4_14_1},\ref{4_14_2}),
    it shows the universal behavior in the range where $1 < e^{\phi^{''}} < (\omega/{\bar g})^3$. In this regime, (i) ${\dot \phi} \approx {\dot \phi}' \approx (2\pi) \omega^{\gamma-1}/{\bar g}^\gamma \gg {\bar g}$, while higher-order derivatives are smaller, and (ii) $D(\omega)$ is small, such that $\sqrt{1-D^2} \approx 1$.   In this situation, one can sum up the full Taylor series for $(2-\gamma)$ term in $C(\omega)$.  We use that $\tan \phi \approx i$ and the $n$-th detivative
     $D^{n} (\omega) \approx i^n (\dot \phi)^n D$. Integrating each term in Talor series over $\Omega$, we obtain the modified l.h.s. of (\ref{4_17}) in the form
     \beq
   {\dot \phi} -i (2-\gamma){\dot \phi} F\left(\frac{\omega}{\bar g} \dot \phi\right)
   \label{e_2}
   \eeq
    where
    \beq
     F(x) = \int_0^x \frac{1- \cos{y}}{y^2} = SI (x) - \frac{1-\cos{x}}{x}
      \label{e_3}
   \eeq
  and $SI(x)$ is SinIntegral.  Eq. (\ref{4_17}) is reproduced if we approximate $1-\cos{y}$ by $y^2/2$.
   Then $F(x) \approx x/2$.
    If we use the full expression, we find that at large $x$, which we are interested in, $F(x) \approx \pi/2$. The equation for $\phi$ then reduces to
  \begin{equation}
    \dot\phi=\frac{2}{\pi\bar{g}^\gamma}\frac{\omega^{\gamma-1}-\frac{Q_\gamma \bar{g}^\gamma}{\omega^2}e^{i\pi\gamma/2}\sin\phi}{1-i(2-\gamma)\pi/2}\label{e_3_1}
  \end{equation}
  We show the solution in Fig.\ref{fig:phi_2} and  present the plots of
    the gap function $\Delta_0(\omega)$, the phase $\eta_0(\omega)$, and the variation of  $m$ with $\gamma$ in Fig.\ref{fig:Delta_eta_m}.

  The behavior of $\phi (\omega)$ is qualitatively similar to the one in Fig.\ref{fig:phi}, i.e., there is a single crossover frequency,
  and the total variation of $\eta (\omega)$ between $\omega =0$ and $\omega = \infty$ is $2\pi m + \pi \gamma/2$.
  However the crossover scale $\omega_{cr}$ has different dependence on $2-\gamma$ compared to Eq. \eqref{4_17}, and also the
  dependence of $m$ on $\gamma$ is different from that in Fig.\ref{fig:phi1_m}(b) (and  there  are no wiggles in $\phi(\omega)$ at $\omega>\omega_{cr}$).
 To obtain the modified crossover frequency, $\omega_{cr}$,  we note that in the universal regime $1 < e^{\phi''} < (\omega/{\bar g})^3$,
  ${\dot \phi}' (\omega)$ is still given by (\ref{4_17_3}), while
  \beq
  {\dot \phi}^{''} (\omega) \approx (2-\gamma) \frac{\omega^{\gamma-1}}{{\bar g}^\gamma}
  \label{4_17_3_1}
  \eeq
  Accordingly, $\phi' (\omega) \approx (2\pi \gamma) (\omega/{\bar g})^\gamma$, $\phi^{''} (\omega) = (2-\gamma)/\gamma (\omega/{\bar g})^\gamma$.  The crossover frequency  is then determined by
  $e^{\phi^{''}} \sim (\omega/{\bar g})^3$. Solving for $\omega$, we obtain, for $\gamma \leq 2$,
     \beq
   \omega_{cr} \sim \bar g \left(\frac{|\log{(2-\gamma)}|}{2-\gamma}\right)^{1/2}
   \label{e_4_4}
   \eeq
   This is the same scale as $\omega_{cr}$ that we obtained  for $n=\infty$, Eq. (\ref{nn_6_1_1}).

    We see therefore that for $n=\infty$ and $n=0$ the new behavior of the gap function emerges in the same frequency range ${\bar g} < \omega < \omega_{cr}$. Moreover, the period of oscillations in this range is the same function of frequency, Eqs. (\ref{nn_6_6_1}) and (\ref{nn_6_6}).

\begin{figure}
  \includegraphics[width=10cm]{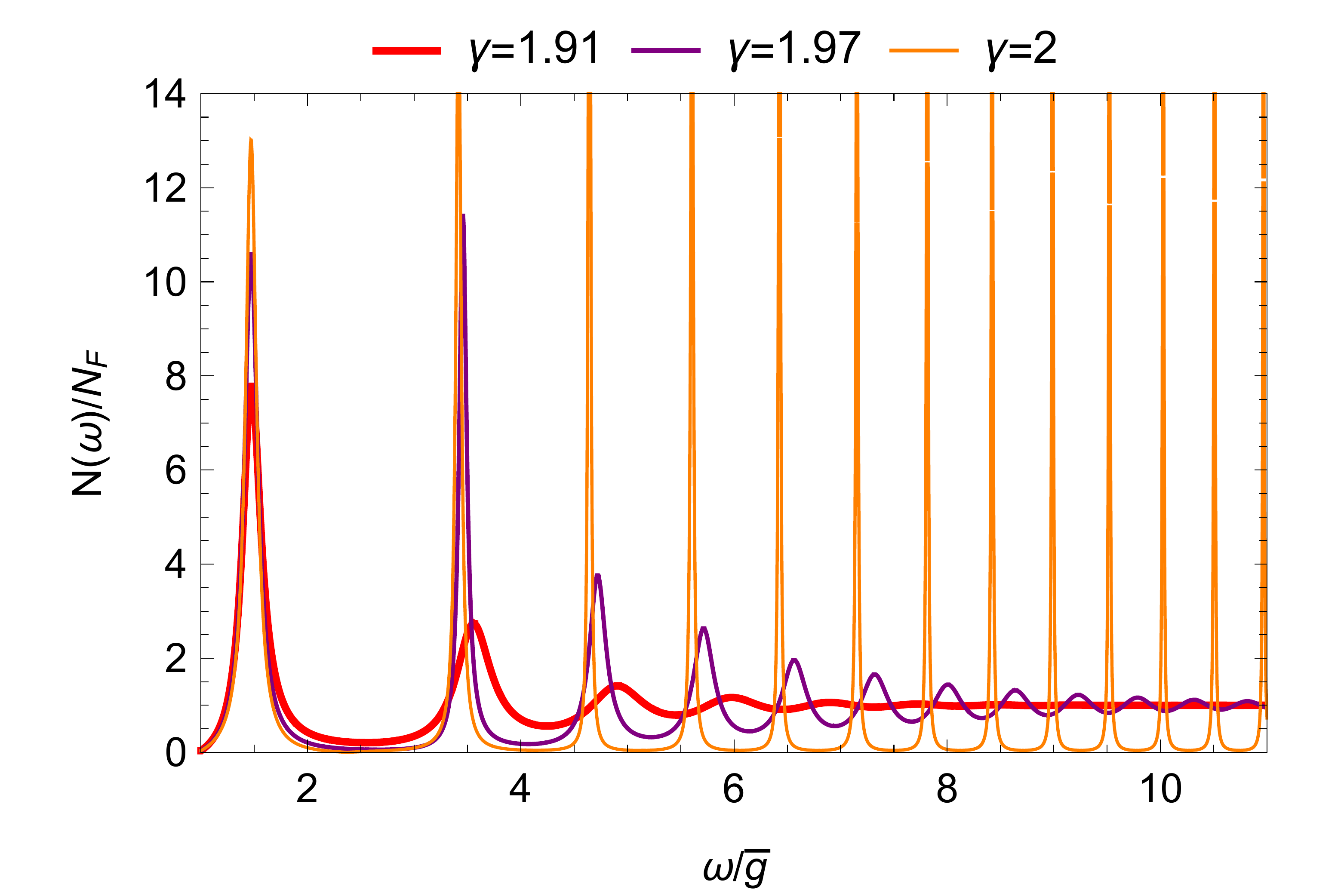}
  \caption{The DoS $N(\omega)$ for the $n=0$ solution, for several $\gamma$. Peaks in $N(\omega)$ sharpen as $\gamma\to2$, while the weight between the peaks is reduced.}\label{fig:Dos}
\end{figure}

 In Fig.\ref{fig:Dos} we plot the DoS
   \beq
   N_0 (\omega) = N_F {\text Re}\sqrt{ \frac{\omega^2}{\omega^2 - \Delta^2_\infty (\omega)}}
   \label{nn_6_6_4}
   \eeq
  for several $\gamma$.
 We see that $N_0(\omega)$  vanishes
   below a certain threshold frequency, and at ${\bar g} < \omega < \omega_{max}$
    has a set of maxima and minima.
   The peaks in $N_0(\omega)$ get progressively sharper as $\gamma$ approaches $2$, while in between the peaks
    $N_0(\omega)$ gets smaller.  The positions of the peaks almost coincide with the maxima of
     $N_\infty(\omega)$.

Before we move to finite $n$, we pause to discuss the validity of neglecting the $\ddot \phi$ term in Eq. \eqref{4_17}.   This is justified if $|\ddot{\phi }|$ is smaller than  $(\dot{\phi })^{2} |\tan \phi| $ for all frequencies. Of particular relevance here are frequencies near $\omega_a$, specified by Eq. (\ref{e_1_1_a}). If we  set $a=0$ in the initial condition (\ref{e_1}) and
use Eq. (\ref{e_1_1}), we find that $\phi^{''}$ initially increases as $(\omega-\omega_0)^{3/2}$.  In this situation, $\ddot{\phi }$ diverges at $\omega = \omega_0 +0$, and the behavior of $\phi (\omega)$, which we found earlier in this Section, is valid outside of the vicinity of $\omega_0$, while near $\omega_0$, $\ddot{\phi }$ and higher derivatives cannot be neglected.
Specifically, evaluating $\ddot \phi$, $\dot \phi$, and $\tan \phi$ near $\omega_0$, we find that for $\gamma \leq 2$,  the condition $|\ddot \phi| < (\dot \phi)^2|\tan \phi|$ is satisfied when
\beq
\phi '' > \pi (2-\gamma)
\eeq
 For $a=0+$, this holds outside a finite range near $\omega_0$. However, if $a = O(1)$, $|\ddot \phi|$ never becomes much larger than  $(\dot \phi)^2|\tan \phi|$, hence neglecting  $|\ddot \phi|$  does not change the results qualitatively even near $\omega_0$.

 \subsection{Frequency dependence of $\Delta_n (\omega)$.  Finite $n$}
\label{sec:real_finite_n}

The behavior of $\Delta_n (\omega)$ at $\omega < {\bar g}$ on the real axis parallels the one on the Matsubara axis.  Namely, the phase varies by $n \pi$ between $\omega =0$ and ${\bar g}$ due to n vortices on the Matsubara axis.  At $\gamma \to 2$, oscillations shift to smaller $\omega$.  At $\omega > {\bar g}$, it is natural to expect oscillations with the period set by (\ref{nn_6_6_1}) at ${\bar g} < \omega < \omega_{cr}$ and $e^{i\pi \gamma/2}/\omega^\gamma$ behavior at $\omega > \omega_{cr}$.
The total phase variation between $\omega =0$ and $\omega = \infty$ is
  \beq
   n \pi + 2\pi m  + \frac{\pi \gamma}{2}
   \label{4_27}
   \eeq

The DoS at a finite $n$ vanishes below a certain threshold frequency. Above the threshold, it  oscillates around $N_F$ with a period set by $\log{\omega}$, and at larger ${\bar g} <\omega < \omega_{cr}$ has a set of peaks and dips at about the same frequencies as $N_\infty (\omega)$ and $N_0 (\omega)$.
\section{Vortices near the real axis}
\label{sec:vortices}

We now use
the Cauchy relation
\beq
\Delta (z) = \frac{2}{\pi}\int_0^\infty dx \frac{\Delta^{''} (x) x}{x^2-z^2}
\label{e_5}
\eeq
  and extend the gap functions $\Delta_n (\omega)$ into the upper frequency half-plane, to $z = \omega' + i \omega^{''}$ ($\omega^{''} >0$).  Earlier, we demonstrated that $\Delta_n (z)$ has $n$ vortices on the Matsubara axis.  Here, we analyze the behavior of $\Delta_n (z)$ between the Matsubara and the real axis.
   We show that as $\gamma$ increases from one to two, new vortices appear one-by-one  in the upper frequency half-plane, near the real axis.
  These vortices are located at $|z| > {\bar g}$, and their number, $m$, is determined by $\gamma$ and is the same for all $n$.  When $\gamma \to 2$, $m$  tends to infinity.
  The emergence of vortices obviously  correlates with the oscillations of $\Delta_n (\omega)$ on the
    real axis, and as such is  another consequence of the change of sign of the real part of
    the interaction on the real axis,  $V' (\Omega)$, which becomes repulsive at $\gamma =1$. The increase of the number of vortices as $\gamma$ approaches $2$ in turn correlates with the
   decrease of $V^{''} (\Omega)$.

\begin{figure}
  \includegraphics[width=10cm]{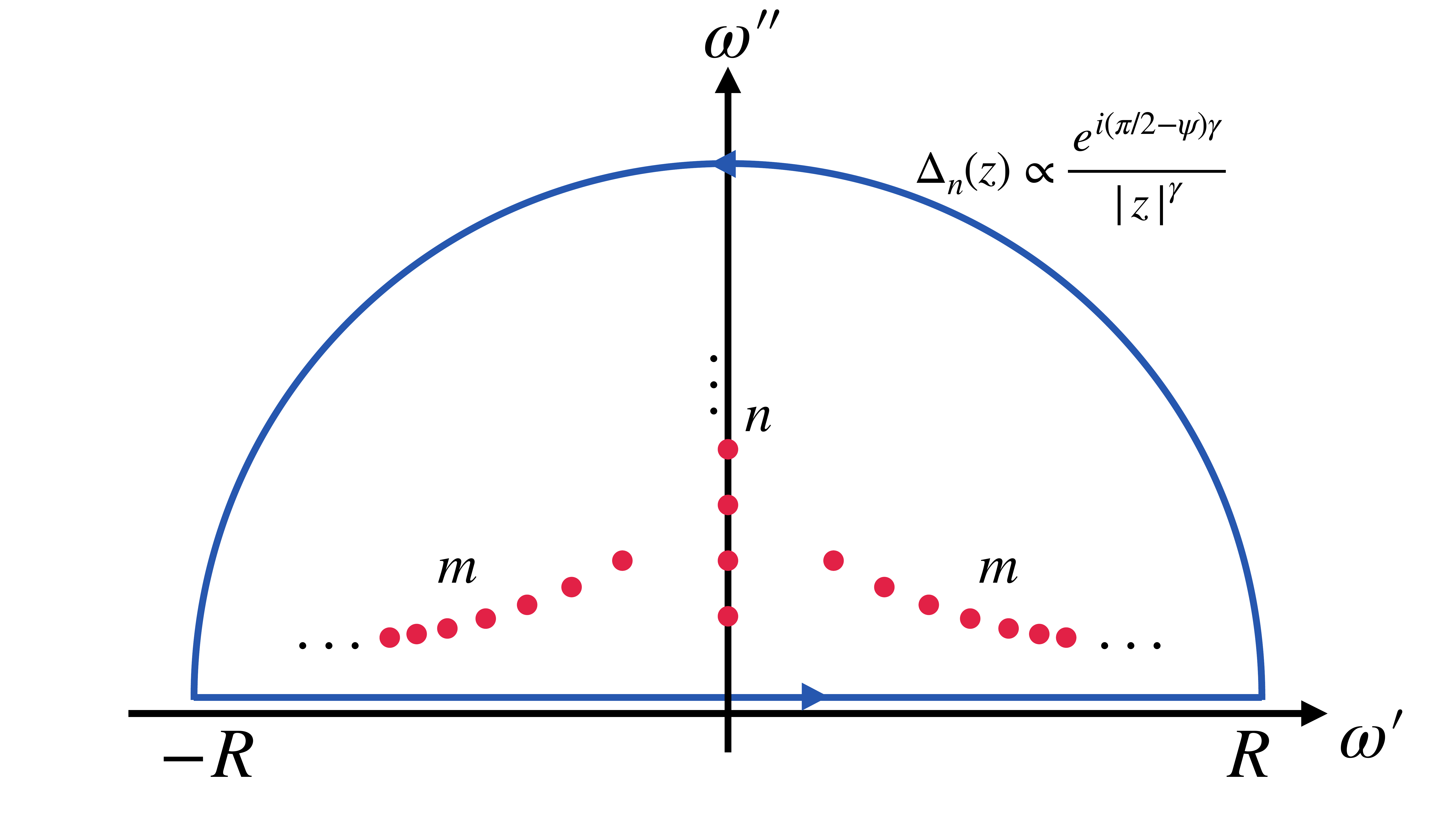}
  \caption{The contour $\Gamma$ used in Eq.\eqref{ex_ex_2}. The contour consists of a real axis and an semi-circle(arc) at $z=|z|e^{i\psi}$, $|z|\to\infty$ and $0<\psi<\pi$. Along the arc, $\Delta_n(z)\propto e^{i(\pi/2-\psi)\gamma}/|z|^\gamma$, so the corresponding contribution to $\delta\eta_L$ is $-\pi\gamma$. Along the real axis, $\delta\eta_\Gamma =2\pi(n+2m)+\pi\gamma$. The total $\delta \eta_\Gamma$ along $\Gamma$ is then $2\pi(n+2m)$. The same $\delta\eta_\Gamma$ must be obtained by counting contributions from the poles inside $\Gamma$. Each vortex is a pole with residue $2\pi$, hence there should be $n+2m$ vortices. This is consistent with our analysis of $\Delta(z)$: there are $n$ vortices on the Matsubara axis, and $2m$ vortices in the upper half plane near the real axis.}\label{fig:contour}
\end{figure}

That  such  vortices  must be present  can be understood by comparing the behavior of $\Delta_n (z)$ for $|z| > {\bar g}$ near the Matsubara axis and the real axis. Along the Matsubara axis,
   $\Delta_n (\omega_m)$ is real and does not change sign at $|\omega_m| > {\bar g}$. By continuity,
    $\Delta' (z)$ near the Matsubra axis should remain sign-preserving,
     hence $\eta (z)$ does not wind.
On the other hand, on the  real axis,  the phase winds by $2\pi m$,
as we found in the previous Section.
This number is
topologically protected against perturbations and can only disappear upon rotation from real to imaginary $z$ if there are vortices at some complex $z$.  Indeed, let's  compute
 \beq
\delta \eta_\Gamma  =\text{Im} \int_{\Gamma }dz \left[\partial \log \Delta (z)/\partial z\right]
\label{ex_ex_2}
\eeq
 along  the path $\Gamma $, which starts at the large negative  real  $z=-R$, goes along the real axis up to $+R$, and then closes along the large arc in the upper half plane (see Fig.\ref{fig:contour})
   The arc is chosen such that along it
 $\Delta_n (z) \propto e^{i(\pi/2-\psi)}/|z|^\gamma$ .  The total phase variation along the arc is
 $-\pi\gamma$,
  and the phase variation along the real axis is $2\pi (n+2m) + \pi \gamma$.
   The contour integral then gives
 \beq
 \delta \eta_\Gamma  = 2\pi (n +2m)
 \eeq
 This $\delta\eta_\Gamma$ should be equal to the contribution from inside the contour.
  Because $\partial \log \Delta (z)/\partial z $ has a  simple pole at each
 point where $\Delta (z)$ vanishes, there must be $n+2m$ vortices inside the contour
 (we recall that $\Delta (z)$ by itself has no poles in the upper half plane).  There are $n$ vortices on the Matsubara axis, the other $2m$ should be located in between the Matsubara and the real axis.  By symmetry, there must then be $m$ vortices in the first quadrant, and another $m$ is in the second one.

To determine the location of the vortices, it is instructive to  again consider separately the cases $n = \infty$, $n=0$, and a finite $n$.

\subsection{$n=\infty$}
\label{sec:vortex_n_infinite}

\begin{figure}
   \includegraphics[width=10cm]{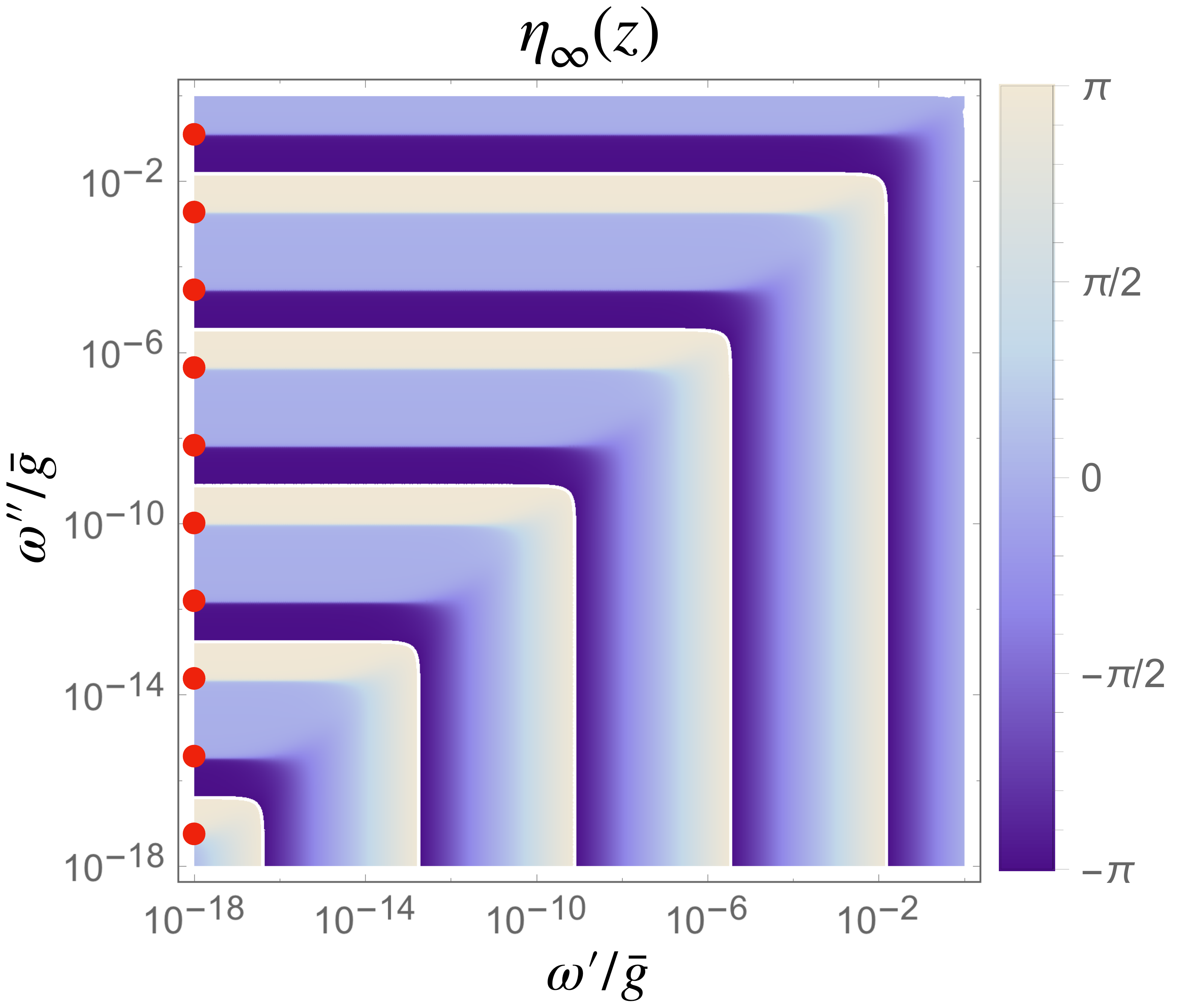}
   \caption{The phase $\eta_\infty(z)$ in the complex plane at $|z|<O(\bar g)$, for $\gamma=1.96$. We set the lower boundary for $|z|$ at $\omega_0\sim 10^{-18}\bar g$, as in
     Fig.\ref{fig:Delta_eta_n16}. Along the white lines, $\Delta^{''}(z)=0$ and $\Delta'(z)<0$.
     The phase of the gap then
changes
by $2\pi$ upon crossing each of these lines.
       Observe that each white line starts in between
         a pair of vortices on the Matsubara axis, and  ends on the real axis. As a result,
          if there are $k$ vortices on the Matsubara axis,
            the phase variation $\delta\eta$ on the positive real axis is exactly $k\pi$. In our case, $k=10$. }\label{fig:semicircle}
 \end{figure}

We first quickly verify that there are no vortices away from Matsubara axis for $|z| < {\bar g}$.
 We express $\Delta(z)$ in terms of  $|z|$ and $\psi$, defined via $z = |z| e^{\psi}$.
  Replacing $\omega_m$ by $-i z$ in Eq. (\ref{nn_2}) we obtain the expressions similar to (\ref{e_4}) and (\ref{e_4_1}):
  \bea
&&\Delta'_{\infty} (z) \propto \cos{\left(\frac{\gamma}{2} \left(\frac{\pi}{2} - \psi\right)\right)} \cos{D_{1}} \cosh{D_{2}} + \sin{\left(\frac{\gamma}{2} \left(\frac{\pi}{2} - \psi\right)\right)}  \sin{D_{1}} \sinh{D_{2}} \nonumber \\
&& \Delta^{''}_{\infty} (z) \propto
 \cos{\left(\frac{\gamma}{2} \left(\frac{\pi}{2} - \psi\right)\right)}\sin{D_{1}} \sinh{D_{2}} - \sin{\left(\frac{\gamma}{2} \left(\frac{\pi}{2} - \psi\right)\right)} \cos{D_{1}} \cosh{D_{2}}
\label{4_24_a}
\eea
but $D_1$ and $D_2$ are now given by
\bea
&&D_1 = \beta \log{(|z|/{\bar g})^\gamma} + \phi, \label{4_24_1}\\
&&D_2 = \beta\left((\pi-2 \psi) \frac{\gamma}{2}\right)
\label{4_24}
\eea
The vortices are the points where $\Delta'_\infty(z) = \Delta^{''}_\infty (z) =0$. For
$\Delta_{\infty} (z)$ given by (\ref{4_24_a}) this is satisfied if $D_1 =\pi/2 + m \pi$
 and $D_2 =0$. The second condition is satisfied only if $\psi = \pi/2$, i.e., on the Matsubara axis.
The first condition coincides with $\Delta_\infty (\omega_m) =0$.
In  Fig.\ref{fig:semicircle} we show that the phase $\eta_\infty (z)$  of $\Delta_\infty (z) = |\Delta_\infty (z)|e^{i \eta_\infty (z)}$ evolves in a "rectangular'' way upon rotation from the Matsubara to the real axis.
 The white curves in this Figure are determined by
    $\Delta^{''} (z) =0$ and $\Delta' (z) <0$. The phase $\eta (z)$ changes by $2\pi$ upon crossing each of these
   curves.  We see that the phase winds five times by $2\pi $ at Re$z >0$, consistent with the presence of 10 vortices on the Matsubara axis.

  \begin{figure}
  \includegraphics[width=17cm]{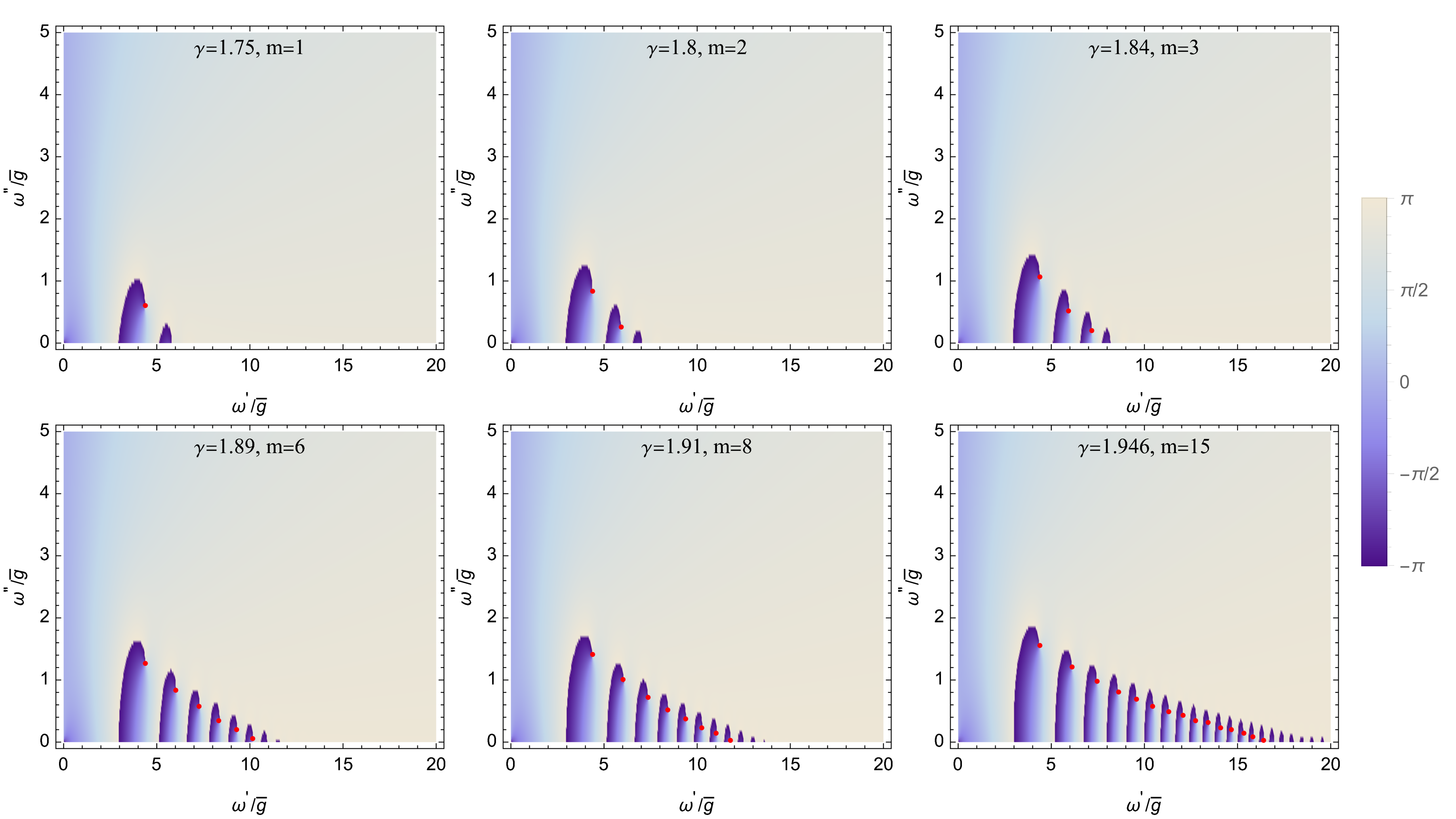}
  \caption{The case $n=\infty$. We plot the phase of the gap function $\eta_{\infty}(z)$ in the upper half plane ($ z=\omega'+i\omega''$) for different $\gamma$ near $\gamma =2$.
   The locations of the vortices are marked by red dots. At the vortex core,  $\Delta_\infty (z)=0$ and $\eta_\infty$ is undefined. This number of vortices $m$ rapidly increases as $\gamma \to 2$.}
      \label{fig:vortex_1}
\end{figure}

  We now consider the range $|z| >  {\bar g} $. We recall that on the real axis,
   $\Delta_\infty (\omega)$ oscillates  as a function of $(|\omega|/{\bar g})^{\gamma/(\gamma-1)}$ between
    ${\bar g}$ and  a larger  $\omega_{cr}$ given by Eq. (\ref{nn_6_1_1}).  We detected the oscillating term
     on the Matsubara axis $(\psi= \pi/2$) , where it is exponentially small,  and converted it to the real axis ($\psi =0$), where it becomes the largest piece in $\Delta_\infty (\omega)$ at ${\bar g} <\omega < \omega_{cr}$. We now analyze this term at arbitrary $\psi$.   Replacing $\omega_m$ by  $|z|
    e^{-i(\pi/2-\psi)}$ in (\ref{nn_3_5})  and using (\ref{nn_6_1}), we obtain the oscillating component of $\Delta_\infty (z)$ in the form
\beq
   \Delta_{\infty} (z) = \int_{k_{min}}^\infty dk \bigg(e^{-(\pi+\theta) k} e^{-ik S_k(|z|)} +
    e^{-(\pi-\theta) k} e^{ik S_k(|z|)} \bigg).
    \eeq
    Here, $\theta = (\pi/2-\psi) \gamma$ and
    $S_k (|z|) = \log{\left[(|z|/{\bar g})^\gamma (e/A_\gamma k)^{\gamma-1}\right]}$,
     where $A_\gamma$ is defined in (\ref{nn_3_3}).  For $\gamma \to 2$, $\theta \to \pi - 2 \psi$, and
 evaluating
the integral over $k$ in the same way as in Sec. \ref{infnMatsubara}, we obtain
 \begin{equation}
  \Delta_{\infty} (z) \propto  |z|^{\frac{1}{2}\frac{\gamma}{\gamma-1}}
\exp \left(i\frac{\gamma-1}{A_\gamma}
  \left(\frac{|z|}{\bar{g}}\right)^{\frac{\gamma}{\gamma-1}}
\left[
  1-\frac{1}{2} \left(\frac{\pi-\theta}{\gamma-1} \right)^{2}
+ i\frac{\pi-\theta}{\gamma -1}
 \right]\right)
+
(\theta \rightarrow -\theta )
 \label{oscillating_cpx}
 \end{equation}
  We plot $\Delta_\infty (z)$  from Eq. (\ref{oscillating_cpx})  in Fig.~\ref{fig:vortex_1} for several $\gamma$ near $\gamma=2$.
   We see that there is an array of vortices near the real axis, at  $\rvert z\rvert > {\bar g}$.
 We now recall that  Eq. (\ref{oscillating_cpx}) is valid in a  range ${\bar g} < |z| < \omega_{cr} (\theta)$. At larger $|z|$,
   the oscillating term becomes smaller than the regular piece $e^{i \theta}/|z|^\gamma$ and there are
   no vortices.  The boundary
   frequency for vortices is $\omega_{cr} (\theta) \sim {\bar g} (|\log{(\pi -\theta)}|/(\pi-\theta))^{1/2}$ for $\gamma \leq 2$. Because $\theta \approx \pi -2\psi$ in this limit, the range of $|z|$, where Eqn. (\ref{oscillating_cpx}) is valid, remains wide for small $\psi$.
   The vortices  move closer to the real axis at larger $|z|$ and for $|z| \geq \omega_{cr} (\theta)$
     escape into the lower half-plane of frequency.
    We see from  Fig. ~\ref{fig:vortex_1} that the number of vortices increases rapidly as $\gamma \to 2$.

\subsection{Case $n=0$}
\label{sec:vortex_n_0}

We now show that the same vortex structure appears for $\Delta_0 (z)$, which, we recall,  is sign-preserving
 along the Matsubata axis.

\begin{figure}
  \includegraphics[width=16cm]{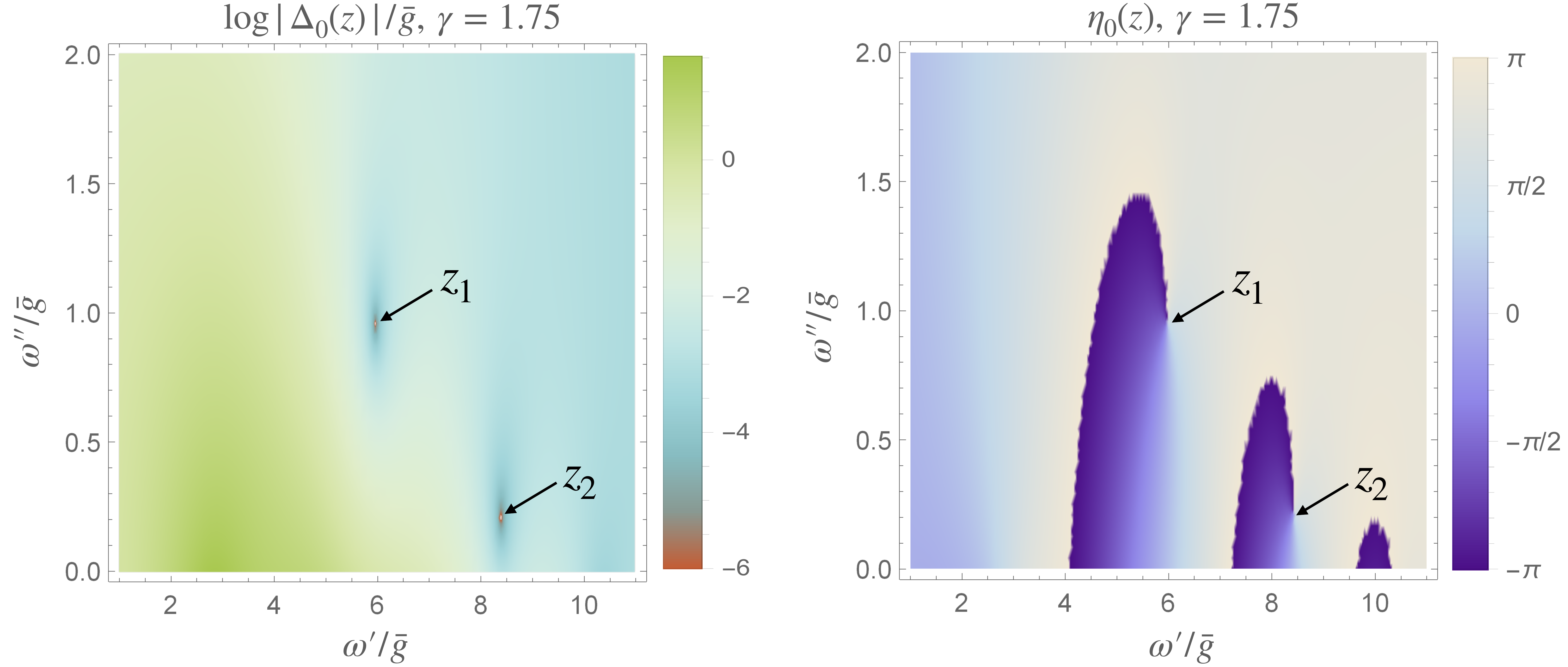}
  \caption{The case $n=0$. Left panel: Plot of  $\log{|\Delta_0(z)|}$ in the upper half plane, for $\gamma=1.75$.
    The gap $\Delta_0(z)$ for a generic $z$ in the upper half plane is obtained by analytic continuation from $\Delta^{''} (\omega)$ on the
     real axis, Eq.\eqref{e_3_1}.
       The two bright points $z_1$ and $z_2$ are the locations of the vortices (points where
      $|\Delta_0(z)|=0$). Right panel: Plot of $\eta_0(z)$ in the same region. }
      \label{fig:node}
\end{figure}

\begin{figure}
  \includegraphics[width=17cm]{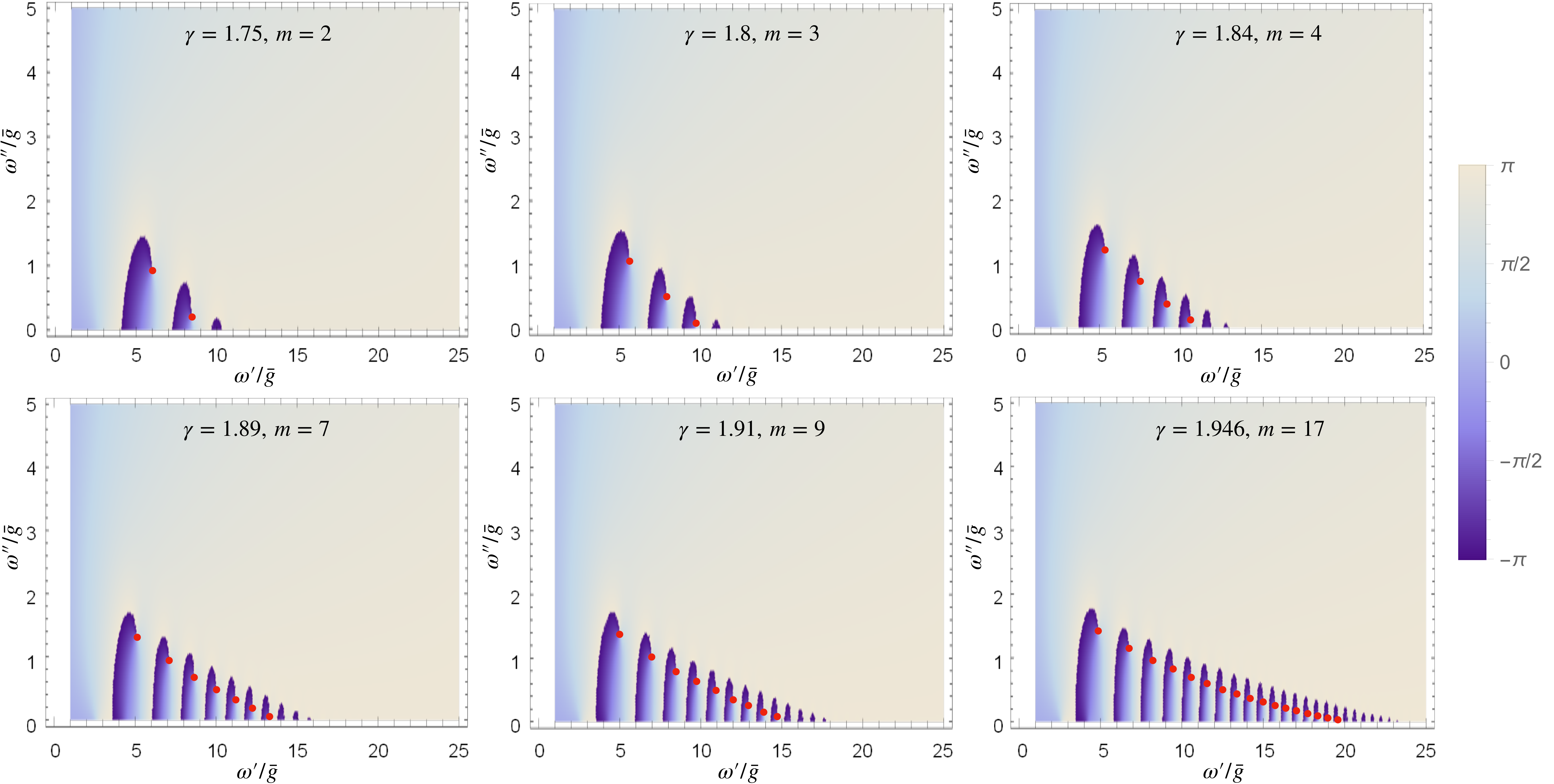}
  \caption{The case $n=0$. The phase of the gap function $\eta_0(z)$ in the upper half plane ($ z=\omega'+i\omega''$) for different $\gamma$ near $\gamma =2$.
   The locations of the vortices are marked by red dots. At the vortex core  $\Delta_0 (z)=0$ and $\eta_0$ is undefined. The  number of vortices is set by  $m$ in \eqref{4_17_1}. This number rapidly increases as
   $\gamma \to 2$.}\label{fig:vortices_gamma}
\end{figure}

We obtain $\Delta_0 (z)$ by Cauchy relation, Eq. (\ref{e_5}),  using as an input $\Delta^{''} (\omega)$ from
 Eq. \eqref{eq:Delta}. In Fig.\ref{fig:node} we show the amplitude of the gap  $|\Delta_0(z)|$ in the
  first quadrant of complex $z = \omega' + i \omega^{''}$ for $|z|>\bar g$ for
   $\gamma=1.75$. We see that there are two vortices at complex $z$. We verified that this is consistent with $m=2$ in Eq. \eqref{4_17_1} for this $\gamma$.
    In Fig.\ref{fig:vortices_gamma} we show that the number of vortices increases
    when $\gamma$ increases towards $2$.

\begin{figure}
  \includegraphics[width=12cm]{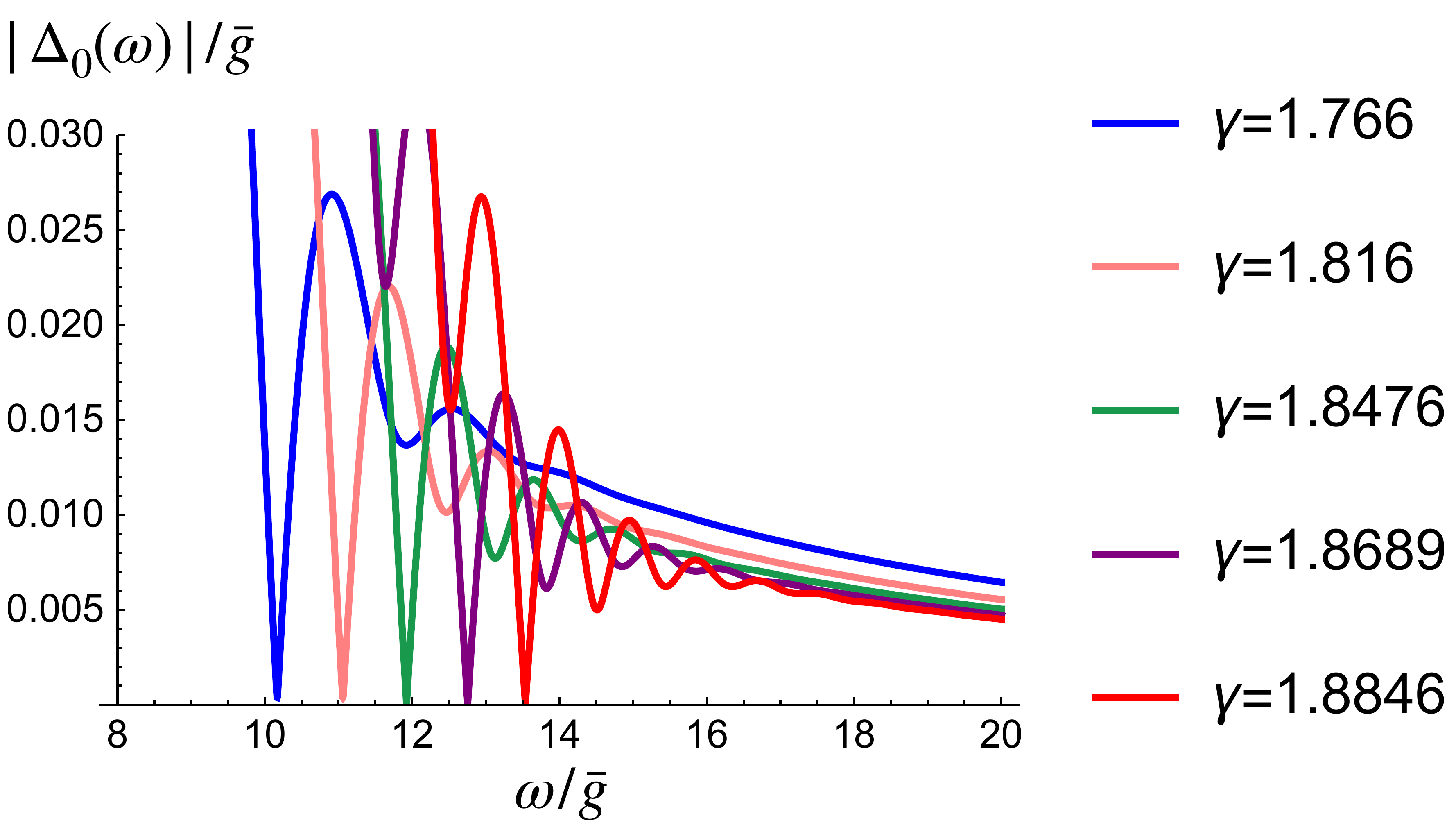}
  \caption{The gap amplitude $|\Delta_0(\omega)|$ as a function of $\omega$ for a set of $\gamma_i$.
    For each of these $\gamma_i$, the gap amplitude vanishes at a particular $\omega$, indicating that
       a vortex crosses the real axis on its way from the lower to the upper frequency half-plane.}
  \label{fig:node_real}
\end{figure}

In Fig.\ref{fig:node_real}
 we show how  vortices emerge in the upper frequency half-plane one-by-one at a set of $\gamma_i$.
  A  given vortex is located in the lower frequency half-plane for $\gamma < \gamma_i$ and moves into the
  upper half-plane at $\gamma > \gamma_i$.  Right at $\gamma = \gamma_i$ it appears on the real axis. At this $\gamma$, both $\Delta'_0(\omega)$ and $\Delta''_0(\omega)$  vanish at some frequency, hence $|\Delta_0(\omega)| =0$. We  see from Fig.\ref{fig:node_real} that this indeed happens for $\gamma_i$ specified in this Figure.
  Once $\gamma$ becomes larger than $\gamma_i$, $m$ increases by one, and $\delta \eta_0$  increases by $2\pi$.
We extended $\omega$ to $\omega \pm i \delta$ and verified that the vortex indeed moves from the lower to the upper frequency half-plane as $\gamma$ increases through $\gamma_i$.

\begin{figure}
	\includegraphics[width=16cm]{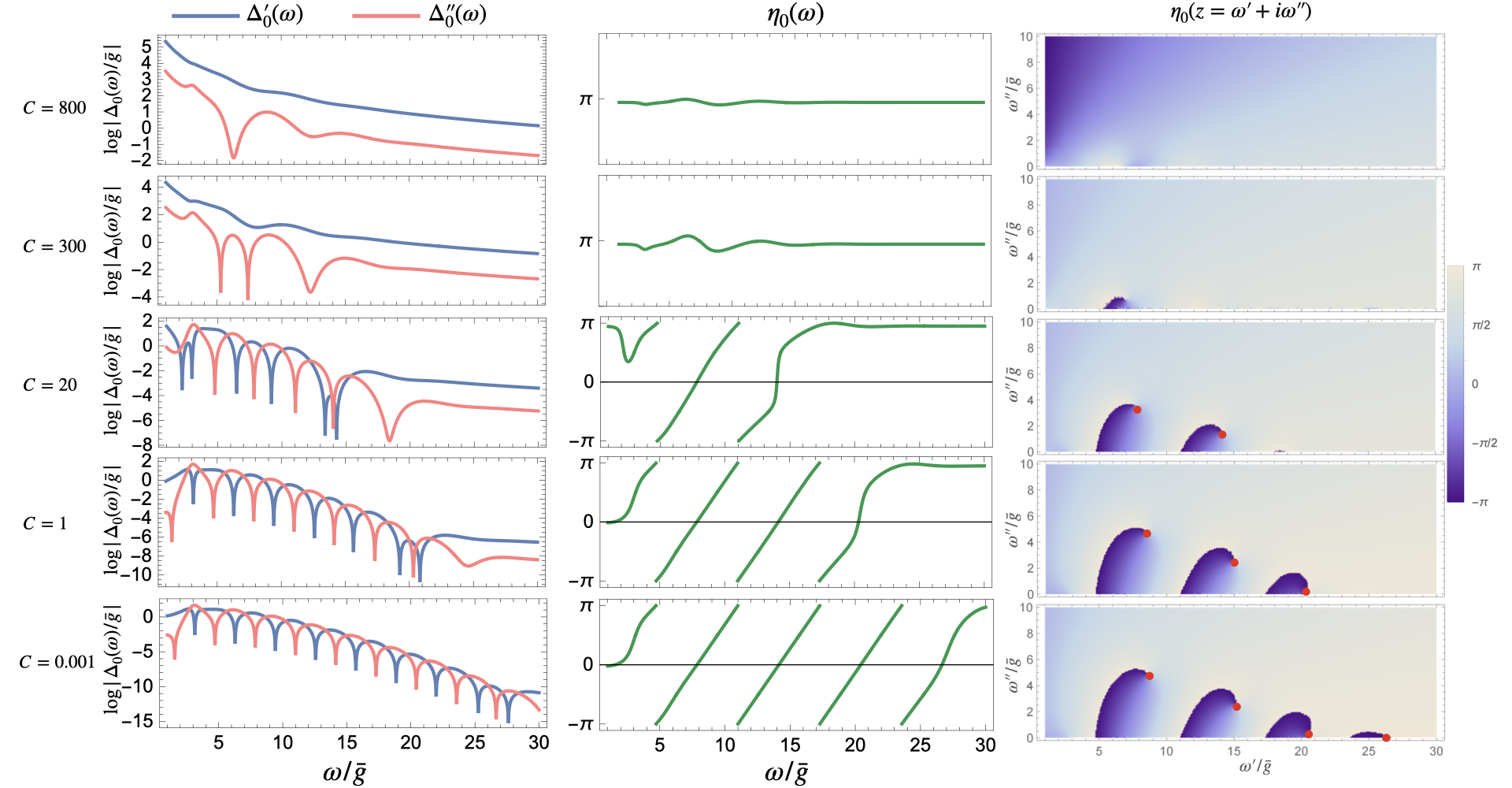}
	\caption{Left panel:
  The plots of $\log|\Delta'_0 (\omega)/\bar g|$ and $\log|\Delta^{''}_0 (\omega)/\bar g|$,  from Eq. \eqref{4_21}.
  The form of $\Delta_0 (\omega)$ is controlled by the parameter $C$.
   At small $C$ (which corresponds to larger $\gamma$), the gap function oscillates, and the range of oscillations increases with decreasing $C$.
 Middle panel. The phase  $\eta_0(\omega)$, constrained to $(-\pi,\pi)$, for different $C$. The
 winding number increases as $C$ decreases.  Right panel. The phase $\eta_0(z)$ in the upper half-plane of frequency.  At large $C$, there are no vortices in the upper half-plane. As $C$ decreases , vortices appear one-by-one . }\label{fig:DeltaC}
\end{figure}

 The emergence of the line of dynamical vortices
 (dynamical in the sense that they are at complex $z$)
 can also be analyzed within a semi-phenomenological analytical model,
  in which we combine oscillating behavior of $\Delta_0 (\omega)$ at ${\bar g} < \omega < \omega_{cr}$
  and non-oscillating $1/\omega^\gamma$ form at $\omega > \omega_{cr}$ by approximating $\Delta_0 (\omega)$
   as the sum of the two terms:
   \beq
 \Delta_0 (\omega) =  2\omega \frac{\sin{\phi'} \cosh{\phi''} - i \cos{\phi'} \sinh{\phi''}}{\cosh{2\phi''} - \cos{2\phi'}} + C {\bar g} \left(\frac{\bar g}{\omega}\right)^\gamma e^{i \pi \gamma/2},
 \label{4_21}
 \eeq
  where $\phi'$ and $\phi'$ are some functions of $\omega$. We  treat $C$ as a phenomenological parameter, which controls the width of the range  where $\Delta_0 (\omega)$ oscillates. We assume that $C$ scales as some power of $2-\gamma$ and vanishes at $\gamma =2$.   The smaller $C$ is, the larger is the width of this range where $\Delta_0 (\omega)$ oscillates.
    The precise forms of $\phi' (\omega)$ and $\phi^{''} (\omega)$ are not essential for this consideration,
     except that $\phi'(\omega)$ must be odd in $\omega$ and $\phi''(\omega)$ must be even, and  KK relations for $\Delta_0 (\omega)$ must be satisfied with enough accuracy within the range
     where $\Delta_0 (\omega)$ oscillates.
   We use $\phi'(\omega)=\omega$ and $\phi''(\omega)=(2-\gamma)|\omega|^{1.5}$ to generate the plots in Fig.\ref{fig:DeltaC}.
  At large $C$,
 $\Delta'_0 (\omega)$ and $\Delta^{''}_0 (\omega)$ gradually decrease  without changing sign (the top panel). This models the behavior of the gap function
 for  $\gamma \sim 1$.  As $C$  decreases,
   $\Delta'_0 (\omega)$ and $\Delta^{''}_0 (\omega)$ evolve and start undergoing sign changes. At first, this does not give rise to phase winding
(second panel from the top).
    Then, at some critical $C=C_1 \approx 110$,
there appears
     a point infinitesimally close to the real axis, where $\Delta'_0 (\omega)$ and $\Delta^{''}_0 (\omega) $ change sign at the same frequency. Once
$C$ becomes smaller than this value, a vortex appears in the upper frequency half-plane, and the phase $\eta_0 (\omega)$ rapidly changes by $2\pi$ in
     a narrow frequency range.
     At a smaller $C = C_2 \approx 25$, another vortex emerges on the real axis and moves to the upper  frequency half-plane at smaller $C$, causing another rapid change of $\eta_0 (\omega)$ by $2\pi$
     (the third panel). The third vortex emerges at $C = C_3 =1.5$  and so on (the lower panels).

   Comparing the results for $n =0$ and $n = \infty$ we see that in both cases
   a line of vortices emerges at complex $z$ in the first quadrant, and equal number of vortices forms in the second quadrant.  The vortices develop for $\gamma >1$. The number of vortices increases with $\gamma$  and tends to infinity at $\gamma \to 2$. In this limit,  the line of vortices extends to $z = \infty$.
Comparing Figs.~\ref{fig:vortex_1}  and \ref{fig:vortices_gamma}, we see that  the phase winding along the  real axis and the number of  vortices in the upper frequency half-plane for any given $\gamma$
 are very likely the same for  $n=0$ and $n=\infty$  A small difference in the values of $\gamma_i$, where the new vortices appear in the two cases,  is most probably related to the fact that the $n=0$ solution is an approximate one.

\subsection{Finite $n$}
\label{sec:vortex_finite_n}

 Given the equivalence between the number of vortices at complex $z$ for $n=0$ and $n=\infty$ and near-equivalence of their position,  we expect that the geometry of the vortices near the real axis is determined solely by $\gamma$ and is the same for all values of $n$.

 The emergence of the array of vortices is another indication
  that  the pairing  at $\gamma >1$, where $V' (\omega)$ is repulsive, is quite special.  The argument for this goes as follows. Take the set  of $m$ points, where $\Delta_n (z) =0$, and make it infinite  by adding points $z_i$ at  larger $z > \omega_{cr}$, for which $\Delta_n (z_i) \propto C_n/|z_i|^\gamma$. As long as
    $\Delta_n (z)$ vanishes at $|z| = \infty$ everywhere in the upper half-plane, one can
   continue analytically from  such set and obtain a unique $\Delta_n (z)$  for all $z$.  We see that
   $\Delta_n (z)$  is non-zero only
   because  we extended the set of vortex points by adding additional $z_i$ with $|z_| > \omega_{cr}$.
   The implication is that $\Delta_n (z)$  for all $z$ is actually  determined by the set of complex frequencies, above the  crossover scale $\omega_{cr}$.
    At $\gamma \to 2$, $\omega_{cr}$ tends to infinity. In this  limit, the set of the gap functions on the Matsubara axis is determined solely by an essential singularity at $z = \infty$.
    This obviously points out that the pairing at $\gamma >1$ is highly unconventional and becomes more so as $\gamma \to 2$.

\section{Conclusions}
\label{sec:conclusions}

In this paper we continued with our analysis of the interplay between the pairing and the non-Fermi liquid behavior
  in a metal for a set of quantum-critical (QC) systems with an effective dynamical electron-electron interaction
 $V(\Omega_m) \propto 1/|\Omega_m|^\gamma$, mediated by a critical massless boson (the $\gamma$-model).
 In previous papers we considered the cases $0<\gamma <1$ and $\gamma \approx 1$. We  argued that
   the pairing by a gapless boson is fundamentally different from BCS/Eliashberg pairing by a massive boson
     as for QC pairing there exists an infinite discrete set of topologically distinct solutions for the gap function $\Delta_n (\omega_m)$ at $T=0$ ($n=0,1,2...$), each with its own condensation energy $E_{c,n}$.
   Here we extended the analysis to larger $1< \gamma <2$.  We argued that the discrete spectrum of $E_{c,n}$ get progressively denser as $\gamma$ increases towards $2$. This increases the strength  of
    "longitudinal" gap fluctuations, which tend to reduce the actual superconducting $T_c$ compared to the onset temperature for the pairing.  We also reported two new features on the real axis, which again become critical at $\gamma \to 2$.  First, the density of states evolves towards a set of discrete $\delta-$functions. Second,  on a real axis, $\Delta_n (\omega)$ with all $n$, including $n=0$, oscillate between ${\bar g}$ and $\omega_{cr} > {\bar g}$, and the phase of the gap $\eta_n (\omega)$ varies in this range by an integer number of $\pi$.
 We associated this variation with the presence of dynamical  vortices in the upper half-plane of frequency, at a complex $z = \omega' + i \omega^{''}$. The vortices appear one-by-one at a discrete set of $\gamma_i >1$, and the number of vortices tends to infinity at
 $\gamma \to 2$.   We related the emergence of oscillations on the real axis and vortices at complex $z$
   with the fact that for real $\omega$,  the interaction  $V' (\Omega) \propto \cos (\pi\gamma/2)=/|\Omega|^\gamma$  becomes repulsive for $\gamma >1$, and the imaginary part $V' (\Omega) \propto \sin (\pi\gamma/2)/|\Omega|^\gamma$ gets progressively smaller as $\gamma \to 2$.  We  speculated that there is the emergence of an infinite number of vortices  and a continuum spectrum of $E_{c,n}$ at $\gamma \to 2$ are related phenomena.

 The case $\gamma =2$ requires separate consideration and will be discussed in the next paper in the series. There we also discuss in detail longitudinal gap fluctuations.

  \acknowledgements
  We thank   I. Aleiner, B. Altshuler, E. Berg, R. Combescot, D. Chowdhury, L. Classen,  K. Efetov, R. Fernandes,  A. Finkelstein, E. Fradkin, A. Georges, S. Hartnol, S. Karchu, S. Kivelson, I. Klebanov, A. Klein, R. Laughlin, S-S. Lee, G. Lonzarich, D. Maslov, F. Marsiglio, M. Metlitski, W. Metzner, A. Millis, D. Mozyrsky,  C. Pepin, V. Pokrovsky,  N. Prokofiev,  S. Raghu,  S. Sachdev,  T. Senthil, D. Scalapino, Y. Schattner, J. Schmalian, D. Son, G. Tarnopolsky, A-M Tremblay, A. Tsvelik,  G. Torroba,  E. Yuzbashyan,  J. Zaanen, and particularly Y. Wang,  for useful discussions.   The work by  YW and AVC
   was supported by the NSF DMR-1834856.

\appendix

\section{The exact form of $\Delta_\infty (\omega_m)$}
\label{app:exact}

\subsection{The Linearized gap equation}

We choose the unit for frequency to be  $\bar{g}$ and re-write the linearized Eliashberg equation at zero temperature in dimensionless variable ${\bar \omega} = \omega/{\bar g}$ as
\begin{equation}
\Delta({\bar \omega})=\frac{1}{2}\int d{\bar \omega}^{\prime}\frac{\Delta({\bar \omega}^{\prime})-\Delta({\bar \omega})\frac{{\bar \omega}^{\prime}}{{\bar \omega}}}{\rvert {\bar \omega}^{\prime}\rvert}\frac{1}{\rvert {\bar \omega}-{\bar \omega}^{\prime}\rvert^{\gamma}},\label{eq:delta}
\end{equation}
  To obtain $\Delta({\bar \omega})$, we will explore the same strategy as in Paper I, where we found the exact solution for $\gamma <1$. For this, we introduce  $S({\bar \omega})=\text{sign}({\bar \omega})\left(\rvert{\bar \omega}\rvert - {1\over \gamma-1} \rvert{\bar \omega}\rvert^{1-\gamma} + i0^{+}\right)$ and
\beq
\tilde{\Phi} ({\bar \omega}) = S({\bar \omega}) \frac{\Delta({\bar \omega})}{{\bar \omega}}.
\label{eq:transf2-1}
\eeq
The choice of the sign of  $i0^{+}$ term in $S({\bar \omega})$ is arbitrary.  The result for $\Delta({\bar \omega})$ is the same for either sign.

The gap equation in terms of $\tilde{\Phi} ({\bar \omega})$ takes the form
\begin{equation}
\tilde{\Phi}({\bar \omega})=\frac{1}{2}\int d{\bar \omega}^{\prime}\left(\frac{\tilde{\Phi}({\bar \omega}^{\prime})}{S({\bar \omega}^{\prime})}-\frac{\tilde{\Phi}({\bar \omega})}{S({\bar \omega})}\right)\frac{\text{sign}({\bar \omega}^{\prime})}{\rvert{\bar \omega}-{\bar \omega}^{\prime}\rvert^{\gamma}}-\frac{{\bar \omega}-S({\bar \omega})}{S({\bar \omega})}\tilde{\Phi}({\bar \omega}).\label{eq:delta-1-1}
\end{equation}
Like for $\gamma <1$, we introduce
 complete and orthogonal basis, specified by
\begin{eqnarray}
\Phi_{\beta}(\Omega) & = & \frac{\rvert\Omega\rvert^{-2i\beta+\delta_{\Omega}}}{\rvert\Omega\rvert^{\gamma/2}},\label{eq:basis}
\end{eqnarray}
where $\beta\in[-\infty,\infty]$ and $\delta_{\Omega}=\delta\mbox{sign}(1-|\Omega|)$.
The functions $\Phi_{\beta}(\Omega)$ satisfy the orthogonality and completeness relations
\begin{eqnarray}
\int_{-\infty}^{\infty}\Phi_{\beta}^{*}(\omega)\Phi_{\beta}(\Omega)\frac{d\beta}{2\pi} & = & \frac{1}{2}\rvert\Omega\rvert^{1-\gamma}\delta(\rvert\Omega\rvert-\rvert\omega\rvert).\label{eq:complete-1}\\
\int_{-\infty}^{\infty}\Phi_{\beta}^{*}(\omega)\Phi_{\beta'}(\omega)\frac{d\omega}{|\omega|^{1-\gamma}} & = & 2\pi\delta(\beta-\beta').\label{eq:orthogonality-1}
\end{eqnarray}
The function $\tilde{\Phi}({\bar \omega})$ is expressed via $\Phi_{\beta}(\Omega)$ as
\beq
\tilde{\Phi}({\bar \omega})=\int_{-\infty}^{\infty}\frac{d\beta}{2\pi}a_{\beta}\Phi_{\beta}({\bar \omega}),
\label{eq:new_new}
\eeq
where
\beq
a_{\beta}=\int\frac{d{\bar \omega}}{\rvert{\bar \omega}\rvert^{1-\gamma}}\Phi_{\beta}^{\ast}({\bar \omega})\tilde{\Phi}({\bar \omega}).\label{eq:exp1}
\eeq
Multiplying both sides of Eq.~\ref{eq:delta-1-1} by $\Phi_{\beta}^{\ast}({\bar \omega})$
and integrating over $d{\bar \omega}/|{\bar \omega}|^{1-\gamma}$, we obtain the integral equation on
$a_{\beta}$ in the form
\beq
 a_{\beta}=\frac{1}{2}\int_{-\infty}^{\infty}\frac{d\beta'}{2\pi}A_{\beta,\beta'}a_{\beta'}-\int_{-\infty}^{\infty}\frac{d\beta'}{2\pi}B_{\beta,\beta'}a_{\beta'}\\
 \eeq
 where
 \begin{eqnarray}
 &  & A_{\beta\beta'}=\int\frac{d{\bar \omega}}{|{\bar \omega}|^{1-\gamma}}\int d{\bar \omega}^{\prime}\left(\frac{\Phi_{\beta^{\prime}}({\bar \omega}^{\prime})}{S({\bar \omega}^{\prime})}-\frac{\Phi_{\beta^{\prime}}({\bar \omega})}{S({\bar \omega})}\right)\frac{\text{sign}({\bar \omega}^{\prime})}{\rvert{\bar \omega}-{\bar \omega}^{\prime}\rvert^{\gamma}}\Phi_{\beta}^{\ast}({\bar \omega})\\
 &  & B_{\beta,\beta'}=\int\frac{d{\bar \omega}}{|{\bar \omega}|^{1-\gamma}}\frac{{\bar \omega}-S({\bar \omega})}{S({\bar \omega})}\Phi_{\beta^{\prime}}({\bar \omega})\Phi_{\beta}^{\ast}({\bar \omega})
\end{eqnarray}
One can verify that
$A_{\beta\beta^{\prime}}=(\gamma-1)B_{\beta\beta^{\prime}}F_{\beta}$, where
\begin{eqnarray}
F_{\beta}&=&\int_{-\infty}^{\infty}dx\frac{1}{\rvert x-1\rvert^{\gamma}}\left(|x|^{2i\beta+\gamma/2-1}-\text{sign}(x)\right) \nonumber \\
& = & \frac{2}{\gamma-1}\left(1-\epsilon_{\beta}\right),
\end{eqnarray}
and
\begin{equation}
\epsilon_{\beta}=\frac{1-\gamma}{2}\frac{\Gamma\left(\gamma/2-2i\beta\right)\Gamma\left(\gamma/2+2i\beta\right)}{\Gamma(\gamma)}\left(1+\frac{\cosh(2\pi\beta)}{\cos(\pi\gamma/2)}\right).
\end{equation}
Evaluating the integral for
$B_{\beta\beta^{\prime}}$, we obtain ($\text{P}\int$ means
Principal Value)
\begin{eqnarray}
 &  & B_{\beta\beta^{\prime}}={2 \over \gamma-1}\text{P}\int_{0}^{\infty}\frac{d{\bar \omega}}{{\bar \omega}}\frac{{\bar \omega}^{2i(\beta-\beta^{\prime})+\delta_{{\bar \omega}}}}{ {(\gamma-1)} {\bar \omega}^{\gamma}- i 0^+} \label{app_1} \\
 &  & =\frac{2i\pi}{\gamma }\frac{e^{-2\pi(\beta-\beta^{\prime}) /\gamma -2i(\beta-\beta^{\prime})/\gamma \log (\gamma-1) } }{\sinh(2\pi(\beta-\beta^{\prime})/\gamma-i0^{+})},
\end{eqnarray}
The convergence factor $\delta_{{\bar \omega}}$ in the first line in (\ref{app_1}) is  only relevant at small ${\bar \omega}$, where it is positive.
 Substituting $A_{\beta\beta^{\prime}}$ and $B_{\beta\beta^{\prime}}$ into the  integral  equation on $a_\beta$, we obtain
\begin{equation}
a_{\beta}=-\frac{2i\pi \epsilon_{\beta} }{\gamma} \int\frac{d\beta^{\prime}}{2\pi} \frac{e^{-2\pi(\beta-\beta^{\prime})/\gamma -2i(\beta-\beta^{\prime})/\gamma \log (\gamma-1) }}{\sinh(2\pi(\beta-\beta^{\prime})/\gamma-i0^{+})}a_{\beta^{\prime}}.
\end{equation}
It is convenient to  change the variables to $\beta=\gamma k/2$, $a_{\beta}=e^{-\pi k - i k \log (\gamma-1)} \epsilon_{\gamma k/2} \tilde{b}_{k}$. The gap equation then takes the form
\begin{equation}
\tilde{b}_{k}=\frac{i}{2}\int\frac{\epsilon_{k'\gamma/2}} {\sinh(\pi(k^{\prime}-k)+i0^{+})}\tilde{b}_{k'}dk',\label{eq:bk1}
\end{equation}
  Once we know $\tilde{b}_{k}$, one can compute $\tilde{\Phi}({ \omega})$ and the gap function $\Delta({ \omega})$ using
Eqs. \eqref{eq:new_new} and \eqref{eq:transf2-1}.

\subsection{Solution of the gap equation}

Let's define the function $B_z$ of a complex argument  $z$ as
\begin{equation}
B_{z}=\frac{i}{2}\int\frac{\epsilon_{k^{\prime}\gamma/2} }{\sinh(\pi (k^{\prime}-z))}b_{k'}dk^{\prime}.
\end{equation}
 For real $k$,
 $\tilde{b}_{k}=B_{k-i0^{+}}$
 The function $B_{z}$
  satisfies the periodicity condition
\begin{equation}
B_{z+in}=(-1)^{n}B_{z},
\label{app_3}
\end{equation}
where  $n$ integer, and has  branch cuts
at $z=x+in$, where  $x$ is real.
 In particular, for $n=0$,
\begin{equation}
B_{k+i0^{+}}=\left(1-\epsilon_{k\gamma/2} \right)B_{k-i0^{+}}.
\label{app_2}
\end{equation}
 We now take the
logarithm of both sides of (\ref{app_2}). For $\gamma>1$, the function $\epsilon_{k\gamma/2}$
increases monotonically with $|k|$, and $\epsilon_{k\gamma/2}=1$ at $k= k_\gamma =\pm \beta$.
Using this, we obtain
\begin{equation}
\log B_{k+in+i0^{+}}-\log B_{k+in-i0^{+}}=\log\rvert1-\epsilon_{k\gamma/2}\rvert+i\pi\chi_{n}\Theta(k-\beta)+
i\pi\xi_{n}\Theta(-k-\beta).
\label{app_4}
\end{equation}
Here $\chi_{n}, \xi_{n}=\pm1$ are two parameters, which reflect the ambiguity of
evaluating  $\log{(1-\epsilon_{k\gamma/2})}$ for $\rvert k\rvert>\beta$ due to a discontinuity of a logarithm across its branch cut.
 Because of the periodicity condition (\ref{app_3}),
the l.h.s. of (\ref{app_4}) is independent to $n$, hence $\chi_{n}\equiv\chi$ and $\xi_{n}\equiv\xi$ are just two numbers, each is either $+1$ or $-1$.
Using the Sokhotski-Plemelj theorem, we can then re-express (\ref{app_4}) as
\begin{eqnarray}
\log B_{k} & = & \frac{1}{2i}\int_{-\infty}^{\infty}\log\rvert1-\epsilon_{k\gamma/2}\rvert\coth\left(\pi(k^{\prime}-k)\right)dk^{\prime}+\frac{\pi\chi}{2}\int_{\beta}^{\infty}\coth\left(\pi(k^{\prime}-k)\right)dk^{\prime}\nonumber \\
 &  & +\frac{\pi\xi}{2}\int_{-\infty}^{-\beta}\coth\left(\pi(k^{\prime}-k)\right)dk^{\prime}+G_{k},
\label{app_5}
\end{eqnarray}
where $G_{k}$ is an analytic function, which satisfies
the periodicity relation
\begin{eqnarray}
G_{k+in} & = & G_{k}+(2l+1)in\pi,~~l,n\in\mathbb{Z}.
\label{app_6}
\end{eqnarray}
The last condition uniquely  specifies $G_k =c+(2l+1)\pi k$, where $c$ is a constant.  The latter is irrelevant as  one can easily verify that it only contributes to an irrelevant overall factor in $\Delta_\infty ({\bar \omega})$.
Using (\ref{app_6}) and absorbing the divergent piece into $c$, we obtain
   from (\ref{app_5}):
   \begin{eqnarray}
\tilde{b}_{k}=B_{k-i0^{+}} & = & \exp\Bigg[\frac{1}{2i}\int_{-\infty}^{\infty}\log\rvert1-\epsilon_{k\gamma/2}\rvert\coth\left[\pi(k^{\prime}-k + i 0^+)\right]dk^{\prime}\nonumber \\
 &  & -\frac{\chi}{2}\log\frac{\sinh[\pi(k_{\gamma}-k+i0^{+})]}{\sinh(\pi k_{\gamma})}+\frac{\xi}{2}\log\frac{\sinh[\pi(k_{\gamma}+k-i0^{+})]}{\sinh(\pi k_{\gamma})}\nonumber \\
 &  & +(2l+1-\frac{\chi+\xi}{2})\pi k \Bigg].
\label{app_7}
\end{eqnarray}
This $\tilde{b}_{k}$ depends on three parameters: $\chi, \xi = \pm 1$ and $l\in\mathbb{Z}$.
Below we select  $\chi,\xi$, and $l$ based on the two conditions: (1) the integrand in the r.h.s. of the gap equation (\ref{eq:bk1})
must be convergent; (2)  $b_{k}$ must give rise to
a gap function $\Delta({\bar \omega})$, which vanishes at ${\bar \omega} \to \infty$.
  The first condition  puts the following constraints:
\begin{itemize}
\item If $\chi=\xi=1$, then $l=0$.
\item If $\chi=\xi=-1$, then $l=-1$.
\item If $\chi=-\xi=1$, then $l=0,-1$.
\item If $\chi=-\xi=-1$, there is no solution for $l$.
\end{itemize}
We verified that the second condition is satisfied only if
 $\chi=-\xi=1$ and $l=0,-1$. The dependence on $l$ in (\ref{app_6}), and the difference between the two choices for $l$ is an irrelevant overall factor for $\Delta({\bar \omega})$.   We then concludes that the two conditions uniquely specify ${\tilde b}_k$.

  Substituting these $\chi, \xi$, and $2l+1$ into (\ref{app_6}) and re-expressing the result back in terms of
 $b_k = e^{-\pi k- i k \log (\gamma-1)}   {\tilde b}_{k}$, we obtain
   \begin{eqnarray}
b_{k} & = &  \exp\left[ -i k \log (\gamma-1) \right] \exp\Bigg[\frac{1}{2i}\int_{-\infty}^{\infty}\log\rvert1-\epsilon_{k^{\prime}\gamma/2}\rvert\coth[\pi(k^{\prime}-k+
i0^{+})]dk^{\prime}\nonumber \\
 &  & -\frac{1}{2}\log\frac{\sinh[\pi(k_{\gamma}-k+i0^{+})]}{\sinh[\pi k_{\gamma}]}-\frac{1}{2}\log\frac{\sinh[\pi(k_{\gamma}+k-i0^{+})]}{\sinh[\pi k_{\gamma}]} \Bigg].
\end{eqnarray}

\subsection{Computation of $\Delta_{\infty}({\omega})$}

We now substitute this $b_{k}$ into
Eq. (\ref{eq:new_new})
and compute
 $\tilde{\Phi}({ \omega})$ and the gap function $\Delta({ \omega})$ using
 Eq. \eqref{eq:transf2-1}.
  It is convenient to
  introduce $y \equiv \rvert {\bar \omega} \rvert^{\gamma}$. In terms of $y$, we obtain
\begin{eqnarray}
\Delta_{\infty}(y) =  y^{1/2} \int_{-\infty}^{\infty} dk b_{k}e^{-i k \log y}.
\end{eqnarray}
The $y^{1/2}$ in the prefactor can be eliminated by shifting the integration contour  away from the real axis, to $k - i/2$. This is a safe procedure because $b_k$ is analytic within the interval $-1 < \text{Im}(k) < 0$. Shifting the integration, we obtain the final expression
 \beq
 \Delta_\infty (y) = \int_{-\infty}^\infty dk b_k e^{-ik \log y },
\label{app_8}
 \eeq
where
  \beq
  b_k = \frac{e^{-i I_k -i k \log (\gamma-1)}}{ \sqrt{ \cosh(\pi (k-\beta))\cosh(\pi (k+\beta)) } }
  \eeq
and
  \beq
  I_k = \frac{1}{2} \int_{-\infty}^\infty dk' \log{|\epsilon_{k'} -1|} \tanh{\pi (k'-k)}.
  \label{eq:ik2}
  \eeq

 \subsection{Series expansion for $\Delta_\infty (y)$}

The function $I_k$ can be expressed  as an infinite product of the Gamma-functions (see Paper I for details). For $b_k$, this yields, up to an overall factor,
 \beq
 b_k = \frac{\Gamma(1-ik)}{\Gamma(1+ik)} \Gamma\left(\frac{1}{2} + i (k+ \beta)\right)\Gamma\left(\frac{1}{2} + i (k- \beta)\right) \prod_{m=1}^{\infty} \frac{\Gamma\left(\frac{1}{2} + i (k -i \beta_m)\right) \Gamma\left(1+ \frac{2m}{\gamma} - ik\right)}{\Gamma\left(\frac{1}{2} - i (k +i \beta_m)\right) \Gamma\left(1+ \frac{2m}{\gamma} + ik\right)}
\label{nn_4}
 \eeq
 Here $\beta_m >0$ are the solutions of $\epsilon_{i\beta_m} =1$ for imaginary argument.  There is  an infinite set of such $\beta_m$, specified by integer $m =0,1,2..$ and located at  $1/2 + 2m/\gamma <b_m < 1/2 + 2(m+1)/\gamma$ (see Fig.~\ref{fig:epsilon} (b) in the main text).

 The integral in (\ref{app_8}) can be evaluated by closing the integration contour along an infinite arc in the complex plane of frequency.  For $y <1$, i.e., $\omega_m < {\bar g}$, the arc must be in the upper half-plane, and for $y >1$, i.e.,  $\omega_m > {\bar g}$, in the lower half-plane.   Viewed as a function of complex $k$, $b_k$ has poles from individual $\Gamma$-functions in the upper frequency half-plane at $k = \pm \beta + i (n+1/2)$, where $n =0,1,2..$ and at
 $k = i \beta_m + i(n+1/2)$, and in the lower half-plane, at $k = -i(n+1)$ and $k = -i(1+2m/\gamma +n)$,
 where $n=0,1,2,...$ and $m=1,2,...$.

 \subsubsection{$y<1$}

\begin{figure}
  \includegraphics[width=15cm]{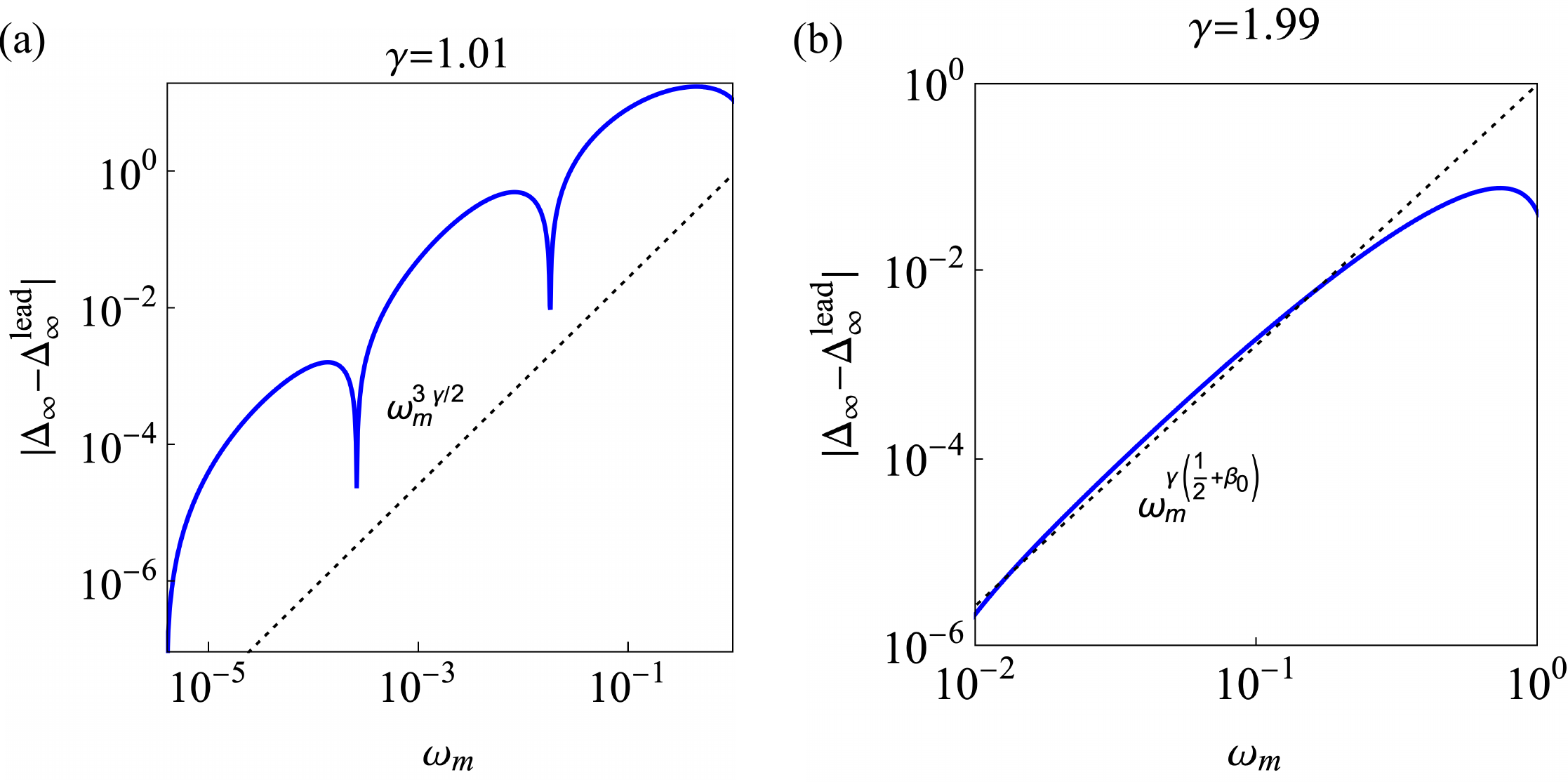}
  \caption{The expansion of the exact $\Delta_{\infty} (\omega_m)$ in powers of $\omega_m$.
   The leading term at small $\omega_m$ is $\Delta^{{\text lead}}_{\infty} |\omega_m|^{\gamma/2} \cos{(\beta \gamma \log{|\omega_m|/{\bar g}} + \phi)}$. The subleading terms form  series of local and non-local terms. Local series hold in powers of
    $|\omega_m|^{n\gamma}$ ($n=1,2...$), and all terms in these series oscillate, like the leading term.  The non-local terms do not oscillate and form series in  $|\omega_m|^{(n' + \beta_m) \gamma}$, where $n' = (0,1,2..)$ and $\beta_m$ ($m=0,1,2..$) are some $\gamma-$dependent numbers.  We show $|\Delta^\infty - \Delta^{{\text lead}}_\infty|$ for $\gamma =1.01$ and $\gamma =1.99$.
      An analytical analysis shows that in the first case, $\beta_0 > 1$, such that the leading correction comes from the local series and oscillates. For $\gamma =1.99$, $\beta_0 \simeq 0.9 < 1$. In this case  the leading correction comes from non-local series and does not oscillate.  The numerical evaluation of $|\Delta^\infty - \Delta^{{\text lead}}_\infty|$ shown here, confirms these results. }
      \label{fig:small_y}
\end{figure}

  For $y<1$, relevant poles are at  $k = \pm \beta + i (n+1/2)$ and at
 $k = i \beta_m + i(n+1/2)$. This yields series expansion for $\Delta_\infty (y)$ in the form
\begin{equation}
    \Delta_\infty (y) =  \text {Re}~ \sum_{n=0}^{\infty } e^{i( \beta \log{y} + \phi)} C^{<}_n~ y^{n+1/2}
+  \sum_{n,m=0}^{\infty }D^{<}_{n,m} y^{n + b_m/\gamma+1/2}
\label{app_9}
\end{equation}
We cited this result in Eq. (\ref{nn_6}) in the main text. Here $\phi$ is some particular, $\gamma-$dependent number, and $ C^{<}_n$ and $D^{<}_{n,m}$ are $\gamma$-dependent coefficients.
 The leading term in (\ref{app_9}) at small $y$ is
 \beq
   \Delta_\infty (y) = y^{1/2} C^{<}_n \cos{\left(\beta \log{y} + \phi\right)}
\label{app_10}
\eeq
The first subleading term scales as $y^{3/2}$ for $\gamma \leq 1.81$, where $\beta_0/\gamma >1$, and as $y^{1/2+\beta_0/\gamma}$ for $\gamma$ closer to  $2$, where $\beta_0/\gamma <1$.
We verified this explicitly by subtracting the leading term,  given by (\ref{app_10}), from the exact solution and identifying the leading term in the leftover
 (see Fig.~\ref{fig:small_y}).

\begin{figure}
	\includegraphics[width=16cm]{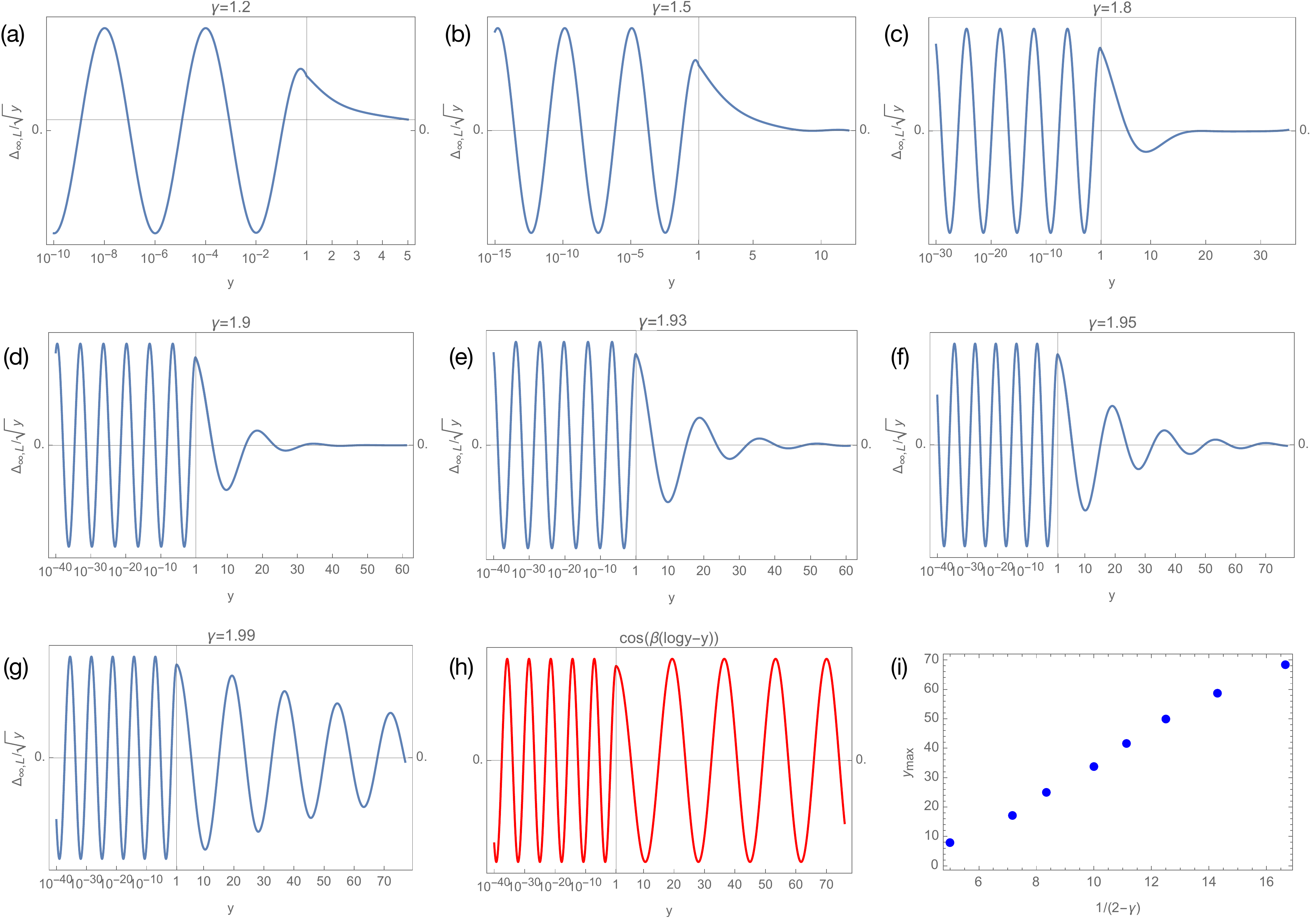}
	\caption{Panels (a)-(g). The gap function $\Delta_{\infty,L}(y)$ as a function of $y = (|\omega_m|/{\bar g})^\gamma$ for various $\gamma$. At $y <1$, $\Delta_{\infty,L}(y)$ oscillates for all $\gamma$  with the period  set by $\log y$. As $\gamma$ increases towards $2$, the new oscillating regime emerges if one extends $\Delta_{\infty,L}(y)$ to $y>1$. In this new regime the period of oscillations is set by $y$. (h) The analytical form of $\Delta_{\infty,L}(y)$ at $\gamma \to 2$.  (i)
 The approximately linear dependence of the upper boundary for new oscillations, $y^*$, on $2-\gamma$.}
\label{fig:solutions}
\end{figure}

 In the direct perturbation expansion in $y$, the series in $y^n$ (the first term in (\ref{app_9}))
  come from fermions with internal $y' \sim y$ and form the "local" series. The second term in (\ref{app_9})) is a sum of  contributions from fermions with  $y' = O(1)$, which for $y \ll 1$ can we regarded as  "non-local".  We can then express $\Delta_\infty (y)  =    \Delta_{\infty,L} (y) +    \Delta_{\infty,NL} (y)$, where
 \begin{eqnarray}
  &&    \Delta_{\infty,L} (y) =  \text{ Re}~ \sum_{n=0}^{\infty } e^{i( \beta \log{y} + \phi)} C^{<}_n~ y^{n+1/2} \label{eq:x<1_1} \\
   &&    \Delta_{\infty,NL} (y) =
 \sum_{n,m=0}^{\infty }D^{<}_{n,m} y^{n + b_m/\gamma+1/2}
\label{eq:x<1_1_1}
\end{eqnarray}
   The coefficients
  $C^{<}_n$ in (\ref{eq:x<1_1}) can be obtained analytically, as we already found in Papers I and III for $\gamma <1$. The computations for $\gamma >1$ are similar, and we only present the result. We obtained
 \beq
  C^{<}_n = C^{<}_0 \displaystyle\prod_{m=1}^{n} \frac{1}{{\bar I}_m},
  \label{dd_10a}
  \eeq
  \bea
&&{\bar I}_m  =  \frac{1}{2} \left[\frac{\Gamma((m+1/2)\gamma +i{\beta} \gamma) \Gamma((1/2-m)\gamma -i{\beta} \gamma)}{\Gamma(\gamma)}  - \frac{\Gamma(\gamma(1/2 +i{\beta})) \Gamma(\gamma(1/2 -i\beta))}{\Gamma(\gamma)}\right] +  \nonumber \\
&&\frac{\Gamma(1-\gamma)}{2} \left(\frac{\Gamma((m+1/2)\gamma +i{\beta} \gamma)}{\Gamma(1-(1/2-m)\gamma +i{\beta} \gamma)} + \frac{\Gamma((1/2-m)\gamma -i{\beta} \gamma)}{\Gamma(1-(m+1/2)\gamma -i{\beta} \gamma)} \right) \nonumber \\
&&-\frac{\Gamma(1-\gamma)}{2} \left(\frac{\Gamma(\gamma(1/2 +i{\beta}))}{\Gamma(1-\gamma(1/2-i{\beta}))} + \frac{\Gamma(\gamma (1/2-i{\beta}))}{\Gamma(1-\gamma(1/2+i\beta))}\right)
\label{dd_7b}
\eea
Using the expansion for $\Gamma$-functions, one can verify that at large $m$, ${\bar I}_m \sim m^{\gamma-1}$, i.e.,
$C^{<}_n \sim 1/n^{n(\gamma-1)}$.
 Interestingly, the sum in (\ref{eq:x<1_1}) then converges for any $y$ and can be obtained numerically by summing up the proper number of terms in (\ref{eq:x<1_1}).
 We show the results for $\Delta_{\infty,L} (y)$ for different $\gamma$ in Fig. (\ref{fig:solutions}), extending also into the range $y >1$.
At $\gamma \leq 2$, one can expand the $\Gamma$ functions in (\ref{dd_7b}) in $2-\gamma$.
 For ${\bar I}_m$ we then obtain, to first order in $2-\gamma$,
\beq
{\bar I}_m = \frac{im}{\beta} \left(1  + i Q(2-\gamma) (m+1)\right)
\label{4_3}
\eeq
 where  $Q\approx 0.7$.
This holds for  $Q (2-\gamma) m  < 1$. For larger $m$, $Q(2-\gamma) m$ becomes $\sin(Q(2-\gamma) m)$.
 Substituting ${\bar I}_m$ from (\ref{4_3}) into  (\ref{dd_10a}), evaluating the product to first order in $(2-\gamma)$, and substituting the result into (\ref{eq:x<1_1}), we obtain, to the same accuracy,
 \beq
\label{eq:x<1_3}
\Delta_{\infty,L} (y) \propto  \sqrt{y} e^{-2Q (2-\gamma) \beta y}
 \cos{(\beta(\log y -y) - Q (2-\gamma) (y \beta)^2/2) + \phi)}
  \eeq
 The expression becomes particularly simple for $\gamma =2$, where
  \beq
\label{eq:x<1_3_1}
\Delta_{\infty,L} (y) \propto  \sqrt{y}
 \cos{(\beta(\log y -y) + \phi)}
  \eeq

\subsubsection{$y >1$}

For $y >1$,
relevant poles are in the lower half-plane, at $k = -i (n+1)$ and $k = -i (n+1+ 2m/\gamma)$.
  This yields the expansion for $\Delta_\infty (y)$ in powers of $1/y$ in the form
  \begin{equation}
 \Delta_\infty (y)= \sum_{n=0}^\infty C^{>}_n~ \left(\frac{1}{y}\right)^{n+1} +
 \sum_{n,m=0}^{\infty }D^{>}_{n,m} \left(\frac{1}{y} \right)^{n+1 + 2(m+1)/\gamma}
\label{app_11}
\end{equation}
 We cited this result in Eq. (\ref{nn_7}) in the main text.
The leading term in the series is $1/y$, i.e., at $\omega_m \gg {\bar g}$,  $\Delta_\infty (\omega) \propto 1/|\omega|^\gamma$. The exact
$\Delta_\infty (\omega)$ clearly shows this behavior (see Fig.~\ref{fig:Delta_infty_Mats})

 Eq. (\ref{app_11}) is formally the same as the  series expansion result for $\gamma <1$, however the
  coefficients  $C^{>}_n$ and $D^{>}_{n,m}$  depend on $\gamma$.
 We now argue that this dependence is qualitatively different for $\gamma <1$ and $\gamma >1$. Namely, we argue below that for $\gamma >1$, there is a universal piece in $ \Delta_\infty (y)$ at large $y$, which does not fit into the power-law series in (\ref{app_11}).  We obtain this piece by analyzing the form of $b_k$ in Eq. (\ref{app_8}) and evaluating the integral over $k$ directly, without closing the itegation contour in the complex plane of $k$. While we didn't compute $C^{>}_n$ and $D^{>}_{n,m}$ explicitly,
 the presence of a piece  that does not fit into power-law series implies that
  the series expansion in (\ref{app_11}) does not converge at $\gamma >1$ (e.g., the coefficients
   $C^{>}_n$  grow fast enough with $n$ and eventually overcome the smallness of $1/y^n$) and would
    yields incorrect results, starting from some critical $n$, which depends on $y$.  In Paper V, where we specifically focus on $\gamma =2$, we show explicitly that this is the case and determine a critical $n$.

Below we focus on the  universal, non-power-law term in $\Delta_\infty (y)$. We show that this term  is present at $\gamma >1$,
 because the complex phase of the integrand in Eq.~(\ref{app_8})  has an extreme  at $k=k_* \sim y^{1/(\gamma-1)}$.
 The universal contribution then comes from the expansion of the integrand around $k = k^*$. There is no such extreme for $\gamma <1$.

   To identify the  universal term, we
consider large $y$
and  analyze the contribution to the integral in (\ref{app_8}) from $k$ above some $k_{min} =  {\cal O}(1)$.
    We label the corresponding term as
$ \Delta_{\infty}^{u} (y)$.
In explicit form,
  \begin{align}
  \Delta_{\infty; u} (y)
& = \int_{k_{\text{min}}}^{\infty} dk {\cos{(I_k + k \log y(\gamma-1))} \over (\cosh \[\pi (k-\beta)\] \cosh \[ \pi (k+\beta) \] )^{1/2}}.
  \label{eq:delta_I}
  \end{align}
 For large $k$,  the integral, which determines $I_k$ in Eq.~(\ref{eq:ik2}), is determined by $k^{\prime}\sim k$.   This integral contains ${\epsilon}_{\gamma k'/2}$.  For $\gamma >1$, ${\epsilon}_{\gamma k'/2}$ is an increasing function of $k'$ and its leading term at large $k'$  is
 ${\epsilon}_{\gamma k'/2} \simeq (\gamma-1)(A_{\gamma}k')^{\gamma-1}$, where $A_{\gamma} = \gamma  \left( \frac{\pi }{2 \Gamma(\gamma)  \cos(\pi (2-\gamma)/2)} \right)^{1/(\gamma-1)}$. Substituting this form into
 Eq.~(\ref{eq:ik2}), we  obtain
\begin{eqnarray}
I(k) + \log{[y(\gamma-1)]} \simeq  (\gamma-1)k\ln\frac{y e}{A_{\gamma} k},
\label{eq:ik}
\end{eqnarray}
Substituting this form into ~(\ref{eq:delta_I}), we find that the argument of the integrand has a maximum at
\begin{equation}
k_{*} =y^{\frac{1}{\gamma-1}} A_{\gamma}^{-1}.
\end{equation}
 Expanding $I(k)+k \log \[y(\gamma-1)\]$ to quadratic order in $u = k/k_{*}-1$, as
\begin{equation}
I(k)+k \log \[ y (\gamma-1)\] \approx  (\gamma-1)k_* \left( 1 - \frac{1}{2} u^{2} \right),
\end{equation}
substituting into (\ref{eq:delta_I}), and approximating the denominator in (\ref{eq:delta_I}) by its large $k$ value, we obtain
\beq
    \Delta_{\infty}^{u} (y) \sim \sqrt{k_*} e^{-\pi k^*} \text{Re}\left[ e^{-i (\gamma-1)k_* \left(1 - \frac{\pi^2}{(\gamma-1)^2}\right)} \int_{-[k^*(\gamma -1)/2]^{1/2}[1- i \pi/(\gamma-1)] }^{\infty} d{\tilde u}  e^{ i {\tilde u}^2} \right].
 \label{app_12}
\eeq
   where ${\tilde u} = u ((\gamma -1) k^*/2)^{1/2}$.
  The universal part of this expression is obtained by taking the lower limit to $-\infty$, in which case
  $\int_{-\infty}^\infty e^{i {\tilde u}^2 }=  \sqrt{\pi}e^{i\pi/4}$.
  Substituting into (\ref{app_12}) and expressing $k^*$ in terms of $|\omega_m|/{\bar g}$,
   we obtain for the universal term
 \begin{align}
\Delta_{\infty}^{u}
(\omega_m)  \sim & \left({\rvert \omega_m \rvert \over {\bar g}}\right)^{{\gamma \over 2(\gamma-1)} } \exp \left[ -\frac{\pi}{A_{\gamma}} \left({\rvert \omega_m \rvert \over {\bar g}}\right)^{{\gamma \over \gamma-1} }  \right] \nonumber \\
& \cos\left[\frac{\gamma-1}{A_{\gamma}} \left(1 - \frac{\pi^2}{(\gamma-1)^2}\right) \left({\rvert \omega_m \rvert \over {\bar g}}\right)^{{\gamma \over \gamma-1} } - \frac{\pi}{4}\right].
\end{align}
Working along the same lines, but extending the analysis to include the subleading terms at large $k$, we obtain
$\Delta_{\infty}^{u}(\omega_m)$
with a more accurate argument of the  $\cos$:
\beq
\Delta_{\infty}^{u}
(\omega_m) \sim  \left({\rvert \omega_m \rvert \over {\bar g}}\right)^{{\gamma \over 2(\gamma-1)} } \exp \left({-\frac{\pi} {A_{\gamma}} \left({\rvert \omega_m \rvert \over {\bar g}}\right)^{{\gamma \over \gamma-1} } } \right) \cos{\Psi_m (\omega_m/{\bar g})}
\label{app_14}
\eeq
 where
 \beq
 \Psi_m (x) = \frac{\gamma-1}{A_{\gamma}}  \left[ 1   -  {1\over 2} \left( {\pi \over \gamma-1} \right)^2 \right] x^{{\gamma \over \gamma-1} } + \frac{\pi^{2}}{2(\gamma-1) A_{\gamma}} x^{\frac{\gamma (2-\gamma)} {\gamma-1}}
+  \frac{ x^{ {\gamma (2-\gamma) \over \gamma-1}} -1 } {2-\gamma} A_{\gamma}^{1-\gamma} - \frac{\pi}{4}.
 \label{app_15}
\eeq
At $\gamma \to 2$, $\Psi (x)$ acquires a simple form $\Psi (x) = (x^2 (1- \pi^2/2) + \log{x^2})/\pi + \pi/4$.

 \subsection{Analytical continuation}

The gap function given by (\ref{app_8}) can be analytically continued away from the Matsubara axis
by a simple rotation of the argument:  $i\omega_m \to z =  |z| e^{i\psi}$.
Under this transformation, $\log y$ in (\ref{app_8}) transforms into
$ \log {y_z}  - i \theta$, where $y_z = (\rvert z\rvert/{\bar g})^{\gamma}$ and
$\theta = (\pi/2 - \psi) \gamma$. The gap function transforms to
 \beq
 \Delta_\infty ( z ) = \int_{-\infty}^\infty dk  \frac{e^{-\theta k} e^{-i I_k - i k \log y_z}}{ \sqrt{ \cosh(\pi (k-\beta ))\cosh(\pi (k+\beta)) } } .
 \label{gap_z}
 \eeq
Applying this transformation, we extend (\ref{app_15}) to
\beq
\Delta_{\infty}^{u}
(z) \sim  \left({\rvert \omega_m \rvert \over {\bar g}}\right)^{{\gamma \over 2(\gamma-1)} } \left(Q(\theta, |z|) + Q(-\theta, |z|)\right)
\label{app_16}
\eeq
where
\beq
Q(\theta,|z|) =  \exp \left({-\frac{\pi-\theta} {A_{\gamma}} \left({\rvert z \rvert \over {\bar g}}\right)^{{\gamma \over \gamma-1} } } \right) e^{i\Psi (|z|/{\bar g}, \theta)}
\eeq
and
\begin{align}
 \Psi (x,\theta) = &
 \frac{\gamma-1}{A_{\gamma}}  \left( 1   -  {1\over 2} \left( {\pi-\theta \over \gamma-1} \right)^2 \right) x^{{\gamma \over \gamma-1} } +  \frac{(\pi-\theta)^{2}}{2(\gamma-1) A_{\gamma} }  x^{ {\gamma (2-\gamma) \over \gamma-1}  }  + \frac{ x^{ {\gamma (2-\gamma) \over \gamma-1}  } -1 } {2-\gamma} A_{\gamma}^{1-\gamma}
\end{align}
The largest value of $\theta$ is on the real axis, where $\psi =0$ and $\theta = \pi \gamma/2$.
 Here, $Q(\theta,|z|) \gg Q(-\theta,|z|)$.
 Keeping only $Q(\theta,|z|)$, we obtain on the real axis,
  \begin{align}
\Delta_{\infty}^{u}
(\omega) \sim & \left({ \rvert \omega \rvert \over {\bar g}}\right)^{{\gamma \over 2(\gamma-1)} } \exp \left( -\frac{\pi(2-\gamma)}{2A_{\gamma}} \left({\rvert \omega \rvert \over {\bar g}}\right)^{{\gamma \over \gamma-1} } \right) e^{i \Psi_r (|\omega|/{\bar g})}
\label{app_17}
\end{align}
where
\begin{align}
\Psi_r (x) = \frac{\gamma-1}{A_{\gamma}} \left(1-\frac{\pi^{2}(2-\gamma)^{2}}{8(\gamma-1)^2}\right) x^{{\gamma \over \gamma-1} } +  \frac{\pi^2 (2-\gamma)^{2}}{8(\gamma-1) A_{\gamma} } x^{ {\gamma (2-\gamma) \over \gamma-1}} + \frac{x^{ {\gamma (2-\gamma) \over \gamma-1}  } -1 } {2-\gamma} A_{\gamma}^{1-\gamma}
\end{align}
 Comparing this  oscillating
 $\Delta_{\infty}^{u}(\omega)$ with the regular term  on the real axis $e^{i\pi \gamma/2} ({\bar g}/|\omega|)^\gamma$, we see that for $\gamma \leq 2$, the oscillating term is larger in a range
 ${\bar g} <|\omega| < \omega_{cr}$, where $\omega_{cr} \sim {\bar g}/(2-\gamma)^{1/2}$ has been defined in
 (\ref{nn_6_1_1}).

  At $\gamma \to 2$, $\omega_{cr}$ diverges  and
$\Delta_{\infty}^{u}(\omega)$
from (\ref{app_17}) remains the dominant  term in $\Delta_{\infty}(\omega)$  at all $\omega > {\bar g}$. In this limit,
$\Delta_{\infty}^{u}(\omega) =\Delta_{\infty}(\omega)$
simplifies to
 \begin{align}
\Delta_{\infty}(\omega) & \sim  { \rvert {\omega} \rvert \over {\bar g}} \exp \left\{ \frac{i}{\pi} \left[ \left( {{\omega} \over {\bar g}}\right)^{2}+ \log  \left( {{\omega} \over {\bar g} }\right)^2 \right] \right\}.
\end{align}

\bibliography{gamma_between_1_and_2}

\end{document}